\documentclass[12pt,a4paper]{article}
\usepackage{graphicx}
\usepackage{placeins}

\newtheorem{theorem}{Theorem}
\newtheorem{acknowledgement}[theorem]{Acknowledgement}

\input{tcilatex}
\begin{document}

\title{Field theoretical approach to the description of the coherent
structures in $2D$ fluids and plasmas}
\author{F. Spineanu and M. Vlad \\
National Institute of Laser, Plasma and Radiation Physics \\
MG-36, Magurele, Bucharest, Romania}
\date{}
\maketitle

\begin{abstract}
Evolving from turbulent states the $2D$ fluids and the plasmas reach states
characterized by a high degree of order, consisting of few vortices. These
asymptotic states represent a small subset in the space of functions and are
characterised by properties that are difficult to identify in a direct
approach. The field theoretical approach to the dynamics and to the
asymptotic states of fluids and plasmas in $2D$ provides a considerable
extension of the usual perspective. The present works discusses a series of
consequences of the field theoretical approach, when it is applied to
particular problems. The discussion is developed around known physical
problems: the current density profiles in cylindrical plasma, the density
pinch in tokamak and the concentration of vorticity.
\end{abstract}

\tableofcontents

\section{Introduction}

The fluids and plasma exhibit in two-dimensions a strong tendency to
self-organisation in the undriven evolution towards stationary states. As
shown by experiments and numerical simulation, the $2D$ fluids and plasmas
reach states of high coherency of the flow, generated by concentration of
vorticity in few large scale vortical flows. The process is essentially
non-dissipative since the energy is almost conserved during this process.
The presence of dissipation is however essential since breaking up of
streamlines and reconnection into larger structures is only possible in the
presence of irreversible resistive-like mechanisms. The vortex merging is
the typical process and a large number of studies have been done both
experimentally and by numerical simulation. When the initial state is
turbulent one may invoke arguments related with the inverse cascade of
energy in $2D$ but this approach has limitted relevance when the process has
evolved to the point where the number of cuasi-coherent structures is large:
the high number of irreducible correlations necessary to describe a
statistical ensemble of turbulence with embedded structures invalidates any
perturbative considerations. 

The problem of evolution towards the asymptotic coherent flow states is very
complex and possibly different approaches must be developed to examine
different aspects of it. For the states close to the final, organised one,
the field theoretical formulation seems adequate. This is confirmed by the
purely analytic derivation of the \emph{sinh}-Poisson equation (describing
the asymptotic states of the Euler fluid) and by the derivation of a new
equation, for plasma and planetary atmosphere vortices, with substantial
practical confirmation.

The particular nature of the physical systems that are investigated is
reflected in construction of the Lagrangian of the field theoretical model:
for point-like vortices we have to adopt an algebraic formulation of the
fields, namely $su\left( 2\right) $. For short range interaction we need the
Higgs mechanism to generate a mass for the photon, and this imposes a
particular form for the scalar matter self-interaction. Finally, in
particular cases it appears possible that the $su\left( 2\right) $ structure
is reduced to the Abelian substructure (this is possibly similar to the 
\emph{Abelian dominance}) and the nature of the extrema of the action
functional changes significantly.

In the present work we apply the results of the models developed before.
Since these models lead to equations describing the stationary states
(flows) of lowest energy of the physical systems, the application should
consists basically in solving these equations and confronting with
experiments or observations. However some of these equations are already
known and eventually they have been derived from different considerations,
like the statistical theory. We will make a comparison between our approach
and the statistical one.

The following partial conclusions seem to be supported by the analysis of
the applications that are presented below.

\begin{enumerate}
\item the field theoretical model of the current profiles in tokamak is
compatible with the Liouville equation. Comparison with the model of J. B.
Taylor gives interesting suggestions for the physical interpretation of the
FT parameters.

\item for the Euler fluid we obtain in FT a possible confirmation of the
existence of a current of vorticity leading to concentration into filaments.

\item for the $2D$ plasma in strong magnetic field we obtain patterns of
vortical flows that confirm previous numerical calculations.

\item for the LH transtion, we obtain, after a renormalisation of the Larmor
radius into an \emph{effective} Larmor radius, profiles of electric fields
at the edge (in H mode) that are compatible with the experiments

\item for the density pinch we are able to build a physical picture that is
consistent with the idea that the pinch of density is due to a pinch of
vorticity.

\item for the $2D$ atmosphere we obtain quantitative results that compares
(very) well with the observations of \emph{tropical cyclone}.

\item for the \emph{Abelian dominance} model the first results show the
existence of ring-type vortices.
\end{enumerate}

\bigskip

The various physically relevant cases are included in the following
classification

\bigskip

\begin{center}
\hspace{-2cm}%
\begin{tabular}{llll}
Current $J$ Abelian & $\Delta \psi +\exp \left( \psi \right) =0$ & conformal
invariant & no $\rho _{s}$ \\ 
Euler Non-Abelian & $\Delta \psi +\sinh \psi =0$ & conformal invariant & no $%
\rho _{s}$ \\ 
Superfluid Abelian & $\Delta \psi =\exp \left( \psi \right) -1$ & n.a.
(Minardi?) & finite $\rho _{s}$ \\ 
CHM Non-Abelian & $\Delta \psi =\pm \sinh \psi \left( \cosh \psi -1\right) $
& non-topological & finite $\rho _{s}$ \\ 
CHM Abelian & $\Delta \psi =e^{\psi }\left( e^{\psi }-1\right) $ & 
topological & finite $\rho _{s}$%
\end{tabular}
\end{center}

\section{The Liouville equation}

\subsection{Applications of the Liouville equation}

There are at least three physical problems that may imply a description in
terms of the Liouville equation:

\begin{enumerate}
\item the \emph{natural current profile} in tokamak plasma

\item the \emph{snake} of density in JET and other tokamaks

\item the \emph{filament} of current density developed on a particular
magnetic surface in tokamak (see \textbf{Huysmans})
\end{enumerate}

The last two phenomena seem to result from a concentration of a scalar field
(density and respectively the current density) and show robustness. The last
property is an indication that the system has reached an equilibrium that is
of a purely nonlinear nature, analogous to the relaxed states, therefore
they could be derived from the extremum of an action functional.

\subsubsection{The current profile}

The statistical theory of J.B. Taylor (and also : Montgomery, etc.) relies
on the principle of \emph{maximum entropy}, keeping the total energy and the
total number of particles constant. It leads to the Liouville equation.

The \emph{constant entropy} principle used by Minardi leads to the equation
for the current and is well verified by statistical analysis of the peaking
factors for current, density, pressure.

The statistical studies carried out on a large set of discharges with the
purpose of testing the prediction of the Turbulent Equipartition theory have
suggested that the current density is given by the equation $\Delta j+\left(
\lambda ^{2}/4\right) j=0$ where $\lambda $ is a constant. Replacing $%
j=\Delta \psi $ and assuming that two space integrations are possible
without introducing new physical effects, we would have the equation%
\[
\Delta \psi +\frac{\lambda ^{2}}{4}\psi =0 
\]%
that can be seen as a small $\psi $ approximation to%
\[
\Delta \psi +\frac{\lambda ^{2}}{4}\left[ \exp \left( \psi \right) -1\right]
=0 
\]%
We just note that this equation is of a similar form as%
\[
\Delta \chi =\exp \left( \chi \right) -1 
\]%
that describes the superfluid streamfunction in the \textbf{Abelian-Higgs}
model. Actually these equations are \emph{not} compatible, except when
taking the Lagrangian multiplier $\lambda ^{2}/4$ negative. This suggests
that the principle of constant entropy used by Minardi cannot be derived
from the topological theory of vortices in superfluids.

We will examine below the Taylor approach to the natural current profile in
tokamak.

\subsubsection{The snake of density}

The \emph{snake} phenomenon has been observed in JET during the experiments
of pellet injection and consisted in formation of persistent density
perturbations at rational-$q$ surfaces. These structures persist over
several sawtooth collapses and are difficult to explain as magnetic
perturbations. On the other hand there are indications that the tokamak
plasma density has an anomalous radial pinch, much larger than that of the
neoclassical origin. Apparently in the class of phenomena one should also
include the persistent impurity accumulation in laser blow-off injected
impurity, observed in experiments in TCV. There are several studies of the
statistical properties of the correlations between the peaking factors for
density, current density or pressure, with plasma parameters and these
studies seem to support the idea of turbulent equipartition of the
theromodynamic invariants. However we should note that these studies involve
quantities expressed as global variables (like averages) and they can hide
other dependences not immediately obvious.

We consider the possibility that the particle density behavior (and
particularly the \emph{snake} phenomenon) can be connected with the
existence of attracting solutions of certain nonlinear integrable equations.

The reason to consider the \emph{sinh}-Poisson equation comes from the
existing proofs that this equation governs the asymptotic states of ideal
fluids, or, more generally, of $2D$ systems that can be reduced to the
dynamics of point-like elements interacting by the potential which is the
inverse of the Laplacean operator (Jackiw and Pi, Spineanu and Vlad). This
equation is however obtained when there are two kinds of elements (like
positive and negative vorticity) and they are of equal number, $n_{+}=n_{-}$%
. Then the \emph{sinh}-Poisson equation is obtained as governing the states
with maximum entropy of the discrete statistical system at negative
temperature (Montgomery \emph{et al.}). Since the equation for the current
density mentioned above is derived under the assumption of turbulent
equipartition, the two descriptions may be related. However, the solutions
for the \emph{unbalanced} system of elements, $\alpha \equiv n_{+}/n_{-}\neq
1$, $\Delta \phi =\left( \lambda ^{2}/8\right) \left[ \exp \left( \phi
\right) /\sqrt{\alpha }-\sqrt{\alpha }\exp \left( -\phi \right) \right] $
have been obtained numerically (Pointin and Lundgren) and have been shown to
have higher entropy and higher stability than those of the \emph{sinh}%
-Poisson, precisely the characteristics we are seeking for. The limiting
form of the unbalanced equation is the Liouville equation, $\Delta \phi
=\left( \lambda ^{2}/8\right) \exp \left( -\phi \right) $.

\section{Classical physical interpretation of natural current profile in
tokamak}

The model is simplified in order to exhibit the essential aspect of current
self-organization in tokamak. The equation is%
\[
\mathbf{\nabla \times }\left( \mathbf{J\times B}\right) =0 
\]%
or%
\[
B_{0}\frac{\partial J}{\partial z}+\left( \mathbf{B}_{\perp }\cdot \mathbf{%
\nabla }_{\perp }\right) J\mathbf{=}0 
\]%
with%
\begin{eqnarray*}
\mathbf{B}_{\perp } &=&-\mathbf{\nabla }\psi \times \widehat{\mathbf{n}} \\
J &\equiv &J_{z}=\nabla _{\perp }^{2}\psi
\end{eqnarray*}

In the work of J.B. Taylor (1993) it is shown that a reasonable assumption
is that the current density should consists of \emph{filaments}, acted upon
by $\mathbf{B}_{\perp }$ as a velocity field and with $z$ as time. The
position of a filament is $\mathbf{r}_{i}\left( z\right) \equiv \left[
x_{i}\left( z\right) ,y_{i}\left( z\right) \right] $ and the equations of
motion are (all filaments are assumed equal $j_{0}$)%
\begin{eqnarray*}
j_{0}\frac{dx_{i}}{dz} &=&\frac{1}{B_{0}}\frac{\partial H}{\partial y_{i}} \\
j_{0}\frac{dy_{i}}{dz} &=&-\frac{1}{B_{0}}\frac{\partial H}{\partial x_{i}}
\end{eqnarray*}%
where%
\begin{eqnarray*}
H &=&\sum\limits_{k<i}j_{0}^{2}U\left( \mathbf{r}_{i},\mathbf{r}_{k}\right) 
\\
\mathbf{\nabla }_{\perp }^{2}U\left( \mathbf{r},\mathbf{r}^{^{\prime
}}\right)  &=&\delta \left( \mathbf{r-r}^{^{\prime }}\right) 
\end{eqnarray*}%
In an infinite region, the Green function $U\left( \mathbf{r,r}^{^{\prime
}}\right) $ of the Laplace operator is%
\[
U\left( \mathbf{r,r}^{^{\prime }}\right) =\ln \left\vert \mathbf{r-r}%
^{^{\prime }}\right\vert 
\]%
The current distribution and the magnetic flux function $\psi $
(streamfunction)%
\begin{eqnarray}
J &=&\sum\limits_{i}j_{0}\delta \left( \mathbf{r-r}_{i}\right) 
\label{psidis} \\
\psi  &=&j_{0}\sum\limits_{i}U\left( \mathbf{r,r}_{i}\right)   \nonumber
\end{eqnarray}%
The statistical description of the system.

The energy of the system is $H$ and it is strictly conserved and a \emph{%
microcanonical} ensemble, where the \emph{joint probability distribution} of
the positions of $N$ filaments is%
\[
\rho \left( \left\{ \mathbf{r}_{i}\right\} \right) \sim \delta \left(
E-H\left\{ \mathbf{r}_{i}\right\} \right) 
\]%
is appropriate. The entropy%
\[
S=-k\int d\mathbf{r}\ n\left( \mathbf{r}\right) \ln \left[ n\left( \mathbf{r}%
\right) \right] 
\]%
is a measure of the number of \emph{microscopic configurations}
corresponding to a \emph{macroscopic configuration} $n\left( \mathbf{r}%
\right) $.

Statistical equilibrium is obtained by maximizing $S$ (the entropy) under
the constraint of energy conservation and fixed total number of filaments%
\begin{eqnarray*}
E\ \text{(fixed)} &=&j_{0}\int d\mathbf{r}\ n\left( \mathbf{r}\right) \psi
\left( \mathbf{r}\right) \\
N\ \text{(fixed)} &=&\int d\mathbf{r}\ n\left( \mathbf{r}\right)
\end{eqnarray*}%
The continuum version of the magnetic flux function ($z$-component of the
magnetic potential), $\psi $ is obtained from Eq.(\ref{psidis})%
\[
\psi \left( \mathbf{r}\right) =j_{0}\int d\mathbf{r}^{^{\prime }}\ U\left( 
\mathbf{r,r}^{^{\prime }}\right) n\left( \mathbf{r}^{^{\prime }}\right) 
\]

The equilibrium current distribution obtained by extremizing%
\[
S-\beta E-\gamma N 
\]%
is%
\begin{eqnarray*}
J\left( \mathbf{r}\right) &=&j_{0}\left\langle n\left( \mathbf{r}\right)
\right\rangle \\
&=&K\exp \left[ -\beta j_{0}\psi \left( \mathbf{r}\right) \right]
\end{eqnarray*}

\textbf{Natural current density profiles}. Take $\psi $ to be zero on the
magnetic axis. Then%
\[
\nabla _{\perp }^{2}\psi =J_{0}\exp \left( -\lambda \psi \right) 
\]%
For circular symmetry in a tokamak of radius $a$,%
\[
\psi \left( r\right) =\frac{2}{\lambda }\ln \left( 1+\alpha \frac{r^{2}}{%
a^{2}}\right) 
\]%
where%
\[
\alpha =J_{0}\lambda \frac{a^{2}}{8\pi } 
\]%
Introducing the total current $I$, 
\[
I=Nj_{0} 
\]%
it is found a relation between the \emph{peaking factor} of the current
density, $\alpha $ , and the \emph{inverse temperature} $\beta $ of the
current filaments%
\[
\beta Nj_{0}^{2}=\frac{8\pi \alpha }{1+\alpha } 
\]%
Uniform current (which means that the whole plasma volume is chaotic) is
obtained for a magnetic temperature $T_{m}\equiv 1/\beta $ of%
\[
\alpha =0\ \text{or}\ T_{m}\rightarrow \infty 
\]%
and it is identified a \emph{critical} magnetic temperature $T_{m}^{c}$
where the totality of the current is concentrated into a singular central
filament,%
\[
\alpha \rightarrow \infty \ \text{or}\ T_{m}\rightarrow T_{m}^{c}\equiv 
\frac{Nj_{0}^{2}}{8\pi } 
\]%
Values of the magnetic temperatures between $0$ and $T_{m}^{c}$ are not
accessible.

\emph{Hollow current profiles} correspond to negative magnetic temperatures.
They are only accessible through the infinite value of the magnetic
temperature, $T_{m}\rightarrow \infty $, which in terms of profiles means
that the current passes first through a state of uniform distribution.

\bigskip

To compare with Field Theory we take the two equations%
\[
\text{FT:}\ \Delta \ln \rho =\pm \frac{2e^{2}}{c\kappa }\rho 
\]%
This is the Liouville equation after the substitution%
\[
\rho =\exp \left( \psi \right) 
\]%
It has nonsingular and \emph{non-negative }solutions for $\rho $ when it has
the form%
\[
\Delta \ln \rho +\left\vert \gamma \right\vert \rho =0 
\]%
Then the convention on the choice of signs in the right hand side follows
from the choice of sign for $\kappa $:

\begin{enumerate}
\item when $\kappa =\left\vert \kappa \right\vert >0$ one has to take the $-$
sign in the right hand side, such that%
\[
\Delta \ln \rho =-\frac{2e^{2}}{c\left\vert \kappa \right\vert }\rho 
\]%
or%
\[
\Delta \psi +\left( \frac{2e^{2}}{c\kappa }\right) \exp \left( \psi \right)
=0 
\]

\item when $\kappa =-\left\vert \kappa \right\vert <0$ one has to take the $%
+ $ sign in the right hand side, such that%
\[
\Delta \ln \rho =+\frac{2e^{2}}{c\kappa }\rho =-\frac{2e^{2}}{c\left\vert
\kappa \right\vert }\rho 
\]%
or%
\[
\Delta \psi +\left( \frac{2e^{2}}{c\left\vert \kappa \right\vert }\right)
\exp \left( \psi \right) =0 
\]
\end{enumerate}

Therefore the equation is always the same, whatever is the sign in front of
the Chern-Simons term.

\bigskip

In Taylor's theory, we have%
\[
\mathbf{\nabla }_{\perp }^{2}\psi =J_{0}\exp \left( -\lambda \psi \right) 
\]%
with%
\begin{eqnarray*}
\lambda &=&\frac{8\pi \alpha }{J_{0}a^{2}} \\
\frac{8\pi \alpha }{1+\alpha } &=&\frac{1}{T_{m}}Nj_{0}^{2}
\end{eqnarray*}%
We change the variable%
\[
\psi \rightarrow \psi ^{^{\prime }}=-\lambda \psi 
\]%
and the Taylor's equation becomes%
\begin{eqnarray*}
\left( -\lambda \right) \mathbf{\nabla }_{\perp }^{2}\psi ^{^{\prime }}
&=&J_{0}\exp \left( \psi ^{^{\prime }}\right) \\
\mathbf{\nabla }_{\perp }^{2}\psi ^{^{\prime }}+\left( \frac{J_{0}}{\lambda }%
\right) \exp \left( \psi ^{^{\prime }}\right) &=&0
\end{eqnarray*}%
or%
\[
\frac{J_{0}}{\lambda }=\frac{J_{0}^{2}a^{2}}{8\pi \alpha }=\frac{2e^{2}}{%
c\left\vert \kappa \right\vert } 
\]

Then we can translate the results obtained by Taylor:

\begin{enumerate}
\item when the peaking factor $\alpha $ goes to $0$, (\emph{i.e.} the
magnetic temperature $T_{m}\rightarrow \infty $) the current profile is
fully relaxed to a uniform distribution; This corresponds of vanishing $%
\kappa $ in the field theory: no Chern-Simons is present.

\item when the peaking factor $\alpha $ goes to $\infty $, (\emph{i.e.} the
magnetic temperature reaches the \emph{critical} value, $T_{m}\rightarrow
T_{m}^{c}$) the current is strongly concentrated on the axis. This
corresponds to infinite value for $\kappa $ in the field theory, $\left\vert
\kappa \right\vert \rightarrow \infty $: the Chern-Simons term is largely
dominating everything else in the Lagrangian.

\item Negative magnetic temperature%
\[
T_{m}<0 
\]%
are obtained in the Taylor's model when 
\[
\alpha <0 
\]%
or, the current profile is \emph{hollow}. In field theory this corresponds
to a change of sign of $\kappa $. But the equation remains the same. The
field theory starts with a certain sign of $\kappa $, then the Chern-Simons
term is suppressed (taking $\left\vert \kappa \right\vert \rightarrow 0$)
(leading to uniform solution for $\psi \rightarrow -\infty $ everywhere,
while $\Delta \psi $ may remain finite). After that the CS term is
re-established but with an effect which is opposite to the previous regime.
\end{enumerate}

Everything should be seen as an evolution \emph{on the manifold of SELF-DUAL
states, or solutions of the Liouville equation.} The parameter that moves
the states on this manifold is $\kappa $.

\bigskip

The following quantities have dimension of inverse distance squared%
\[
\frac{J_{0}}{\lambda }=\frac{2e^{2}}{c\left\vert \kappa \right\vert }=\frac{1%
}{\rho ^{2}} 
\]%
where $\rho $ is a distance. This distance will be the \emph{natural} unit
of space-like quantities in the problem. For example if our physical problem
is localised spatially in the disk of radius $a$, the adimensional space
range is%
\[
L\equiv \frac{a}{\rho } 
\]%
We note that the space unit $\rho $ is proportional with $\kappa $. We can
say that the passage of the system from a concentrated current profile to a
hollow current profile includes a state of strong localisation, where the
natural space unit is extremely small, which means that different parts of
the system are separated and non-interacting (physically this means chaos
and uniform current everywhere).

\section{The equation for the velocity of the fluid of point-like vortices
in static configurations}

In the field theoretical models of developed for the $2D$ current density
distribution and for the vorticity in ring-type structures in fluids or
plasmas, it is found that the magnetic potential (that carries the
interaction between the point-like vortices) has spatial components given by
the equations:%
\[
\mathbf{A}=\nabla \theta \pm \Lambda \nabla \times \ln \rho 
\]%
where $c$ is a dimensional constant (in the quantum theroies where the
objective is to describe the Abrikosov Nielsen Olesen vortices, $\Lambda
\equiv \hbar c/\left( 2e\right) $).

We want to calculate the contribution of the physical velocity in the
balance equation at stationarity%
\[
\left( \mathbf{v\cdot \nabla }\right) \mathbf{v=-\nabla }p 
\]%
where $p$ is the scalar pressure in the fluid. The physical velocity is just
the last term in the expression of the magnetic potential $\mathbf{A}$.

Using the definition of $\mathbf{v}$ we have 
\[
\left( \mathbf{v\cdot \nabla }\right) \mathbf{v=}\left[ \left( \Lambda
\nabla \times \ln \rho \right) \cdot \mathbf{\nabla }\right] \left( \Lambda
\nabla \times \ln \rho \right) 
\]%
or 
\[
\nabla \times \ln \rho \rightarrow \widehat{\mathbf{e}}_{z}\times \mathbf{%
\nabla }\left( \ln \rho \right) 
\]%
We extract for the next calculations the physical coefficient $\Lambda $.
This will leave some apparent incompatibilities in dimensions, but at the
end $\Lambda $ should be re-introduced.%
\begin{eqnarray*}
&&\left( \mathbf{v\cdot \nabla }\right) \mathbf{v} \\
&=&\left\{ \left[ \widehat{\mathbf{e}}_{z}\times \mathbf{\nabla }\left( \ln
\rho \right) \right] \cdot \mathbf{\nabla }\right\} \left[ \widehat{\mathbf{e%
}}_{z}\times \mathbf{\nabla }\left( \ln \rho \right) \right] \\
&=&\left[ \varepsilon ^{ik}\partial _{k}\left( \ln \rho \right) \right]
\partial _{i}\left[ \varepsilon ^{jl}\partial _{l}\left( \ln \rho \right) %
\right] \\
&=&\varepsilon ^{ik}\varepsilon ^{jl}\partial _{k}\left( \ln \rho \right)
\partial _{i}\partial _{l}\left( \ln \rho \right)
\end{eqnarray*}%
This is a vector and we have to calculate the two components in plane. For $%
j=x$, 
\begin{eqnarray*}
&&\left[ \left( \mathbf{v\cdot \nabla }\right) \mathbf{v}\right] _{x} \\
&=&\varepsilon ^{xl}\left\{ \varepsilon ^{ik}\partial _{k}\left( \ln \rho
\right) \partial _{i}\partial _{l}\left( \ln \rho \right) \right\} \\
&=&\varepsilon ^{xy}\left\{ \varepsilon ^{ik}\partial _{k}\left( \ln \rho
\right) \partial _{i}\partial _{y}\left( \ln \rho \right) \right\}
\end{eqnarray*}%
the summation over $l$ consists of just one term, $l=y$. We have considered
the convention $\varepsilon ^{xy}=1$. We make the summations over $i$ and $k$%
: 
\begin{eqnarray*}
&&\left[ \left( \mathbf{v\cdot \nabla }\right) \mathbf{v}\right] _{x} \\
&=&\varepsilon ^{ik}\partial _{k}\left( \ln \rho \right) \partial
_{i}\partial _{y}\left( \ln \rho \right) \\
&=&\varepsilon ^{xy}\partial _{y}\left( \ln \rho \right) \partial
_{x}\partial _{y}\left( \ln \rho \right) +\varepsilon ^{yx}\partial
_{x}\left( \ln \rho \right) \partial _{y}\partial _{y}\left( \ln \rho \right)
\\
&=&\partial _{x}\left\{ \frac{1}{2}\left[ \partial _{y}\left( \ln \rho
\right) \right] ^{2}\right\} -\partial _{x}\left( \ln \rho \right) \partial
_{y}^{2}\left( \ln \rho \right)
\end{eqnarray*}%
The other component, $j=y$%
\begin{eqnarray*}
&&\left[ \left( \mathbf{v\cdot \nabla }\right) \mathbf{v}\right] _{y} \\
&=&\varepsilon ^{yl}\left\{ \varepsilon ^{ik}\partial _{k}\left( \ln \rho
\right) \partial _{i}\partial _{l}\left( \ln \rho \right) \right\} \\
&=&\varepsilon ^{yx}\left\{ \varepsilon ^{ik}\partial _{k}\left( \ln \rho
\right) \partial _{i}\partial _{x}\left( \ln \rho \right) \right\} \\
&=&-\varepsilon ^{ik}\partial _{k}\left( \ln \rho \right) \partial
_{i}\partial _{x}\left( \ln \rho \right)
\end{eqnarray*}%
with the convention $\varepsilon ^{yx}=-1$. Now we make the summation over $%
i $ and $k$: 
\begin{eqnarray*}
&&\left[ \left( \mathbf{v\cdot \nabla }\right) \mathbf{v}\right] _{y} \\
&=&-\varepsilon ^{ik}\partial _{k}\left( \ln \rho \right) \partial
_{i}\partial _{x}\left( \ln \rho \right) \\
&=&-\varepsilon ^{xy}\partial _{y}\left( \ln \rho \right) \partial
_{x}\partial _{x}\left( \ln \rho \right) -\varepsilon ^{yx}\partial
_{x}\left( \ln \rho \right) \partial _{y}\partial _{x}\left( \ln \rho \right)
\\
&=&-\partial _{y}\left( \ln \rho \right) \partial _{x}^{2}\left( \ln \rho
\right) +\partial _{y}\left\{ \frac{1}{2}\left[ \partial _{x}\left( \ln \rho
\right) \right] ^{2}\right\}
\end{eqnarray*}

In the expression of the first component we take into account the
differential equation verified by $\rho $%
\[
\Delta \ln \rho =\alpha \rho 
\]%
where $\alpha $ is a constant. Then 
\[
\partial _{x}^{2}\left( \ln \rho \right) +\partial _{y}^{2}\left( \ln \rho
\right) =\alpha \rho 
\]%
and replace 
\[
\partial _{y}^{2}\left( \ln \rho \right) =\alpha \rho -\partial
_{x}^{2}\left( \ln \rho \right) 
\]%
in the expression of the $x$ component 
\begin{eqnarray*}
&&\left[ \left( \mathbf{v\cdot \nabla }\right) \mathbf{v}\right] _{x} \\
&=&\partial _{x}\left\{ \frac{1}{2}\left[ \partial _{y}\left( \ln \rho
\right) \right] ^{2}\right\} -\partial _{x}\left( \ln \rho \right) \partial
_{y}^{2}\left( \ln \rho \right) \\
&=&\partial _{x}\left\{ \frac{1}{2}\left[ \partial _{y}\left( \ln \rho
\right) \right] ^{2}\right\} \\
&&-\partial _{x}\left( \ln \rho \right) \left[ \alpha \rho -\partial
_{x}^{2}\left( \ln \rho \right) \right] \\
&=&\partial _{x}\left\{ \frac{1}{2}\left[ \partial _{y}\left( \ln \rho
\right) \right] ^{2}\right\} \\
&&-\partial _{x}\left( \ln \rho \right) \alpha \rho \\
&&+\partial _{x}\left( \ln \rho \right) \partial _{x}^{2}\left( \ln \rho
\right) \\
&=&\partial _{x}\left\{ \frac{1}{2}\left[ \partial _{y}\left( \ln \rho
\right) \right] ^{2}\right\} \\
&&-\left( \partial _{x}\rho \right) \alpha \\
&&+\partial _{x}\left\{ \frac{1}{2}\left[ \partial _{x}\left( \ln \rho
\right) \right] ^{2}\right\} \\
&=&\partial _{x}\Xi
\end{eqnarray*}%
where 
\begin{eqnarray*}
\Xi &\equiv &\frac{1}{2}\left[ \partial _{y}\left( \ln \rho \right) \right]
^{2}+\frac{1}{2}\left[ \partial _{x}\left( \ln \rho \right) \right]
^{2}-\alpha \rho \\
&=&\frac{1}{2\rho ^{2}}\left[ \left( \partial _{x}\rho \right) ^{2}+\left(
\partial _{y}\rho \right) ^{2}\right] -\alpha \rho
\end{eqnarray*}%
Another expression can be obtained if we substitute 
\begin{eqnarray*}
\partial _{x}^{2}\left( \ln \rho \right) &=&\partial _{x}\left( \frac{%
\partial _{x}\rho }{\rho }\right) \\
&=&\frac{\partial _{x}^{2}\rho }{\rho }-\frac{\left( \partial _{x}\rho
\right) ^{2}}{\rho ^{2}}
\end{eqnarray*}%
\begin{eqnarray*}
&&\frac{\left( \partial _{x}\rho \right) ^{2}}{\rho ^{2}}+\frac{\left(
\partial _{y}\rho \right) ^{2}}{\rho ^{2}} \\
&=&\frac{\partial _{x}^{2}\rho }{\rho }-\partial _{x}^{2}\left( \ln \rho
\right) +\frac{\partial _{y}^{2}\rho }{\rho }-\partial _{y}^{2}\left( \ln
\rho \right) \\
&=&\frac{\Delta \rho }{\rho }-\Delta \left( \ln \rho \right) \\
&=&\frac{\Delta \rho }{\rho }-\alpha \rho
\end{eqnarray*}%
Then 
\begin{eqnarray*}
\Xi &=&\frac{1}{2\rho ^{2}}\left[ \left( \partial _{x}\rho \right)
^{2}+\left( \partial _{y}\rho \right) ^{2}\right] -\alpha \rho \\
&=&\frac{1}{2}\left( \frac{\Delta \rho }{\rho }-\alpha \rho \right) -\alpha
\rho \\
&=&\frac{1}{2}\frac{\Delta \rho }{\rho }-\frac{3}{2}\alpha \rho
\end{eqnarray*}%
In conclusion the $x$ component is 
\begin{eqnarray*}
&&\left[ \left( \mathbf{v\cdot \nabla }\right) \mathbf{v}\right] _{x} \\
&=&-\partial _{x}\left\{ \frac{3}{2}\alpha \rho -\frac{1}{2}\frac{\Delta
\rho }{\rho }\right\}
\end{eqnarray*}%
and the pressure is 
\begin{eqnarray*}
p &=&\frac{3}{2}\alpha \rho -\frac{1}{2}\frac{\Delta \rho }{\rho } \\
&=&\frac{3}{2}\Delta \left( \ln \rho \right) -\frac{1}{2}\frac{\Delta \rho }{%
\rho }
\end{eqnarray*}%
If we take 
\[
\rho =\exp \psi 
\]%
we have 
\begin{eqnarray*}
\frac{\Delta \rho }{\rho } &=&\left( \mathbf{\nabla }\psi \right)
^{2}+\Delta \psi \\
&=&v^{2}+\omega
\end{eqnarray*}%
and 
\begin{eqnarray*}
p &=&\frac{3}{2}\omega -\frac{1}{2}\left( v^{2}+\omega \right) \\
&=&\omega -\frac{1}{2}v^{2}
\end{eqnarray*}%
where $\omega $ and $v$ are normalized.

This is the equivalent physical pressure that exists in a fluid for which
the Liouville equation is fullfilled. We used the field theoretical
formulation in order to calculate the pressure from the equation of momentum
conservation, at stationarity.

\bigskip

\textbf{NOTE}

We remark that the quantity%
\[
-\frac{1}{2}v^{2}+\omega 
\]%
looks similar to the square 
\[
\mathcal{A}^{\dagger }\mathcal{A} 
\]%
where%
\[
\mathcal{A}_{\mu }\equiv A_{\mu }\pm \sqrt{\frac{\kappa }{2}}\phi 
\]%
as a " vector potential" of a flat curvature space.

\section{The physical pressure for the EULER fluid}

We take the part of the velocity that is determined by the gradient of the
vorticity density. The balance equation at stationarity is 
\[
\left( \mathbf{v}^{\omega }\cdot \mathbf{\nabla }\right) \mathbf{v}^{\omega
}\sim -\frac{1}{\rho _{0}}\mathbf{\nabla }p 
\]%
and 
\begin{eqnarray*}
\mathbf{v}^{\omega } &\equiv &\mathbf{\nabla }\omega =\mathbf{\nabla }\left(
-\sinh \psi \right) \\
&=&-\cosh \psi \mathbf{\nabla }\psi
\end{eqnarray*}%
Then 
\[
\left( \mathbf{v}^{\omega }\cdot \mathbf{\nabla }\right) \mathbf{v}^{\omega
}=\cosh \psi \left( \mathbf{\nabla }\psi \cdot \mathbf{\nabla }\right)
\left( \cosh \psi \mathbf{\nabla }\psi \right) 
\]%
\begin{eqnarray*}
&&\left( \partial _{i}\psi \right) \partial _{i}\left[ \cosh \psi \left(
\partial _{j}\psi \right) \right] \\
&=&\left( \partial _{i}\psi \right) \sinh \psi \left( \partial _{i}\psi
\right) \left( \partial _{j}\psi \right) +\left( \partial _{i}\psi \right)
\cosh \psi \left( \partial _{i}\partial _{j}\psi \right)
\end{eqnarray*}%
\begin{eqnarray*}
\left( \mathbf{v}^{\omega }\cdot \mathbf{\nabla }\right) \mathbf{v}^{\omega
}|_{j} &=&\sinh \psi \cosh \psi \left( \partial _{i}\psi \right) ^{2}\left(
\partial _{j}\psi \right) +\left( \cosh \psi \right) ^{2}\left( \partial
_{i}\psi \right) \left( \partial _{i}\partial _{j}\psi \right) \\
&=&\cosh \psi \left( \mathbf{\nabla }\psi \right) \cdot \left[ \sinh \psi
\left( \mathbf{\nabla }\psi \right) \left( \partial _{j}\psi \right) +\cosh
\psi \mathbf{\nabla }\left( \partial _{j}\psi \right) \right]
\end{eqnarray*}

We use the formula 
\[
\left( \mathbf{v\cdot \nabla }\right) \mathbf{v=\nabla }\left( \frac{v^{2}}{2%
}\right) -\mathbf{v\times }\left( \mathbf{\nabla \times v}\right) 
\]%
Then 
\begin{eqnarray*}
\left( \mathbf{v}^{\omega }\cdot \mathbf{\nabla }\right) \mathbf{v}^{\omega
} &=&\mathbf{\nabla }\left( \frac{v^{\omega 2}}{2}\right) -\mathbf{v}%
^{\omega }\mathbf{\times }\left( \mathbf{\nabla \times v}^{\omega }\right) 
\\
&=&\mathbf{\nabla }\left[ \frac{\left( \mathbf{\nabla }\omega \right) ^{2}}{2%
}\right] -\mathbf{\nabla }\omega \times \left( \mathbf{\nabla \times \nabla }%
\omega \right)  \\
&=&\mathbf{\nabla }\left[ \frac{\left( \mathbf{\nabla }\omega \right) ^{2}}{2%
}\right]  \\
&=&\frac{1}{2}\mathbf{\nabla }\left[ \mathbf{\nabla }\left( \sinh \psi
\right) \right] ^{2}=\frac{1}{2}\mathbf{\nabla }\left[ \cosh \psi \left( 
\mathbf{\nabla }\psi \right) \right] ^{2} \\
&=&\frac{1}{2}\mathbf{\nabla }\left[ \left( \cosh \psi \right) ^{2}\left( 
\mathbf{\nabla }\psi \right) ^{2}\right] 
\end{eqnarray*}%
or 
\[
\left( \mathbf{v}^{\omega }\cdot \mathbf{\nabla }\right) \mathbf{v}^{\omega
}=-\frac{1}{\rho _{0}}\mathbf{\nabla }p
\]%
for 
\[
-\frac{1}{\rho _{0}}\mathbf{\nabla }p\equiv \mathbf{\nabla }\left[ \frac{%
\left( \mathbf{\nabla }\omega \right) ^{2}}{2}\right] 
\]%
Take 
\begin{eqnarray*}
\rho _{0} &=&\rho _{1}+\rho _{2} \\
&=&2\cosh \psi 
\end{eqnarray*}%
Then 
\begin{eqnarray*}
\mathbf{\nabla }p &=&-2\cosh \psi \mathbf{\nabla }\left[ \frac{\left( 
\mathbf{\nabla }\omega \right) ^{2}}{2}\right]  \\
&=&\left( -2\cosh \psi \right) \frac{1}{2}\mathbf{\nabla }\left[ \left(
\cosh \psi \right) ^{2}\left( \mathbf{\nabla }\psi \right) ^{2}\right] 
\end{eqnarray*}%
This should be expressed as a gradient of a scalar function.

In order that the expression above to be written as the gradient of a scalar
function (pressure) we start from thesecond part of the velocity expressed as%
\begin{eqnarray*}
v_{x}^{\omega } &\sim &\frac{\partial }{\partial x}\left( \rho _{1}-\rho
_{2}\right) \\
v_{y}^{\omega } &\sim &\frac{\partial }{\partial y}\left( \rho _{1}-\rho
_{2}\right)
\end{eqnarray*}%
A typical gradient along this velocity is 
\begin{eqnarray*}
\mathbf{v}^{\omega }\cdot \mathbf{\nabla } &=&v_{x}^{\omega }\frac{\partial 
}{\partial x}+v_{y}^{\omega }\frac{\partial }{\partial y} \\
&=&\frac{\partial }{\partial x}\left( \rho _{1}-\rho _{2}\right) \frac{%
\partial }{\partial x}+\frac{\partial }{\partial y}\left( \rho _{1}-\rho
_{2}\right) \frac{\partial }{\partial y}
\end{eqnarray*}%
This may be applied in particular on one of the components of this velocity 
\begin{eqnarray*}
\left( \mathbf{v}^{\omega }\cdot \mathbf{\nabla }\right) v_{x}^{\omega } &=&%
\left[ \frac{\partial }{\partial x}\left( \rho _{1}-\rho _{2}\right) \frac{%
\partial }{\partial x}+\frac{\partial }{\partial y}\left( \rho _{1}-\rho
_{2}\right) \frac{\partial }{\partial y}\right] \frac{\partial }{\partial x}%
\left( \rho _{1}-\rho _{2}\right) \\
&\sim &\frac{\partial \omega }{\partial x}\frac{\partial ^{2}\omega }{%
\partial x^{2}}+\frac{\partial \omega }{\partial y}\frac{\partial ^{2}\omega 
}{\partial y\partial x} \\
&=&\frac{\partial }{\partial x}\left[ \frac{1}{2}\left( \frac{\partial
\omega }{\partial x}\right) ^{2}+\frac{1}{2}\left( \frac{\partial \omega }{%
\partial y}\right) ^{2}\right]
\end{eqnarray*}%
and 
\begin{eqnarray*}
\left( \mathbf{v}^{\omega }\cdot \mathbf{\nabla }\right) v_{y}^{\omega } &=&%
\left[ \frac{\partial }{\partial x}\left( \rho _{1}-\rho _{2}\right) \frac{%
\partial }{\partial x}+\frac{\partial }{\partial y}\left( \rho _{1}-\rho
_{2}\right) \frac{\partial }{\partial y}\right] \frac{\partial }{\partial y}%
\left( \rho _{1}-\rho _{2}\right) \\
&\sim &\frac{\partial \omega }{\partial x}\frac{\partial ^{2}\omega }{%
\partial x\partial y}+\frac{\partial \omega }{\partial y}\frac{\partial
^{2}\omega }{\partial y^{2}} \\
&=&\frac{\partial }{\partial y}\left[ \frac{1}{2}\left( \frac{\partial
\omega }{\partial x}\right) ^{2}+\frac{1}{2}\left( \frac{\partial \omega }{%
\partial y}\right) ^{2}\right]
\end{eqnarray*}%
We can conclude that a certain quantity is suggested: 
\[
\left( \mathbf{v}^{\omega }\cdot \mathbf{\nabla }\right) \mathbf{v}^{\omega
}\sim -\frac{1}{\rho _{0}}\mathbf{\nabla }p 
\]%
with the pressure 
\[
p\sim -\left[ \frac{1}{2}\left( \frac{\partial \omega }{\partial x}\right)
^{2}+\frac{1}{2}\left( \frac{\partial \omega }{\partial y}\right) ^{2}\right]
\]

This pressure is negative, as expected. Since we are approximately at 
\textbf{self-duality for the Euler fluid} we have 
\begin{eqnarray*}
p &\sim &-\left[ \frac{1}{2}\left( \frac{\partial \omega }{\partial x}%
\right) ^{2}+\frac{1}{2}\left( \frac{\partial \omega }{\partial y}\right)
^{2}\right]  \\
&=&-\frac{1}{2}\left( \cosh \psi \right) ^{2}\left[ \left( \frac{\partial
\psi }{\partial x}\right) ^{2}+\left( \frac{\partial \psi }{\partial y}%
\right) ^{2}\right]  \\
&=&-\frac{1}{2}\left( \cosh \psi \right) ^{2}\left\vert \mathbf{\nabla }\psi
\right\vert ^{2}
\end{eqnarray*}%
We use the formulas 
\begin{eqnarray*}
\left( \cosh \psi \right) ^{2} &=&1+\left( \sinh \psi \right) ^{2} \\
&=&1+\omega ^{2}
\end{eqnarray*}%
and 
\[
\left\vert \mathbf{\nabla }\psi \right\vert ^{2}=\mathbf{v}^{2}
\]%
Then 
\begin{eqnarray*}
p &\sim &-\frac{1}{2}\left( \cosh \psi \right) ^{2}\left\vert \mathbf{\nabla 
}\psi \right\vert ^{2} \\
&=&-\frac{1}{2}\left( 1+\omega ^{2}\right) \mathbf{v}^{2} \\
&=&-\frac{1}{2}\mathbf{v}^{2}+\omega ^{2}\mathbf{v}^{2}
\end{eqnarray*}

This part of the pressure of the fluid of point-like vortices arises from
the part of the velocity that comes from the gradient of the density of
point-like vortices. The calculation is also valid in the case of the
Charney-Hasegawa-Mima (CHM) fluids, \emph{i.e.} plasma and planetary
atmosphere. We see that the pressure is most negative there where $\omega $
is higher, \emph{i.e.} at the center of the vortex. However, there $\mathbf{v%
}$ is zero. It appears that the pressure is most negative somewhere between
the maximum $\mathbf{v}$ and the center (maximum of $\omega $). Since the
variation of the $\omega $-velocity $v^{\omega }$ is governed by the force
resulting from the gradient of the pressure and the pressure has a negative
minimum somewhere between the radius of maximum azimuthal wind and the
center, it results that the pressure gradient acts such as to push the
\textquotedblright matter\textquotedblright\ = vorticity toward this point.
This is the reason for the creation of the cyclone eye.

\section{The curvature and the Self-dual state for the Euler fluid}

We take two variables identical with those which at SD give us the \emph{%
zero-curvature} condition. 
\begin{eqnarray*}
\mathcal{A}_{+} &=&A_{+}-\lambda \phi  \\
\mathcal{A}_{-} &=&A_{-}+\lambda \phi ^{\dagger }
\end{eqnarray*}%
where 
\[
\lambda \equiv \text{ real constant}
\]%
and calculate (cf. Eq.(59) and followings of \textbf{Dunne}) 
\[
K_{\pm }\equiv \partial _{\pm }\mathcal{A}_{\mp }-\partial _{\mp }\mathcal{A}%
_{\pm }+\left[ \mathcal{A}_{\pm },\mathcal{A}_{\mp }\right] 
\]

\bigskip

From the definitions we have 
\begin{eqnarray*}
\phi &=&\phi _{1}E_{+}+\phi _{2}E_{-} \\
\phi ^{\dagger } &=&\phi _{1}^{\ast }E_{-}+\phi _{2}^{\ast }E_{+}
\end{eqnarray*}
and 
\begin{eqnarray*}
A_{+} &=&aH \\
A_{-} &=&-a^{\ast }H
\end{eqnarray*}
then 
\begin{eqnarray*}
\mathcal{A}_{+} &=&A_{+}-\lambda \phi \\
&=&aH-\lambda \left( \phi _{1}E_{+}+\phi _{2}E_{-}\right)
\end{eqnarray*}
\begin{eqnarray*}
\mathcal{A}_{-} &=&A_{-}+\lambda \phi ^{\dagger } \\
&=&-a^{\ast }H+\lambda \left( \phi _{1}^{\ast }E_{-}+\phi _{2}^{\ast
}E_{+}\right)
\end{eqnarray*}

\subsubsection{The expression of the curvature $K_{+}$}

Let us take the $+$ sign: 
\begin{eqnarray*}
&&K_{+}= \\
&=&\partial _{+}\mathcal{A}_{-}-\partial _{-}\mathcal{A}_{+}+\left[ \mathcal{%
A}_{+},\mathcal{A}_{-}\right] \\
&=&\partial _{+}\left( A_{-}+\lambda \phi ^{\dagger }\right) -\partial
_{-}\left( A_{+}-\lambda \phi \right) +\left[ A_{+}-\lambda \phi
,A_{-}+\lambda \phi ^{\dagger }\right] \\
&=&\partial _{+}\left[ -a^{\ast }H+\lambda \left( \phi _{1}^{\ast
}E_{-}+\phi _{2}^{\ast }E_{+}\right) \right] \\
&&-\partial _{-}\left[ aH-\lambda \left( \phi _{1}E_{+}+\phi
_{2}E_{-}\right) \right] \\
&&+\left[ aH-\lambda \left( \phi _{1}E_{+}+\phi _{2}E_{-}\right) ,-a^{\ast
}H+\lambda \left( \phi _{1}^{\ast }E_{-}+\phi _{2}^{\ast }E_{+}\right) %
\right]
\end{eqnarray*}%
The first two lines are 
\begin{eqnarray}
&&-\partial _{+}a^{\ast }H+\lambda \partial _{+}\phi _{1}^{\ast
}E_{-}+\lambda \partial _{+}\phi _{2}^{\ast }E_{+}  \label{7800} \\
&&-\partial _{-}aH+\lambda \partial _{-}\phi _{1}E_{+}+\lambda \partial
_{-}\phi _{2}E_{-}  \nonumber \\
&=&\lambda \left( \partial _{+}\phi _{2}^{\ast }+\partial _{-}\phi
_{1}\right) E_{+}  \nonumber \\
&&+\lambda \left( \partial _{+}\phi _{1}^{\ast }+\partial _{-}\phi
_{2}\right) E_{-}  \nonumber \\
&&-\left( \partial _{+}a^{\ast }+\partial _{-}a\right) H  \nonumber
\end{eqnarray}%
and the last line (the commutator $\left[ \mathcal{A}_{+},\mathcal{A}_{-}%
\right] $) 
\begin{eqnarray*}
&&-aa^{\ast }\left[ H,H\right] \ \ \left( \text{this is }0\right) \\
&&+\lambda a\phi _{1}^{\ast }\left[ H,E_{-}\right] \ \ \left( \text{this is }%
-2E_{-}\right) \\
&&+\lambda a\phi _{2}^{\ast }\left[ H,E_{+}\right] \ \ \left( \text{this is }%
2E_{+}\right) \\
&&+\lambda \phi _{1}a^{\ast }\left[ E_{+},H\right] \ \ \left( \text{this is }%
-2E_{+}\right) \\
&&-\lambda ^{2}\phi _{1}\phi _{1}^{\ast }\left[ E_{+},E_{-}\right] \ \
\left( \text{this is }H\right) \\
&&-\lambda ^{2}\phi _{1}\phi _{2}^{\ast }\left[ E_{+},E_{+}\right] \ \
\left( \text{this is }0\right) \\
&&+\lambda \phi _{2}a^{\ast }\left[ E_{-},H\right] \ \ \left( \text{this is }%
2E_{-}\right) \\
&&-\lambda ^{2}\phi _{2}\phi _{1}^{\ast }\left[ E_{-},E_{-}\right] \ \
\left( \text{this is }0\right) \\
&&-\lambda ^{2}\phi _{2}\phi _{2}^{\ast }\left[ E_{-},E_{+}\right] \ \
\left( \text{this is }-H\right)
\end{eqnarray*}%
It results 
\begin{eqnarray*}
&&\left( 2\lambda a\phi _{2}^{\ast }-2\lambda \phi _{1}a^{\ast }\right) E_{+}
\\
&&\left( -2\lambda a\phi _{1}^{\ast }+2\lambda \phi _{2}a^{\ast }\right)
E_{-} \\
&&\left( -\lambda ^{2}\phi _{1}\phi _{1}^{\ast }+\lambda ^{2}\phi _{2}\phi
_{2}^{\ast }\right) H
\end{eqnarray*}%
\begin{eqnarray}
&\left[ \mathcal{A}_{+},\mathcal{A}_{-}\right] =&2\lambda \left( a\phi
_{2}^{\ast }-a^{\ast }\phi _{1}\right) E_{+} \\
&&+2\lambda \left( -a\phi _{1}^{\ast }+a^{\ast }\phi _{2}\right) E_{-} 
\nonumber \\
&&-\lambda ^{2}\left( \rho _{1}-\rho _{2}\right) H  \nonumber
\end{eqnarray}

Adding the two equations 
\begin{eqnarray*}
K_{+} &=&\left[ \lambda \left( \partial _{+}\phi _{2}^{\ast }+\partial
_{-}\phi _{1}\right) +2\lambda \left( a\phi _{2}^{\ast }-a^{\ast }\phi
_{1}\right) \right] E_{+} \\
&&+\left[ \lambda \left( \partial _{+}\phi _{1}^{\ast }+\partial _{-}\phi
_{2}\right) +2\lambda \left( -a\phi _{1}^{\ast }+a^{\ast }\phi _{2}\right) %
\right] E_{-} \\
&&\left[ -\left( \partial _{+}a^{\ast }+\partial _{-}a\right) -\lambda
^{2}\left( \rho _{1}-\rho _{2}\right) \right] H
\end{eqnarray*}
This can be written using notations 
\[
K_{+}\equiv K_{0}^{+}H+K_{1}^{+}E_{+}+K_{2}^{+}E_{-} 
\]
with 
\begin{eqnarray*}
K_{1}^{+} &\equiv &\lambda \left( \partial _{+}\phi _{2}^{\ast }+\partial
_{-}\phi _{1}\right) +2\lambda \left( a\phi _{2}^{\ast }-a^{\ast }\phi
_{1}\right) \\
K_{2}^{+} &\equiv &\lambda \left( \partial _{+}\phi _{1}^{\ast }+\partial
_{-}\phi _{2}\right) +2\lambda \left( -a\phi _{1}^{\ast }+a^{\ast }\phi
_{2}\right) \\
K_{0}^{+} &\equiv &-\left( \partial _{+}a^{\ast }+\partial _{-}a\right)
-\lambda ^{2}\left( \rho _{1}-\rho _{2}\right)
\end{eqnarray*}

If this curvature should be zero we have 
\begin{eqnarray*}
\partial _{+}\phi _{2}^{\ast }+\partial _{-}\phi _{1}+2a\phi _{2}^{\ast
}-2a^{\ast }\phi _{1} &=&0 \\
\partial _{+}\phi _{1}^{\ast }+\partial _{-}\phi _{2}-2a\phi _{1}^{\ast
}+2a^{\ast }\phi _{2} &=&0 \\
-\left( \partial _{+}a^{\ast }+\partial _{-}a\right) -\lambda ^{2}\left(
\rho _{1}-\rho _{2}\right) &=&0
\end{eqnarray*}

This is to be examined as condition for integrability.

\subsubsection{The expression of the curvature $K_{-}$}

Let us take the $-$ sign: 
\begin{eqnarray*}
&&K_{-}= \\
&=&\partial _{-}\mathcal{A}_{+}-\partial _{+}\mathcal{A}_{-}+\left[ \mathcal{%
A}_{-},\mathcal{A}_{+}\right] \\
&=&\partial _{-}\left( A_{+}-\lambda \phi \right) -\partial _{+}\left(
A_{-}+\lambda \phi ^{\dagger }\right) +\left[ A_{-}+\lambda \phi ^{\dagger
},A_{+}-\lambda \phi \right] \\
&=&\partial _{-}\left[ aH-\lambda \left( \phi _{1}E_{+}+\phi
_{2}E_{-}\right) \right] \\
&&-\partial _{+}\left[ -a^{\ast }H+\lambda \left( \phi _{1}^{\ast
}E_{-}+\phi _{2}^{\ast }E_{+}\right) \right] \\
&&+\left[ -a^{\ast }H+\lambda \left( \phi _{1}^{\ast }E_{-}+\phi _{2}^{\ast
}E_{+}\right) ,aH-\lambda \left( \phi _{1}E_{+}+\phi _{2}E_{-}\right) \right]
\end{eqnarray*}%
The first two lines are 
\begin{eqnarray}
&&\partial _{-}aH-\lambda \partial _{-}\phi _{1}E_{+}-\lambda \partial
_{-}\phi _{2}E_{-}  \label{kmin2} \\
&&+\partial _{+}a^{\ast }H-\lambda \partial _{+}\phi _{1}^{\ast
}E_{-}-\lambda \partial _{+}\phi _{2}^{\ast }E_{+}  \nonumber \\
&=&-\lambda \left( \partial _{+}\phi _{2}^{\ast }+\partial _{-}\phi
_{1}\right) E_{+}  \nonumber \\
&&-\lambda \left( \partial _{+}\phi _{1}^{\ast }+\partial _{-}\phi
_{2}\right) E_{-}  \nonumber \\
&&+\left( \partial _{+}a^{\ast }+\partial _{-}a\right) H  \nonumber
\end{eqnarray}%
and the last line (the commutator $\left[ \mathcal{A}_{-},\mathcal{A}_{+}%
\right] $) 
\begin{eqnarray*}
&&-a^{\ast }a\left[ H,H\right] \ \ \left( \text{this is }0\right) \\
&&+\lambda \phi _{1}^{\ast }a\left[ E_{-},H\right] \ \ \left( \text{this is }%
2E_{-}\right) \\
&&+\lambda \phi _{2}^{\ast }a\left[ E_{+},H\right] \ \ \left( \text{this is }%
-2E_{+}\right) \\
&&+a^{\ast }\lambda \phi _{1}\left[ H,E_{+}\right] \ \ \left( \text{this is }%
2E_{+}\right) \\
&&-\lambda ^{2}\phi _{1}^{\ast }\phi _{1}\left[ E_{-},E_{+}\right] \ \
\left( \text{this is }-H\right) \\
&&-\lambda ^{2}\phi _{2}^{\ast }\phi _{1}\left[ E_{+},E_{+}\right] \ \
\left( \text{this is }0\right) \\
&&+a^{\ast }\lambda \phi _{2}\left[ H,E_{-}\right] \ \ \left( \text{this is }%
-2E_{-}\right) \\
&&-\lambda ^{2}\phi _{1}^{\ast }\phi _{2}\left[ E_{-},E_{-}\right] \ \
\left( \text{this is }0\right) \\
&&-\lambda ^{2}\phi _{2}^{\ast }\phi _{2}\left[ E_{+},E_{-}\right] \ \
\left( \text{this is }H\right)
\end{eqnarray*}%
It results 
\begin{eqnarray*}
&&\left( -2\lambda a\phi _{2}^{\ast }+2\lambda \phi _{1}a^{\ast }\right)
E_{+} \\
&&\left( 2\lambda a\phi _{1}^{\ast }-2\lambda \phi _{2}a^{\ast }\right) E_{-}
\\
&&\left( \lambda ^{2}\phi _{1}\phi _{1}^{\ast }-\lambda ^{2}\phi _{2}\phi
_{2}^{\ast }\right) H
\end{eqnarray*}%
\begin{eqnarray}
&&-2\lambda \left( a\phi _{2}^{\ast }-a^{\ast }\phi _{1}\right) E_{+}
\label{kmin20} \\
&&+2\lambda \left( a\phi _{1}^{\ast }-a^{\ast }\phi _{2}\right) E_{-} 
\nonumber \\
&&+\lambda ^{2}\left( \rho _{1}-\rho _{2}\right) H  \nonumber
\end{eqnarray}

Adding the two equations (\ref{kmin2}) and (\ref{kmin20}) 
\begin{eqnarray*}
K_{-} &=&\left[ -\lambda \left( \partial _{+}\phi _{2}^{\ast }+\partial
_{-}\phi _{1}\right) -2\lambda \left( a\phi _{2}^{\ast }-a^{\ast }\phi
_{1}\right) \right] E_{+} \\
&&+\left[ -\lambda \left( \partial _{+}\phi _{1}^{\ast }+\partial _{-}\phi
_{2}\right) +2\lambda \left( a\phi _{1}^{\ast }-a^{\ast }\phi _{2}\right) %
\right] E_{-} \\
&&+\left[ \left( \partial _{+}a^{\ast }+\partial _{-}a\right) +\lambda
^{2}\left( \rho _{1}-\rho _{2}\right) \right] H
\end{eqnarray*}
This can be written using notations 
\[
K_{-}\equiv K_{0}^{-}H+K_{1}^{-}E_{+}+K_{2}^{-}E_{-} 
\]
with 
\begin{eqnarray*}
K_{1}^{-} &\equiv &-\lambda \left( \partial _{+}\phi _{2}^{\ast }+\partial
_{-}\phi _{1}\right) -2\lambda \left( a\phi _{2}^{\ast }-a^{\ast }\phi
_{1}\right) \\
K_{2}^{-} &\equiv &-\lambda \left( \partial _{+}\phi _{1}^{\ast }+\partial
_{-}\phi _{2}\right) +2\lambda \left( a\phi _{1}^{\ast }-a^{\ast }\phi
_{2}\right) \\
K_{0}^{-} &\equiv &\left( \partial _{+}a^{\ast }+\partial _{-}a\right)
+\lambda ^{2}\left( \rho _{1}-\rho _{2}\right)
\end{eqnarray*}

If this curvature should be zero we have 
\begin{eqnarray*}
\partial _{+}\phi _{2}^{\ast }+\partial _{-}\phi _{1}-2a\phi _{2}^{\ast
}+2a^{\ast }\phi _{1} &=&0 \\
\partial _{+}\phi _{1}^{\ast }+\partial _{-}\phi _{2}+2a\phi _{1}^{\ast
}-2a^{\ast }\phi _{2} &=&0 \\
\left( \partial _{+}a^{\ast }+\partial _{-}a\right) +\lambda ^{2}\left( \rho
_{1}-\rho _{2}\right) &=&0
\end{eqnarray*}

\subsubsection{Powers of products of these curvatures}

The fact that these three expression are NOT zero is the signature that the
fluid is NOT at self-duality.

However these expressions only contain the space derivatives and there is no
possibility to derive a time evolution.

We also have, for the powers of the \emph{curvature}, 
\begin{eqnarray*}
\mathrm{tr}\left( E_{+}^{n}\right) &=&0 \\
\mathrm{tr}\left( E_{-}^{n}\right) &=&0 \\
\mathrm{tr}\left( HE_{+}^{m}\right) &=&0 \\
\mathrm{tr}\left( HE_{-}^{m}\right) &=&0 \\
\mathrm{tr}\left( H^{n}E_{+}^{m}\right) &=&0 \\
\mathrm{tr}\left( H^{n}E_{-}^{m}\right) &=&0 \\
\mathrm{tr}\left( E_{+}^{n}E_{-}^{m}\right) &=&\mathrm{tr}\left(
E_{-}^{n}E_{+}^{m}\right) =1 \\
\mathrm{tr}\left( H^{2n}\right) &=&2 \\
\mathrm{tr}\left( H^{2n+1}\right) &=&0
\end{eqnarray*}

Let us consider the product of the two curvatures 
\begin{eqnarray*}
&&\mathrm{tr}\left\{ \left( K_{0}^{+}H+K_{1}^{+}E_{+}+K_{2}^{+}E_{-}\right)
\left( K_{0}^{-}H+K_{1}^{-}E_{+}+K_{2}^{-}E_{-}\right) \right\} \\
&=&\mathrm{tr}\left\{ K_{0}^{+}K_{0}^{-}H^{2}\right\} \;\;\text{the trace is 
}2 \\
&&+\mathrm{tr}\left\{ K_{0}^{+}K_{1}^{-}HE_{+}\right\} \;\;\text{the trace
is }0 \\
&&+\mathrm{tr}\left\{ K_{0}^{+}K_{2}^{-}HE_{-}\right\} \;\;\text{the trace
is }0 \\
&&+\mathrm{tr}\left\{ K_{1}^{+}K_{0}^{-}E_{+}H\right\} \;\;\text{the trace
is }0 \\
&&+\mathrm{tr}\left\{ K_{1}^{+}K_{1}^{-}E_{+}E_{+}\right\} \;\;\text{the
trace is }0 \\
&&+\mathrm{tr}\left\{ K_{1}^{+}K_{2}^{-}E_{+}E_{-}\right\} \;\;\text{the
trace is }1 \\
&&+\mathrm{tr}\left\{ K_{2}^{+}K_{0}^{-}E_{-}H\right\} \;\;\text{the trace
is }0 \\
&&+\mathrm{tr}\left\{ K_{2}^{+}K_{1}^{-}E_{-}E_{+}\right\} \;\;\text{the
trace is }1 \\
&&+\mathrm{tr}\left\{ K_{2}^{+}K_{2}^{-}E_{-}E_{-}\right\} \;\;\text{the
trace is }0
\end{eqnarray*}
Then 
\begin{eqnarray*}
&&\mathrm{tr}\left\{ K_{+}K_{-}\right\} \\
&=&2K_{0}^{+}K_{0}^{-} \\
&&+K_{1}^{+}K_{2}^{-}+K_{2}^{+}K_{1}^{-}
\end{eqnarray*}
or 
\begin{eqnarray*}
&&\mathrm{tr}\left\{ K_{+}K_{-}\right\} \\
&=&2\left[ -\left( \partial _{+}a^{\ast }+\partial _{-}a\right) -\lambda
^{2}\left( \rho _{1}-\rho _{2}\right) \right] \left[ \left( \partial
_{+}a^{\ast }+\partial _{-}a\right) +\lambda ^{2}\left( \rho _{1}-\rho
_{2}\right) \right] \\
&&+\left[ \lambda \left( \partial _{+}\phi _{2}^{\ast }+\partial _{-}\phi
_{1}\right) +2\lambda \left( a\phi _{2}^{\ast }-a^{\ast }\phi _{1}\right) %
\right] \left[ -\lambda \left( \partial _{+}\phi _{1}^{\ast }+\partial
_{-}\phi _{2}\right) +2\lambda \left( a\phi _{1}^{\ast }-a^{\ast }\phi
_{2}\right) \right] \\
&&+\left[ \lambda \left( \partial _{+}\phi _{1}^{\ast }+\partial _{-}\phi
_{2}\right) +2\lambda \left( -a\phi _{1}^{\ast }+a^{\ast }\phi _{2}\right) %
\right] \left[ -\lambda \left( \partial _{+}\phi _{2}^{\ast }+\partial
_{-}\phi _{1}\right) -2\lambda \left( a\phi _{2}^{\ast }-a^{\ast }\phi
_{1}\right) \right]
\end{eqnarray*}
We note that the last two lines are identical. 
\begin{eqnarray*}
&&\mathrm{tr}\left\{ K_{+}K_{-}\right\} \\
&=&-2\left[ \left( \partial _{+}a^{\ast }+\partial _{-}a\right) +\lambda
^{2}\left( \rho _{1}-\rho _{2}\right) \right] ^{2} \\
&&-\lambda ^{2}\left[ \left( \partial _{+}\phi _{2}^{\ast }+\partial
_{-}\phi _{1}\right) +2\left( a\phi _{2}^{\ast }-a^{\ast }\phi _{1}\right) %
\right] \left[ \left( \partial _{+}\phi _{1}^{\ast }+\partial _{-}\phi
_{2}\right) -2\left( a\phi _{1}^{\ast }-a^{\ast }\phi _{2}\right) \right]
\end{eqnarray*}
Now, since we have 
\[
\partial _{+}^{\ast }=\partial _{-} 
\]
we see that the second paranthesis is the complex conjugate of the first. 
\begin{eqnarray*}
&&\left[ \left( \partial _{+}\phi _{2}^{\ast }+\partial _{-}\phi _{1}\right)
+2\left( a\phi _{2}^{\ast }-a^{\ast }\phi _{1}\right) \right] ^{\ast } \\
&=&\left( \partial _{+}^{\ast }\phi _{2}+\partial _{-}^{\ast }\phi
_{1}^{\ast }\right) +2\left( a^{\ast }\phi _{2}-a\phi _{1}^{\ast }\right) \\
&=&\partial _{-}\phi _{2}+\partial _{+}\phi _{1}^{\ast }+2\left( a^{\ast
}\phi _{2}-a\phi _{1}^{\ast }\right)
\end{eqnarray*}
Thus this relationship has been verified. We then have 
\begin{eqnarray*}
&&\mathrm{tr}\left\{ K_{+}K_{-}\right\} \\
&=&-2\left[ \left( \partial _{+}a^{\ast }+\partial _{-}a\right) +\lambda
^{2}\left( \rho _{1}-\rho _{2}\right) \right] ^{2} \\
&&-\lambda ^{2}\left| \left( \partial _{+}\phi _{2}^{\ast }+\partial
_{-}\phi _{1}\right) +2\left( a\phi _{2}^{\ast }-a^{\ast }\phi _{1}\right)
\right| ^{2}
\end{eqnarray*}

Naturally, we obtain that 
\[
-\mathrm{tr}\left\{ K_{+}K_{-}\right\} \geq 0 
\]
since it is a sum of squares and the equality with zero is precisely the SD
equations.

\bigskip

In the powers 
\[
\mathrm{tr}\left( \left( K^{-}\right) ^{n}\right) 
\]
only the powers of $K_{0}^{-}$ remains and the products $\left(
K_{1}^{-}\right) ^{n}\left( K_{2}^{-}\right) ^{m}$.

Then we have 
\begin{eqnarray*}
\mathrm{tr}\left( \left( K^{-}\right) ^{2n}\right) &=&\mathrm{tr}\left\{
\left( \left( \partial _{+}a^{\ast }+\partial _{-}a\right) +\lambda
^{2}\left( \rho _{1}-\rho _{2}\right) \right) H\right\} ^{2n} \\
&=&2\left[ \left( \partial _{+}a^{\ast }+\partial _{-}a\right) +\lambda
^{2}\left( \rho _{1}-\rho _{2}\right) \right] ^{2n} \\
&&+...
\end{eqnarray*}

\bigskip

We can however continue this calculation using equations in close proximity
of the Self-duality.

From the first equation at SD 
\[
D_{-}\phi =0
\]%
we derive  
\begin{eqnarray*}
2\frac{\partial \phi _{1}}{\partial z}-2a^{\ast }\phi _{1} &=&0 \\
2\frac{\partial \phi _{1}^{\ast }}{\partial z^{\ast }}-2a\phi _{1}^{\ast }
&=&0
\end{eqnarray*}%
and 
\begin{eqnarray*}
2\frac{\partial \phi _{2}}{\partial z}+2\phi _{2}a^{\ast } &=&0 \\
2\frac{\partial \phi _{2}^{\ast }}{\partial z^{\ast }}+2a\phi _{2}^{\ast }
&=&0
\end{eqnarray*}%
with the notations 
\begin{eqnarray*}
\partial _{+} &=&\partial _{x}+i\partial _{y}=2\frac{\partial }{\partial
z^{\ast }} \\
\partial _{-} &=&\partial _{x}-i\partial _{y}=2\frac{\partial }{\partial z}
\end{eqnarray*}%
from which we have 
\[
\partial _{+}\partial _{-}\equiv \Delta 
\]%
and%
\[
4\frac{\partial }{\partial z^{\ast }}\frac{\partial }{\partial z}\equiv
\Delta 
\]%
We can obtain the two potentials at SD 
\begin{eqnarray*}
a^{\ast } &=&\frac{1}{2}\partial _{-}\ln \phi _{1} \\
a &=&\frac{1}{2}\partial _{+}\ln \phi _{1}^{\ast } \\
a^{\ast } &=&-\frac{1}{2}\partial _{-}\ln \phi _{2} \\
a &=&-\frac{1}{2}\partial _{+}\ln \phi _{2}^{\ast }
\end{eqnarray*}%
Then 
\begin{eqnarray*}
2a\phi _{2}^{\ast } &=&-2\frac{\partial \phi _{2}^{\ast }}{\partial z^{\ast }%
}=-\partial _{+}\phi _{2}^{\ast } \\
2a^{\ast }\phi _{2} &=&-2\frac{\partial \phi _{2}}{\partial z}=-\partial
_{-}\phi _{2} \\
2a^{\ast }\phi _{1} &=&2\frac{\partial \phi _{1}}{\partial z}=\partial
_{-}\phi _{1} \\
2a\phi _{1}^{\ast } &=&2\frac{\partial \phi _{1}^{\ast }}{\partial z^{\ast }}%
=\partial _{+}\phi _{1}^{\ast }
\end{eqnarray*}%
We have 
\begin{eqnarray*}
K_{1}/\lambda  &=&\partial _{+}\phi _{2}^{\ast }+\partial _{-}\phi
_{1}+2a\phi _{2}^{\ast }-2a^{\ast }\phi _{1} \\
&=&\partial _{+}\phi _{2}^{\ast }+\partial _{-}\phi _{1}+\left( -\partial
_{+}\phi _{2}^{\ast }\right) -\left( \partial _{-}\phi _{1}\right)  \\
&=&0
\end{eqnarray*}%
\begin{eqnarray*}
K_{2}/\lambda  &=&\partial _{+}\phi _{1}^{\ast }+\partial _{-}\phi
_{2}-2a\phi _{1}^{\ast }+2a^{\ast }\phi _{2} \\
&=&\partial _{+}\phi _{1}^{\ast }+\partial _{-}\phi _{2}-\left( \partial
_{+}\phi _{1}^{\ast }\right) +\left( -\partial _{-}\phi _{2}\right)  \\
&=&0
\end{eqnarray*}%
\[
K_{0}=-\left( \partial _{+}a^{\ast }+\partial _{-}a\right) -\lambda
^{2}\left( \rho _{1}-\rho _{2}\right) 
\]%
where we have 
\begin{eqnarray*}
K_{0} &=&-\partial _{+}\left[ \frac{1}{2}\partial _{-}\ln \left( \phi
_{1}\right) \right] -\partial _{-}\left[ \frac{1}{2}\partial _{+}\ln \left(
\phi _{1}^{\ast }\right) \right] -\lambda ^{2}\left( \rho _{1}-\rho
_{2}\right)  \\
&=&-\partial _{+}\partial _{-}\frac{1}{2}\left[ \ln \left( \phi _{1}\right)
+\ln \left( \phi _{1}^{\ast }\right) \right] -\lambda ^{2}\left( \rho
_{1}-\rho _{2}\right)  \\
&=&-\frac{1}{2}\Delta \ln \rho _{1}-\lambda ^{2}\left( \rho _{1}-\rho
_{2}\right) 
\end{eqnarray*}%
which is real. This expression is also zero at SD since it is the equation 
\begin{eqnarray*}
-\frac{1}{2}\Delta \ln \rho _{1}-\lambda ^{2}\left( \rho _{1}-\rho
_{2}\right)  &=&0 \\
\frac{1}{2}\Delta \psi +\lambda ^{2}2\sinh \psi  &=&0
\end{eqnarray*}%
The normal choice for $\lambda $ is 
\[
\lambda ^{2}=\frac{1}{L^{2}}
\]%
with $L$ the length of the box.

\bigskip

We \textbf{NOTE} however that it is NOT necessary that the expression of $%
K_{0}/H$ to go to zero at SD. We just need it to become a numerical factor
multiplying the local vorticity $\omega $. Because in this case we have 
\[
-\frac{1}{2}\Delta \ln \rho _{1}-\lambda ^{2}\left( \rho _{1}-\rho
_{2}\right) =\mu \omega 
\]
and this just produces another number in front of $\omega $ and a scaling of
the space axis.

However it looks better to say that the SD coincides with $K=0$. \textbf{END}

\section{A minimizer for Euler}

We can use the expression of the energy, after applying the Bogomolnyi
procedure, 
\[
E=\frac{1}{2m}\mathrm{tr}\left( \left( D_{-}\phi \right) ^{\dagger }\left(
D_{-}\phi \right) \right) 
\]%
which leads to the equation for the states realizing the lowest energy $%
D_{-}\phi =0$.

We have to express the detailed form of the energy. The only factor is
defined 
\begin{equation}
D_{-}\equiv D_{1}-iD_{2}  \label{610}
\end{equation}%
then 
\begin{equation}
D_{-}\phi =\frac{\partial \phi }{\partial x}+\left[ A_{x},\phi \right] -i%
\frac{\partial \phi }{\partial y}-i\left[ A_{y},\phi \right]  \label{611}
\end{equation}%
To continue we express the components of the potential 
\begin{eqnarray}
A_{+} &=&A_{x}+iA_{y}=aH  \label{612} \\
A_{-} &=&A_{x}-iA_{y}=-a^{\ast }H  \nonumber
\end{eqnarray}%
Then 
\begin{eqnarray}
A_{x} &=&\frac{1}{2}\left( a-a^{\ast }\right) H  \label{613} \\
A_{y} &=&\frac{1}{2i}\left( a+a^{\ast }\right) H  \nonumber
\end{eqnarray}%
Then 
\begin{eqnarray}
D_{-}\phi &=&\left( \frac{\partial \phi _{1}}{\partial x}-i\frac{\partial
\phi _{1}}{\partial y}\right) E_{+}+\left( \frac{\partial \phi _{2}}{%
\partial x}-i\frac{\partial \phi _{2}}{\partial y}\right) E_{-}  \label{614}
\\
&&+\frac{1}{2}\left( a-a^{\ast }\right) \phi _{1}\left[ H,E_{+}\right] 
\nonumber \\
&&+\frac{1}{2}\left( a-a^{\ast }\right) \phi _{2}\left[ H,E_{-}\right] 
\nonumber \\
&&-i\frac{1}{2i}\left( a+a^{\ast }\right) \phi _{1}\left[ H,E_{+}\right] 
\nonumber \\
&&-i\frac{1}{2i}\left( a+a^{\ast }\right) \phi _{2}\left[ H,E_{-}\right] 
\nonumber
\end{eqnarray}%
\begin{eqnarray}
D_{-}\phi &=&\left( \frac{\partial \phi _{1}}{\partial x}-i\frac{\partial
\phi _{1}}{\partial y}+2\frac{1}{2}\left( a-a^{\ast }\right) \phi _{1}-2%
\frac{1}{2}\left( a+a^{\ast }\right) \phi _{1}\right) E_{+}  \label{615} \\
&&+\left( \frac{\partial \phi _{2}}{\partial x}-i\frac{\partial \phi _{2}}{%
\partial y}-2\frac{1}{2}\left( a-a^{\ast }\right) \phi _{2}+2\frac{1}{2}%
\left( a+a^{\ast }\right) \phi _{2}\right) E_{-}  \nonumber
\end{eqnarray}%
or%
\begin{eqnarray}
D_{-}\phi &=&\left( \frac{\partial \phi _{1}}{\partial x}-i\frac{\partial
\phi _{1}}{\partial y}-2a^{\ast }\phi _{1}\right) E_{+}  \label{616} \\
&&+\left( \frac{\partial \phi _{2}}{\partial x}-i\frac{\partial \phi _{2}}{%
\partial y}+2a^{\ast }\phi _{2}\right) E_{-}  \nonumber
\end{eqnarray}

\bigskip

The algebraic ansatz is used for the Hermitean conjugate, $\left( D_{-}\phi
\right) ^{\dagger }$. We have 
\begin{equation}
D_{-}^{\dagger }=\frac{\partial }{\partial x}+\left[ ,A_{x}^{\dagger }\right]
+i\frac{\partial }{\partial y}+i\left[ ,A_{y}^{\dagger }\right]  \label{66}
\end{equation}%
where the adjoint is taken for any matrix as the transpose complex
conjugated. The change of the order in the commutators is due to the
property that for any two matrices $R_{1}$ and $R_{2}$ the Hermitian
conjugate of their commutator is 
\begin{eqnarray}
\left[ R_{1},R_{2}\right] ^{\dagger } &=&\left( R_{1}R_{2}-R_{2}R_{1}\right)
^{\dagger }  \label{67} \\
&=&\left( R_{2}^{T}R_{1}^{T}-R_{1}^{T}R_{2}^{T}\right) ^{\ast }  \nonumber \\
&=&R_{2}^{\dagger }R_{1}^{\dagger }-R_{1}^{\dagger }R_{2}^{\dagger } 
\nonumber \\
&=&\left[ R_{2}^{\dagger },R_{1}^{\dagger }\right]  \nonumber
\end{eqnarray}%
($^{\ast }$ is complex conjugate and $^{T}$ is the transpose operators) and
we take into account that in the expression of $\phi ^{\dagger }$ we have
already used the Hermitian conjugated matrices of $E_{\pm }$.

The Hermitian conjugates of the gauge field matrices are 
\begin{eqnarray}
A_{x}^{\dagger } &=&\frac{1}{2}\left( a^{\ast }-a\right) H^{\dagger }=\frac{1%
}{2}\left( a^{\ast }-a\right) H  \label{68} \\
A_{y}^{\dagger } &=&-\frac{1}{2i}\left( a^{\ast }+a\right) H^{\dagger }=-%
\frac{1}{2i}\left( a^{\ast }+a\right) H  \nonumber
\end{eqnarray}%
Then 
\begin{equation}
D_{-}^{\dagger }\equiv \frac{\partial }{\partial x}+i\frac{\partial }{%
\partial y}+\frac{1}{2}\left( a^{\ast }-a\right) \left[ ,H\right] -\frac{1}{2%
}\left( a^{\ast }+a\right) \left[ ,H\right]  \label{69}
\end{equation}%
We recall that 
\begin{equation}
\phi ^{\dagger }=\phi _{1}^{\ast }E_{-}+\phi _{2}^{\ast }E_{+}  \label{70}
\end{equation}%
The we have 
\begin{eqnarray}
\left( D_{-}\phi \right) ^{\dagger } &=&\left\{ \frac{\partial }{\partial x}%
+i\frac{\partial }{\partial y}+\frac{1}{2}\left( a^{\ast }-a\right) \left[ ,H%
\right] -\frac{1}{2}\left( a^{\ast }+a\right) \left[ ,H\right] \right\}
\label{71} \\
&&\times \left( \phi _{1}^{\ast }E_{-}+\phi _{2}^{\ast }E_{+}\right) 
\nonumber \\
&=&\left( \frac{\partial \phi _{1}^{\ast }}{\partial x}+i\frac{\partial \phi
_{1}^{\ast }}{\partial y}\right) E_{-}+\left( \frac{\partial \phi _{2}^{\ast
}}{\partial x}+i\frac{\partial \phi _{2}^{\ast }}{\partial y}\right) E_{+} 
\nonumber \\
&&+\frac{1}{2}\left( a^{\ast }-a\right) \phi _{1}^{\ast }\left[ E_{-},H%
\right]  \nonumber \\
&&+\frac{1}{2}\left( a^{\ast }-a\right) \phi _{2}^{\ast }\left[ E_{+},H%
\right]  \nonumber \\
&&-\frac{1}{2}\left( a^{\ast }+a\right) \phi _{1}^{\ast }\left[ E_{-},H%
\right]  \nonumber \\
&&-\frac{1}{2}\left( a^{\ast }+a\right) \phi _{2}^{\ast }\left[ E_{+},H%
\right]  \nonumber
\end{eqnarray}%
or 
\begin{eqnarray}
\left( D_{-}\phi \right) ^{\dagger } &=&2\frac{\partial \phi _{1}^{\ast }}{%
\partial z^{\ast }}E_{-}+2\frac{\partial \phi _{2}^{\ast }}{\partial z^{\ast
}}E_{+}  \label{72} \\
&&+\frac{1}{2}\left( a^{\ast }-a\right) \phi _{1}^{\ast }\left( 2E_{-}\right)
\nonumber \\
&&+\frac{1}{2}\left( a^{\ast }-a\right) \phi _{2}^{\ast }\left(
-2E_{+}\right)  \nonumber \\
&&-\frac{1}{2}\left( a^{\ast }+a\right) \phi _{1}^{\ast }\left( 2E_{-}\right)
\nonumber \\
&&-\frac{1}{2}\left( a^{\ast }+a\right) \phi _{2}^{\ast }\left(
-2E_{+}\right)  \nonumber
\end{eqnarray}%
The equation becomes 
\begin{eqnarray}
\left( D_{-}\phi \right) ^{\dagger } &=&\left( 2\frac{\partial \phi
_{1}^{\ast }}{\partial z^{\ast }}+\left( a^{\ast }-a\right) \phi _{1}^{\ast
}-\left( a^{\ast }+a\right) \phi _{1}^{\ast }\right) E_{-}  \label{73} \\
&&+\left( 2\frac{\partial \phi _{2}^{\ast }}{\partial z^{\ast }}-\left(
a^{\ast }-a\right) \phi _{2}^{\ast }+\left( a^{\ast }+a\right) \phi
_{2}^{\ast }\right) E_{+}  \nonumber
\end{eqnarray}%
Then%
\begin{eqnarray}
\left( D_{-}\phi \right) ^{\dagger } &=&\left( 2\frac{\partial \phi
_{1}^{\ast }}{\partial z^{\ast }}-2a\phi _{1}^{\ast }\right) E_{-}
\label{731} \\
&&+\left( 2\frac{\partial \phi _{2}^{\ast }}{\partial z^{\ast }}+2a\phi
_{2}^{\ast }\right) E_{+}  \nonumber
\end{eqnarray}%
Here we have made use of the identifications 
\begin{equation}
\frac{\partial }{\partial x}+i\frac{\partial }{\partial y}\equiv 2\frac{%
\partial }{\partial z^{\ast }}  \label{74}
\end{equation}%
and 
\begin{equation}
\frac{\partial }{\partial x}-i\frac{\partial }{\partial y}\equiv 2\frac{%
\partial }{\partial z}  \label{75}
\end{equation}%
The resulting equations are 
\begin{equation}
\left( D_{-}\phi \right) ^{\dagger }=\left( 2\frac{\partial \phi _{1}^{\ast }%
}{\partial z^{\ast }}-2a\phi _{1}^{\ast }\right) E_{-}+\left( 2\frac{%
\partial \phi _{2}^{\ast }}{\partial z^{\ast }}+2a\phi _{2}^{\ast }\right)
E_{+}  \label{76}
\end{equation}%
which represent the adjoints of the first set, as expected.

\bigskip

We can now calculate the energy%
\begin{eqnarray*}
E &=&\frac{1}{2m}\mathrm{tr}\left( \left( D_{-}\phi \right) ^{\dagger
}\left( D_{-}\phi \right) \right) \\
&=&\frac{1}{2m}\mathrm{tr}\left\{ \left[ \left( \frac{\partial \phi
_{1}^{\ast }}{\partial x}+i\frac{\partial \phi _{1}^{\ast }}{\partial y}%
-2a\phi _{1}^{\ast }\right) E_{-}+\left( \frac{\partial \phi _{2}^{\ast }}{%
\partial x}+i\frac{\partial \phi _{2}^{\ast }}{\partial y}+2a\phi _{2}^{\ast
}\right) E_{+}\right] \right. \\
&&\left. \times \left[ \left( \frac{\partial \phi _{1}}{\partial x}-i\frac{%
\partial \phi _{1}}{\partial y}-2a^{\ast }\phi _{1}\right) E_{+}+\left( 
\frac{\partial \phi _{2}}{\partial x}-i\frac{\partial \phi _{2}}{\partial y}%
+2a^{\ast }\phi _{2}\right) E_{-}\right] \right\}
\end{eqnarray*}%
\[
E=\frac{1}{2m}\left( \left\vert \frac{\partial \phi _{1}}{\partial x}-i\frac{%
\partial \phi _{1}}{\partial y}-2a^{\ast }\phi _{1}\right\vert
^{2}+\left\vert \frac{\partial \phi _{2}}{\partial x}-i\frac{\partial \phi
_{2}}{\partial y}+2a^{\ast }\phi _{2}\right\vert ^{2}\right) 
\]%
\[
E=\frac{1}{2m}\left( \left\vert \phi _{1}\right\vert ^{2}\left\vert \frac{%
\partial }{\partial x_{-}}\ln \phi _{1}-2a^{\ast }\right\vert
^{2}+\left\vert \phi _{2}\right\vert ^{2}\left\vert \frac{\partial }{%
\partial x_{-}}\ln \phi _{2}+2a^{\ast }\right\vert ^{2}\right) 
\]%
We take%
\begin{eqnarray*}
\phi _{1} &=&\sqrt{\rho _{1}}\exp \left( i\chi \right) \\
\phi _{2} &=&\sqrt{\rho _{2}}\exp \left( i\eta \right)
\end{eqnarray*}%
and we have%
\begin{eqnarray*}
\frac{\partial }{\partial x_{-}}\ln \phi _{1} &=&\frac{1}{2}\frac{\partial }{%
\partial x_{-}}\ln \rho _{1}+i\frac{\partial \chi }{\partial x_{-}} \\
&=&\frac{1}{2\rho _{1}}\frac{\partial \rho _{1}}{\partial x_{-}}+i\frac{%
\partial \chi }{\partial x_{-}}
\end{eqnarray*}%
The energy becomes%
\[
E=\frac{1}{2m}\left( \rho _{1}\left\vert \frac{1}{2\rho _{1}}\frac{\partial
\rho _{1}}{\partial x_{-}}+i\frac{\partial \chi }{\partial x_{-}}-2a^{\ast
}\right\vert ^{2}+\rho _{2}\left\vert \frac{1}{2\rho _{2}}\frac{\partial
\rho _{2}}{\partial x_{-}}+i\frac{\partial \eta }{\partial x_{-}}+2a^{\ast
}\right\vert ^{2}\right) 
\]

\bigskip

Consider the Self-Duality.

Now we can take $\rho _{1}=1/\rho _{2}$ . This is not in contradiction with
the fact that we try to NOT investigate the self-dual state, but its
neighborhood. We take 
\begin{eqnarray*}
\rho _{1} &=&\frac{1}{\rho _{2}}=\rho =\exp \left( \psi \right)  \\
\chi  &=&-\eta 
\end{eqnarray*}%
we have%
\[
E=\frac{1}{2m}\left( \exp \left( \psi \right) \left\vert \frac{1}{2}\frac{%
\partial \psi }{\partial x_{-}}+i\frac{\partial \chi }{\partial x_{-}}%
-2a^{\ast }\right\vert ^{2}+\exp \left( -\psi \right) \left\vert -\frac{1}{2}%
\frac{\partial \psi }{\partial x_{-}}-i\frac{\partial \chi }{\partial x_{-}}%
+2a^{\ast }\right\vert ^{2}\right) 
\]%
Or%
\[
E=\frac{1}{2m}\left[ \exp \left( \psi \right) +\exp \left( -\psi \right) %
\right] \left\vert \frac{1}{2}\frac{\partial \psi }{\partial x_{-}}+i\frac{%
\partial \chi }{\partial x_{-}}-2a^{\ast }\right\vert ^{2}
\]%
or%
\[
E=\frac{1}{2m}\left[ \exp \left( \psi \right) +\exp \left( -\psi \right) %
\right] \left\vert \frac{\partial }{\partial x_{-}}\left( \psi /2+i\chi
\right) -2a^{\ast }\right\vert ^{2}
\]

This form of the energy clearly shows in what consists the approach to
stationarity and the formation of structure:

\begin{enumerate}
\item a constant $\rho $ on the equilines, defined as the zero of the
derivatives on $x_{-}$;

\item the potentials $a$ and $a^{\ast }$ become velocities and they become
equal with the derivatives along the equilines of the angle $\chi $.
\end{enumerate}

The fact that the physical velocity is the derivative of the angle of the
complex function $\phi $ is already known from the Abelian-Higgs model and
is valid outside the positions of the centres of the vortices. This also
shows that $\chi $ must have the same physical dimension like $\psi $, 
\[
\left\langle \chi \right\rangle =\frac{m^{2}}{s}
\]%
and that the phase $\chi $ is normalized to a quantity that has the
dimensions of $\left\langle \psi \right\rangle $.

\section{Detailed expression of the Euler current}

\subsection{Introduction}

We note that in the derivation of the Bogomolnyi form of the energy it was
not necessary to impose the static states. Then at this moment the states
may still have a time evolution, assuming that the fluid evolves towards the
SD 
\[
D_{-}\phi \approx 0
\]%
In this case we can combine the spatial components of the current density 
\begin{eqnarray*}
J^{+} &=&J^{x}+iJ^{y} \\
&=&-\frac{i}{2m}\left( \left[ \phi ^{\dagger },\left( D^{+}\phi \right) %
\right] -\left[ \left( D^{-}\phi \right) ^{\dagger },\phi \right] \right) 
\end{eqnarray*}%
and inserting the equation written above we get 
\[
J^{+}=-\frac{i}{2m}\left( \left[ \phi ^{\dagger },\left( D^{+}\phi \right) %
\right] \right) 
\]

We return to the expression of the current in the second (gauge-field)
equation of motion 
\begin{eqnarray*}
\kappa \varepsilon ^{x\mu \nu }F_{\mu \nu } &=&-iJ^{x} \\
\kappa \varepsilon ^{y\mu \nu }F_{\mu \nu } &=&-iJ^{y}
\end{eqnarray*}%
\begin{eqnarray*}
\kappa \left( \varepsilon ^{xy0}F_{y0}+\varepsilon ^{x0y}F_{0y}\right) 
&=&-iJ^{x} \\
2\kappa F_{y0} &=&-iJ^{x} \\
2\kappa \left( \partial _{y}A_{0}-\partial _{0}A_{y}+\left[ A_{y},A_{0}%
\right] \right)  &=&-iJ^{x}
\end{eqnarray*}%
and analogous 
\begin{eqnarray*}
\kappa \left( \varepsilon ^{yx0}F_{x0}+\varepsilon ^{y0x}F_{0x}\right) 
&=&-iJ^{y} \\
-2\kappa F_{x0} &=&-iJ^{y} \\
-2\kappa \left( \partial _{x}A_{0}-\partial _{0}A_{x}+\left[ A_{x},A_{0}%
\right] \right)  &=&-iJ^{y}
\end{eqnarray*}%
Now we combine them 
\begin{eqnarray*}
-i\left( J^{x}+iJ^{y}\right)  &=&2\kappa \left( \partial _{y}A_{0}-\partial
_{0}A_{y}+\left[ A_{y},A_{0}\right] \right.  \\
&&\left. -i\left( \partial _{x}A_{0}-\partial _{0}A_{x}+\left[ A_{x},A_{0}%
\right] \right) \right)  \\
&=&2\kappa \left( \left( \partial _{y}-i\partial _{x}\right) A_{0}\right.  \\
&&-\partial _{0}\left( A_{y}+iA_{x}\right)  \\
&&\left. +\left[ A_{y}-iA_{x},A_{0}\right] \right)  \\
&=&\frac{2\kappa }{i}\left( \left( \partial _{x}+i\partial _{y}\right)
A_{0}\right.  \\
&&-\partial _{0}\left( A_{x}-iA_{y}\right)  \\
&&\left. +\left[ A_{x}+iA_{y},A_{0}\right] \right) 
\end{eqnarray*}%
\begin{eqnarray*}
J^{+} &=&2\kappa \left( \partial ^{+}A_{0}-\partial _{0}A_{-}+\left[
A_{+},A_{0}\right] \right)  \\
&=&2\kappa \left( D^{+}A_{0}-\partial _{0}A_{-}\right) 
\end{eqnarray*}%
where we have introduced the notation 
\[
D^{+}\equiv \partial ^{+}+\left[ A^{+},\right] 
\]%
Now we have two expressions for the current density $J^{+}$%
\begin{eqnarray*}
J^{+} &=&-\frac{i}{2m}\left( \left[ \phi ^{\dagger },\left( D^{+}\phi
\right) \right] \right)  \\
J^{+} &=&2\kappa \left( D^{+}A_{0}-\partial _{0}A_{-}\right) 
\end{eqnarray*}%
At stationarity 
\[
\partial _{0}A_{-}=0
\]%
and 
\[
A_{0}=\frac{i}{4m\kappa }\left[ \phi ,\phi ^{\dagger }\right] 
\]%
This shows that the zero component of the potential of interaction has
algebraic content reduced to the Cartan generator 
\[
A_{0}\sim H
\]%
The magnitude of $A_{0}$ is given by the \emph{charge} $\rho $ or by the
magnetic field $B$ (connected via the Gauss law).

However at stationarity we still have%
\[
\frac{\partial A_{-}}{\partial t}=\frac{\partial }{\partial t}\partial
_{-}\chi 
\]%
since in the expression of $\mathbf{A}$ there is a gradient of an angle.

Then we can say that the system evolves toward this minimum.

\bigskip

\subsubsection{A note on the possible meaning of the equation of motion for
the \textbf{LIOUVILLE }equation}

The equation for $\rho $ is very complicated and has the nature of a
constraint that determines a family of functions $\rho $. For some of them
the current is NOT divergenceless and there is a time variation of $\rho $
from the equation of continuity. But the new function $\rho $ resulted from
advancing in time must be again a solution of the second equation =
constraint.

\bigskip

Let us take the expression of the current%
\[
\mathbf{j}=-g\rho \mathbf{A}
\]%
where $g$ is a dimensional constant factor.%
\[
\mathbf{A}\left( \mathbf{r}\right) =\mathbf{\nabla }\times \frac{1}{\kappa }%
\int d^{2}r\left[ \frac{1}{2\pi }\ln \left( \frac{\left\vert \mathbf{r-r}%
^{\prime }\right\vert }{L}\right) \right] \rho 
\]%
We note that%
\[
\mathbf{\nabla \cdot A}=0
\]%
The current is 
\[
\mathbf{j}=-g\rho \left( \mathbf{\nabla }\times \frac{1}{\kappa }\int
d^{2}r^{\prime }\left[ \frac{1}{2\pi }\ln \left( \frac{\left\vert \mathbf{r-r%
}^{\prime }\right\vert }{L}\right) \right] \rho \left( \mathbf{r}^{\prime
}\right) \right) 
\]%
The divergence of the current is%
\begin{eqnarray*}
\mathbf{\nabla \cdot j} &=&-g\mathbf{\nabla \cdot }\left( \rho \mathbf{A}%
\right)  \\
&=&-g\left[ \left( \mathbf{A\cdot \nabla }\right) \rho +\rho \left( \mathbf{%
\nabla \cdot A}\right) \right] 
\end{eqnarray*}%
and using the Coulomb gauge, as shown above, we have%
\[
\mathbf{\nabla \cdot j}=-g\left( \mathbf{A\cdot \nabla }\right) \rho 
\]%
The equation of continuity becomes%
\begin{eqnarray*}
\frac{\partial \rho }{\partial t}+\mathbf{\nabla \cdot J} &=&0 \\
\frac{\partial \rho }{\partial t}-g\left( \mathbf{A\cdot \nabla }\right)
\rho  &=&0
\end{eqnarray*}%
This should look like%
\[
\frac{\partial \rho }{\partial t}+\left( \mathbf{v\cdot \nabla }\right) \rho
=0
\]%
with $\mathbf{v}$ the divergenceless field of velocities associated (but
only at \emph{self-duality}) with $\mathbf{A}$.

We should identify%
\[
\mathbf{v\equiv }-g\mathbf{A}
\]

\bigskip

The density $\rho $, which for the ABELIAN Jackiw Pi case, or LIOUVILLE
case, is the $0$-th component of the current $J^{\mu }$ , is constant along
the trajectories of the vector field $\mathbf{A}$, which at the self-dual
limit will be $\mathbf{v}$.

Then $\rho $ will evolve adiabatically as the lines of flow of $\mathbf{A}$.
If these lines are converging to the center then we will have a
concentration of \emph{matter} field.

\subsection{The non-covariant charge of the FT Euler}

The paper \textbf{9410065 Dunne} identifies the Abelian, non-covariant
charges%
\begin{eqnarray*}
Q^{0} &=&\mathrm{tr}\left( \Psi ^{\dagger }\Psi \right) \\
Q^{k} &=&-\frac{i}{2m}\mathrm{tr}\left[ \Psi ^{\dagger }\left( D^{k}\Psi
\right) -\left( D^{k}\Psi \right) ^{\dagger }\Psi \right]
\end{eqnarray*}

The $0$ component of this charge vector is calculated%
\[
Q^{0}=\mathrm{tr}\left( \Psi ^{\dagger }\Psi \right) =\rho _{1}+\rho _{2}
\]

The $k$ component of the charge must be calculated. We take into account that%
\begin{eqnarray}
A_{x} &=&\frac{1}{2}\left( a-a^{\ast }\right) H  \label{702} \\
A_{y} &=&\frac{1}{2i}\left( a+a^{\ast }\right) H  \nonumber
\end{eqnarray}%
\begin{eqnarray*}
D^{x}\Psi &=&\left( \frac{\partial }{\partial x}+\left[ A^{x},\right]
\right) \left( \phi _{1}E_{+}+\phi _{2}E_{-}\right) \\
&=&\frac{\partial \phi _{1}}{\partial x}E_{+}+\frac{1}{2}\left( a-a^{\ast
}\right) \phi _{1}\left[ H,E_{+}\right] +\frac{\partial \phi _{2}}{\partial x%
}E_{-}+\frac{1}{2}\left( a-a^{\ast }\right) \phi _{2}\left[ H,E_{-}\right] \\
&=&\left[ \frac{\partial \phi _{1}}{\partial x}+\left( a-a^{\ast }\right)
\phi _{1}\right] E_{+}+\left[ \frac{\partial \phi _{2}}{\partial x}+\left(
a-a^{\ast }\right) \phi _{2}\right] E_{-}
\end{eqnarray*}%
For the Hermitean conjugate%
\begin{eqnarray*}
\left( D^{x}\Psi \right) ^{\dagger } &=&\left\{ \left( \frac{\partial }{%
\partial x}+\left[ A^{x},\right] \right) \left( \phi _{1}E_{+}+\phi
_{2}E_{-}\right) \right\} ^{\dagger } \\
&=&\left\{ \left[ \frac{\partial \phi _{1}}{\partial x}+\left( a-a^{\ast
}\right) \phi _{1}\right] E_{+}+\left[ \frac{\partial \phi _{2}}{\partial x}%
+\left( a-a^{\ast }\right) \phi _{2}\right] E_{-}\right\} ^{\dagger } \\
&=&\left[ \frac{\partial \phi _{1}^{\ast }}{\partial x}+\left( a^{\ast
}-a\right) \phi _{1}^{\ast }\right] E_{-}+\left[ \frac{\partial \phi
_{2}^{\ast }}{\partial x}+\left( a^{\ast }-a\right) \phi _{2}^{\ast }\right]
E_{+}
\end{eqnarray*}%
The same calculation is made for the $y$ component%
\begin{eqnarray*}
D^{y}\Psi &=&\left( \frac{\partial }{\partial y}+\left[ A^{y},\right]
\right) \left( \phi _{1}E_{+}+\phi _{2}E_{-}\right) \\
&=&\frac{\partial \phi _{1}}{\partial y}E_{+}+\frac{1}{2i}\left( a+a^{\ast
}\right) \phi _{1}\left[ H,E_{+}\right] +\frac{\partial \phi _{2}}{\partial y%
}E_{-}+\frac{1}{2i}\left( a+a^{\ast }\right) \phi _{2}\left[ H,E_{-}\right]
\\
&=&\left[ \frac{\partial \phi _{1}}{\partial y}+\frac{1}{i}\left( a+a^{\ast
}\right) \phi _{1}\right] E_{+}+\left[ \frac{\partial \phi _{2}}{\partial y}+%
\frac{1}{i}\left( a+a^{\ast }\right) \phi _{2}\right] E_{-}
\end{eqnarray*}%
\begin{eqnarray*}
\left( D^{y}\Psi \right) ^{\dagger } &=&\left\{ \left[ \frac{\partial \phi
_{1}}{\partial y}+\frac{1}{i}\left( a+a^{\ast }\right) \phi _{1}\right]
E_{+}+\left[ \frac{\partial \phi _{2}}{\partial y}+\frac{1}{i}\left(
a+a^{\ast }\right) \phi _{2}\right] E_{-}\right\} ^{\dagger } \\
&=&\left[ \frac{\partial \phi _{1}^{\ast }}{\partial y}-\frac{1}{i}\left(
a+a^{\ast }\right) \phi _{1}^{\ast }\right] E_{-}+\left[ \frac{\partial \phi
_{2}^{\ast }}{\partial y}-\frac{1}{i}\left( a+a^{\ast }\right) \phi
_{2}^{\ast }\right] E_{+}
\end{eqnarray*}%
Using these formulas we can write in detail the charges%
\[
Q^{x}=-\frac{i}{2m}\mathrm{tr}\left[ \Psi ^{\dagger }\left( D^{x}\Psi
\right) -\left( D^{x}\Psi \right) ^{\dagger }\Psi \right] 
\]%
and consider the first term, 
\[
Q_{I}^{x}\equiv -\frac{i}{2m}\mathrm{tr}\left[ \Psi ^{\dagger }\left(
D^{x}\Psi \right) \right] 
\]%
\begin{eqnarray*}
&&Q_{I}^{x} \\
\hspace{-2cm} &=&-\frac{i}{2m}\mathrm{tr}\left\{ \left( \left[ \frac{%
\partial \phi _{1}^{\ast }}{\partial y}-\frac{1}{i}\left( a+a^{\ast }\right)
\phi _{1}^{\ast }\right] E_{-}+\left[ \frac{\partial \phi _{2}^{\ast }}{%
\partial y}-\frac{1}{i}\left( a+a^{\ast }\right) \phi _{2}^{\ast }\right]
E_{+}\right) \right\} \\
&=&-\frac{i}{2m}\mathrm{tr}\left\{ \phi _{1}^{\ast }\left[ \frac{\partial
\phi _{1}}{\partial x}+\left( a-a^{\ast }\right) \phi _{1}\right]
E_{-}E_{+}\right. \ \text{trace is }1 \\
&&+\phi _{2}^{\ast }\left[ \frac{\partial \phi _{1}}{\partial x}+\left(
a-a^{\ast }\right) \phi _{1}\right] E_{+}E_{+}\ \ \text{trace is }0 \\
&&+\phi _{1}^{\ast }\left[ \frac{\partial \phi _{2}}{\partial x}+\left(
a-a^{\ast }\right) \phi _{2}\right] E_{-}E_{-}\ \ \text{trace is }0 \\
&&\left. +\phi _{2}^{\ast }\left[ \frac{\partial \phi _{2}}{\partial x}%
+\left( a-a^{\ast }\right) \phi _{2}\right] E_{+}E_{-}\ \ \text{trace is }%
1\right\}
\end{eqnarray*}%
Then%
\begin{eqnarray*}
Q_{I}^{x} &=&-\frac{i}{2m}\left\{ \phi _{1}^{\ast }\left[ \frac{\partial
\phi _{1}}{\partial x}+\left( a-a^{\ast }\right) \phi _{1}\right] +\phi
_{2}^{\ast }\left[ \frac{\partial \phi _{2}}{\partial x}+\left( a-a^{\ast
}\right) \phi _{2}\right] \right\} \\
&=&-\frac{i}{2m}\left[ \phi _{1}^{\ast }\frac{\partial \phi _{1}}{\partial x}%
+\phi _{2}^{\ast }\frac{\partial \phi _{2}}{\partial x}+\left( \rho
_{1}+\rho _{2}\right) \left( a-a^{\ast }\right) \right]
\end{eqnarray*}%
Similarly we calculate the second term%
\[
Q_{II}^{x}\equiv \frac{i}{2m}\mathrm{tr}\left[ \left( D^{x}\Psi \right)
^{\dagger }\Psi \right] 
\]%
and obtain%
\begin{eqnarray*}
&&Q_{II}^{x} \\
\hspace{-2cm} &=&\frac{i}{2m}\mathrm{tr}\left\{ \left( \left[ \frac{\partial
\phi _{1}^{\ast }}{\partial x}+\left( a^{\ast }-a\right) \phi _{1}^{\ast }%
\right] E_{-}+\left[ \frac{\partial \phi _{2}^{\ast }}{\partial x}+\left(
a^{\ast }-a\right) \phi _{2}^{\ast }\right] E_{+}\right) \left( \phi
_{1}E_{+}+\phi _{2}E_{-}\right) \right\} \\
&=&\frac{i}{2m}\mathrm{tr}\left\{ \left[ \frac{\partial \phi _{1}^{\ast }}{%
\partial x}+\left( a^{\ast }-a\right) \phi _{1}^{\ast }\right] \phi
_{1}E_{-}E_{+}\right. \ \text{trace is }1 \\
&&+\left[ \frac{\partial \phi _{1}^{\ast }}{\partial x}+\left( a^{\ast
}-a\right) \phi _{1}^{\ast }\right] \phi _{2}E_{-}E_{-}\ \ \text{trace is }0
\\
&&+\left[ \frac{\partial \phi _{2}^{\ast }}{\partial x}+\left( a^{\ast
}-a\right) \phi _{2}^{\ast }\right] \phi _{1}E_{+}E_{+}\ \ \text{trace is }0
\\
&&\left. +\ \left[ \frac{\partial \phi _{2}^{\ast }}{\partial x}+\left(
a^{\ast }-a\right) \phi _{2}^{\ast }\right] \phi _{2}E_{+}E_{-}\ \text{trace
is }1\right\}
\end{eqnarray*}%
which further gives%
\begin{eqnarray*}
Q_{II}^{x} &=&\frac{i}{2m}\left\{ \left[ \frac{\partial \phi _{1}^{\ast }}{%
\partial x}+\left( a^{\ast }-a\right) \phi _{1}^{\ast }\right] \phi _{1}+\ %
\left[ \frac{\partial \phi _{2}^{\ast }}{\partial x}+\left( a^{\ast
}-a\right) \phi _{2}^{\ast }\right] \phi _{2}\right\} \\
&=&\frac{i}{2m}\left[ \frac{\partial \phi _{1}^{\ast }}{\partial x}\phi _{1}+%
\frac{\partial \phi _{2}^{\ast }}{\partial x}\phi _{2}+\left( \rho _{1}+\rho
_{2}\right) \left( a^{\ast }-a\right) \right]
\end{eqnarray*}

The preceding results can be combined into%
\begin{eqnarray*}
Q^{x} &=&Q_{I}^{x}+Q_{II}^{x} \\
&=&-\frac{i}{2m}\left[ \phi _{1}^{\ast }\frac{\partial \phi _{1}}{\partial x}%
+\phi _{2}^{\ast }\frac{\partial \phi _{2}}{\partial x}+\left( \rho
_{1}+\rho _{2}\right) \left( a-a^{\ast }\right) \right] \\
&&+\frac{i}{2m}\left[ \frac{\partial \phi _{1}^{\ast }}{\partial x}\phi _{1}+%
\frac{\partial \phi _{2}^{\ast }}{\partial x}\phi _{2}+\left( \rho _{1}+\rho
_{2}\right) \left( a^{\ast }-a\right) \right] \\
&=&-\frac{i}{2m}\left[ \phi _{1}^{\ast }\frac{\partial \phi _{1}}{\partial x}%
-\frac{\partial \phi _{1}^{\ast }}{\partial x}\phi _{1}+\phi _{2}^{\ast }%
\frac{\partial \phi _{2}}{\partial x}-\frac{\partial \phi _{2}^{\ast }}{%
\partial x}\phi _{2}\right. \\
&&\left. +2\left( \rho _{1}+\rho _{2}\right) \left( a-a^{\ast }\right) 
\right]
\end{eqnarray*}%
and we note that%
\begin{eqnarray*}
\phi _{1}^{\ast }\frac{\partial \phi _{1}}{\partial x}-\frac{\partial \phi
_{1}^{\ast }}{\partial x}\phi _{1} &=&\left\vert \phi _{1}^{\ast
}\right\vert ^{2}\frac{\partial }{\partial x}\left( \frac{\phi _{1}}{\phi
_{1}^{\ast }}\right) \\
&=&\rho _{1}\frac{\partial }{\partial x}\left[ 2\times \text{phase of }\phi
_{1}\right] \\
&=&2\rho _{1}\frac{\partial \chi }{\partial x}
\end{eqnarray*}%
and 
\[
\phi _{2}^{\ast }\frac{\partial \phi _{2}}{\partial x}-\frac{\partial \phi
_{2}^{\ast }}{\partial x}\phi _{2}=2\rho _{2}\frac{\partial \eta }{\partial x%
} 
\]

Then%
\[
Q^{x}=-\frac{i}{m}\left[ \rho _{1}\frac{\partial \chi }{\partial x}+\rho _{2}%
\frac{\partial \eta }{\partial x}+\left( \rho _{1}+\rho _{2}\right) \left(
a-a^{\ast }\right) \right] 
\]

We can write for $Q^{y}$ an analogous form%
\[
Q^{y}=-\frac{i}{m}\left[ \rho _{1}\frac{\partial \chi }{\partial y}+\rho _{2}%
\frac{\partial \eta }{\partial y}+\frac{1}{i}\left( \rho _{1}+\rho
_{2}\right) \left( a+a^{\ast }\right) \right] 
\]

\bigskip

This charge is ordinarly conserved%
\[
\frac{\partial Q^{\mu }}{\partial x^{\mu }}=0.
\]

\subsection{The current of the Euler FT}

The formula for the FT current in the Euler case is (\textbf{Dunne}) 
\begin{eqnarray*}
J^{0} &=&\left[ \Psi ^{\dagger },\Psi \right]  \\
J^{i} &=&-\frac{i}{2}\left( \left[ \Psi ^{\dagger },D_{i}\Psi \right] -\left[
\left( D_{i}\Psi \right) ^{\dagger },\Psi \right] \right) 
\end{eqnarray*}

We note few differences: the CS in the Euler case is defined with the factor 
\[
\kappa 
\]
which we cannot associate to the sound speed since there is no $\rho _{s}$
and $\Omega _{ci}$ for Euler fluid.

The two \emph{similar} quantities are 
\begin{eqnarray*}
L &\equiv &\text{length of the space box} \\
\omega _{tot} &\equiv &\text{the total amount of vorticity, an invariant for
Euler} \\
&=&\frac{1}{L^{2}}\int d^{2}r\omega \left( x,y\right)
\end{eqnarray*}

\textbf{NOTE}

On the units in the Abelian CS (Abelian Euler)  
\[
L_{CS}=\frac{\kappa }{2c}\int d^{2}r\frac{\partial \mathbf{A}}{\partial t}%
\times \mathbf{A-}\kappa \int d^{2}rA^{0}B
\]%
This will further be integrated over time to give the adimensional \emph{%
action} functional. 
\[
\frac{\left[ \kappa \right] }{c}L^{2}\frac{1}{T}\left[ A\right] ^{2}=\frac{1%
}{T}
\]%
Here units of $\kappa $ is physically that of speed, then 
\begin{eqnarray*}
\left[ A\right]  &=&L^{-1} \\
\left[ B\right]  &=&\frac{1}{L}\frac{1}{L}
\end{eqnarray*}

\bigskip

The current for $\mu \equiv k$ (space components) is, after the calculations
presented in more detail for CHM (next Section) 
\begin{eqnarray*}
J^{k} &=&-\frac{i}{2}\left\{ \Psi ^{\dagger }\left( \partial ^{k}\Psi
\right) -\left( \partial ^{k}\Psi \right) \Psi ^{\dagger }-\left( \partial
^{k}\Psi ^{\dagger }\right) \Psi +\Psi \left( \partial ^{k}\Psi ^{\dagger
}\right) \right. \\
&&\left. +\left[ \Psi ^{\dagger },\left[ A^{k},\Psi \right] \right] +\left[
\Psi ,\left[ \Psi ^{\dagger },A^{k\dagger }\right] \right] \right\} \\
&\equiv &\Lambda _{1}^{k}+\Lambda _{2}^{k}
\end{eqnarray*}
where 
\begin{eqnarray*}
\Lambda _{1}^{k} &\equiv &-\frac{i}{2}\left\{ \Psi ^{\dagger }\left(
\partial ^{k}\Psi \right) -\left( \partial ^{k}\Psi \right) \Psi ^{\dagger
}-\left( \partial ^{k}\Psi ^{\dagger }\right) \Psi +\Psi \left( \partial
^{k}\Psi ^{\dagger }\right) \right\} \\
\Lambda _{2}^{k} &\equiv &-\frac{i}{2}\left( \left[ \Psi ^{\dagger },\left[
A^{k},\Psi \right] \right] +\left[ \Psi ,\left[ \Psi ^{\dagger },A^{k\dagger
}\right] \right] \right)
\end{eqnarray*}

\subsubsection{The expression of the first part of the current, $\Lambda
_{1} $}

The terms containing space and time derivatives (here the symbol $\Psi $ is
replaced temporarly by $\phi $) 
\[
\Lambda _{1}^{k}=-\frac{i}{2}\left[ \phi ^{\dagger }\left( \partial ^{k}\phi
\right) -\left( \partial ^{k}\phi \right) \phi ^{\dagger }-\left( \partial
^{k}\phi ^{\dagger }\right) \phi +\phi \left( \partial ^{k}\phi ^{\dagger
}\right) \right] 
\]
where we have to insert 
\begin{eqnarray*}
\phi &=&\phi _{1}E_{+}+\phi _{2}E_{-} \\
\phi ^{\dagger } &=&\phi _{1}^{\ast }E_{-}+\phi _{2}^{\ast }E_{+}
\end{eqnarray*}
This consists of two commutators.

The first commutator is 
\begin{eqnarray*}
\left[ \phi ^{\dagger },\partial ^{k}\phi \right] &=&\phi ^{\dagger }\left(
\partial ^{k}\phi \right) -\left( \partial ^{k}\phi \right) \phi ^{\dagger }
\\
&=&\left( \phi _{1}^{\ast }E_{-}+\phi _{2}^{\ast }E_{+}\right) \left( \frac{%
\partial \phi _{1}}{\partial x_{k}}E_{+}+\frac{\partial \phi _{2}}{\partial
x_{k}}E_{-}\right) \\
&&-\left( \frac{\partial \phi _{1}}{\partial x_{k}}E_{+}+\frac{\partial \phi
_{2}}{\partial x_{k}}E_{-}\right) \left( \phi _{1}^{\ast }E_{-}+\phi
_{2}^{\ast }E_{+}\right) \\
&=&\phi _{1}^{\ast }\frac{\partial \phi _{1}}{\partial x_{k}}E_{-}E_{+}+\phi
_{1}^{\ast }\frac{\partial \phi _{2}}{\partial x_{k}}E_{-}E_{-}+\phi
_{2}^{\ast }\frac{\partial \phi _{1}}{\partial x_{k}}E_{+}E_{+}+\phi
_{2}^{\ast }\frac{\partial \phi _{2}}{\partial x_{k}}E_{+}E_{-} \\
&&-\phi _{1}^{\ast }\frac{\partial \phi _{1}}{\partial x_{k}}E_{+}E_{-}-\phi
_{2}^{\ast }\frac{\partial \phi _{1}}{\partial x_{k}}E_{+}E_{+}-\phi
_{1}^{\ast }\frac{\partial \phi _{2}}{\partial x_{k}}E_{-}E_{-}-\phi
_{2}^{\ast }\frac{\partial \phi _{2}}{\partial x_{k}}E_{-}E_{+}
\end{eqnarray*}
The coefficients of $E_{-}E_{-}$ and of $E_{+}E_{+}$ cancel. The result is 
\begin{eqnarray*}
&&\left[ \phi ^{\dagger },\partial ^{k}\phi \right] \\
&=&\phi _{1}^{\ast }\frac{\partial \phi _{1}}{\partial x_{k}}\left[
E_{-},E_{+}\right] +\phi _{2}^{\ast }\frac{\partial \phi _{2}}{\partial x_{k}%
}\left[ E_{+},E_{-}\right]
\end{eqnarray*}
Here we can use 
\begin{eqnarray*}
\left[ E_{+},E_{-}\right] &=&H \\
\left[ H,E_{\pm }\right] &=&\pm 2E_{\pm } \\
\mathrm{tr}\left( E_{+}E_{-}\right) &=&1 \\
\mathrm{tr}\left( H^{2}\right) &=&2
\end{eqnarray*}
and obtain 
\[
\left[ \phi ^{\dagger },\partial ^{k}\phi \right] =\left( \phi _{1}^{\ast }%
\frac{\partial \phi _{1}}{\partial x_{k}}-\phi _{2}^{\ast }\frac{\partial
\phi _{2}}{\partial x_{k}}\right) H 
\]

The second commutator in $\Lambda _{1}^{k}$ is 
\begin{eqnarray*}
\left[ \phi ,\partial ^{k}\phi ^{\dagger }\right] &=&\phi \left( \partial
^{k}\phi ^{\dagger }\right) -\left( \partial ^{k}\phi ^{\dagger }\right) \phi
\\
&=&\left( \phi _{1}E_{+}+\phi _{2}E_{-}\right) \left( \frac{\partial \phi
_{1}^{\ast }}{\partial x_{k}}E_{-}+\frac{\partial \phi _{2}^{\ast }}{%
\partial x_{k}}E_{+}\right) \\
&&-\left( \frac{\partial \phi _{1}^{\ast }}{\partial x_{k}}E_{-}+\frac{%
\partial \phi _{2}^{\ast }}{\partial x_{k}}E_{+}\right) \left( \phi
_{1}E_{+}+\phi _{2}E_{-}\right) \\
&=&\phi _{1}\frac{\partial \phi _{1}^{\ast }}{\partial x_{k}}E_{+}E_{-}+\phi
_{2}\frac{\partial \phi _{1}^{\ast }}{\partial x_{k}}E_{-}E_{-}+\phi _{1}%
\frac{\partial \phi _{2}^{\ast }}{\partial x_{k}}E_{+}E_{+}+\phi _{2}\frac{%
\partial \phi _{2}^{\ast }}{\partial x_{k}}E_{-}E_{+} \\
&&-\phi _{1}\frac{\partial \phi _{1}^{\ast }}{\partial x_{k}}E_{-}E_{+}-\phi
_{1}\frac{\partial \phi _{2}^{\ast }}{\partial x_{k}}E_{+}E_{+}-\phi _{2}%
\frac{\partial \phi _{1}^{\ast }}{\partial x_{k}}E_{-}E_{-}-\phi _{2}\frac{%
\partial \phi _{2}^{\ast }}{\partial x_{k}}E_{+}E_{-}
\end{eqnarray*}
As above, the coefficients of the terms $E_{+}E_{+}$ and respectively $%
E_{-}E_{-}$ cancel. The other represent commutators that can be expressed by 
$H$: 
\begin{eqnarray*}
&&\left[ \phi ,\partial ^{k}\phi ^{\dagger }\right] \\
&=&\phi _{1}\frac{\partial \phi _{1}^{\ast }}{\partial x_{k}}\left[
E_{+},E_{-}\right] -\phi _{2}\frac{\partial \phi _{2}^{\ast }}{\partial x_{k}%
}\left[ E_{+},E_{-}\right] \\
&=&\left( \phi _{1}\frac{\partial \phi _{1}^{\ast }}{\partial x_{k}}-\phi
_{2}\frac{\partial \phi _{2}^{\ast }}{\partial x_{k}}\right) H
\end{eqnarray*}

Putting together these results we have

\begin{eqnarray*}
\Lambda _{1}^{k} &=&-\frac{i}{2}\left[ \phi ^{\dagger }\left( \partial
^{k}\phi \right) -\left( \partial ^{k}\phi \right) \phi ^{\dagger }-\left(
\partial ^{k}\phi ^{\dagger }\right) \phi +\phi \left( \partial ^{k}\phi
^{\dagger }\right) \right] \\
&=&-\frac{i}{2}\left\{ \left[ \phi ^{\dagger },\partial ^{k}\phi \right] +%
\left[ \phi ,\partial ^{k}\phi ^{\dagger }\right] \right\} \\
&=&-\frac{i}{2}\left[ \left( \phi _{1}^{\ast }\frac{\partial \phi _{1}}{%
\partial x_{k}}-\phi _{2}^{\ast }\frac{\partial \phi _{2}}{\partial x_{k}}%
\right) H+\left( \phi _{1}\frac{\partial \phi _{1}^{\ast }}{\partial x_{k}}%
-\phi _{2}\frac{\partial \phi _{2}^{\ast }}{\partial x_{k}}\right) H\right]
\\
&=&-\frac{i}{2}\left( \frac{\partial }{\partial x_{k}}\phi _{1}\phi
_{1}^{\ast }-\frac{\partial }{\partial x_{k}}\phi _{2}\phi _{2}^{\ast
}\right) H \\
&=&-\frac{i}{2}\frac{\partial }{\partial x_{k}}\left( \rho _{1}-\rho
_{2}\right) H
\end{eqnarray*}

\subsubsection{The expression of the second part of the current, $\Lambda
_{2}$}

The second line, $\Lambda _{2}$, is calculated in the text \textbf{xxx}. 
\[
\Lambda _{2}^{k}\equiv -\frac{i}{2}\left( \left[ \phi ^{\dagger },\left[
A^{k},\phi \right] \right] +\left[ \phi ,\left[ \phi ^{\dagger },A^{k\dagger
}\right] \right] \right) 
\]

We have to give detailed expressions for all components of the current.

The $x$ component 
\begin{eqnarray*}
\Lambda _{2}^{x} &=&-\frac{i}{2}\left\{ -\left( a-a^{\ast }\right) \left(
\rho _{1}+\rho _{2}\right) H+\left( a^{\ast }-a\right) \left( \rho _{1}+\rho
_{2}\right) H\right\} \\
&=&2\frac{i}{2}(a-a^{\ast })\left( \rho _{1}+\rho _{2}\right) H
\end{eqnarray*}

The $y$ component 
\begin{eqnarray*}
\Lambda _{2}^{y} &=&-\frac{i}{2}\left\{ i\left( a+a^{\ast }\right) \left(
\rho _{1}+\rho _{2}\right) H+i\left( a^{\ast }+a\right) \left( \rho
_{1}+\rho _{2}\right) H\right\} \\
&=&2\frac{1}{2}(a+a^{\ast })\left( \rho _{1}+\rho _{2}\right) H
\end{eqnarray*}

\subsubsection{The time component of the Euler current}

This is given by 
\begin{eqnarray*}
J^{0} &=&\left[ \Psi ,\Psi ^{\dagger }\right] \\
&=&\left[ \phi _{1}E_{+}+\phi _{2}E_{-},\phi _{1}^{\ast }E_{-}+\phi
_{2}^{\ast }E_{+}\right] \\
&=&\phi _{1}\phi _{1}^{\ast }\left[ E_{+},E_{-}\right] +\phi _{2}\phi
_{2}^{\ast }\left[ E_{-},E_{+}\right] \\
&=&\left| \phi _{1}\right| ^{2}H-\left| \phi _{2}\right| ^{2}H
\end{eqnarray*}
or 
\[
J^{0}=\left( \rho _{1}-\rho _{2}\right) H 
\]

\subsubsection{The total expression of the EULER current $J^{\protect\mu }$}

Finally 
\[
J^{\mu }=\Lambda _{1}^{\mu }+\Lambda _{2}^{\mu } 
\]
gives 
\begin{eqnarray*}
J^{x} &=&\frac{1}{2}\left[ 2i(a-a^{\ast })\left( \rho _{1}+\rho _{2}\right)
-i\frac{\partial }{\partial x}\left( \rho _{1}-\rho _{2}\right) \right] H \\
J^{y} &=&\frac{1}{2}\left[ 2(a+a^{\ast })\left( \rho _{1}+\rho _{2}\right) -i%
\frac{\partial }{\partial y}\left( \rho _{1}-\rho _{2}\right) \right] H \\
J^{0} &=&\left( \rho _{1}-\rho _{2}\right) H
\end{eqnarray*}

\bigskip

We note that 
\begin{eqnarray*}
A_{x} &=&\frac{1}{2}\left( a-a^{\ast }\right) H \\
A_{y} &=&\frac{1}{2i}\left( a+a^{\ast }\right) H
\end{eqnarray*}

\subsubsection{Expression of the Euler current at \textbf{SELF-DUALITY}}

At self-duality (and only at self-duality) we can replace the functions $a$
and $a^{\ast }$ that define the potentials $A_{\pm }$ with expressions of
the functions $\phi _{1,2}$ and $\phi _{1,2}^{\ast }$ coming from the first
equation of self-duality, $D_{-}\phi =0$.

We have 
\begin{eqnarray*}
J^{x} &=&\frac{1}{2}\left[ 2i(a-a^{\ast })\left( \rho _{1}+\rho _{2}\right)
-i\frac{\partial }{\partial x}\left( \rho _{1}-\rho _{2}\right) \right] \\
&=&\frac{1}{2}2i\frac{i}{2}\left[ \frac{\partial \psi }{\partial y}-\frac{%
\partial \left( 2\chi \right) }{\partial x}\right] \left( \rho _{1}+\rho
_{2}\right) -\frac{1}{2}i\frac{\partial }{\partial x}\left( \rho _{1}-\rho
_{2}\right) \\
&=&-\frac{1}{2}\left[ \frac{\partial \psi }{\partial y}-\frac{\partial
\left( 2\chi \right) }{\partial x}\right] \left( \rho _{1}+\rho _{2}\right) -%
\frac{1}{2}i\frac{\partial }{\partial x}\left( \rho _{1}-\rho _{2}\right)
\end{eqnarray*}%
\begin{eqnarray*}
J^{y} &=&\frac{1}{2}\left[ 2(a+a^{\ast })\left( \rho _{1}+\rho _{2}\right) -i%
\frac{\partial }{\partial y}\left( \rho _{1}-\rho _{2}\right) \right] \\
&=&\frac{1}{2}2\frac{1}{2}\left[ \frac{\partial \psi }{\partial x}+\frac{%
\partial \left( 2\chi \right) }{\partial y}\right] \left( \rho _{1}+\rho
_{2}\right) -\frac{1}{2}i\frac{\partial }{\partial y}\left( \rho _{1}-\rho
_{2}\right) \\
&=&\frac{1}{2}\left[ \frac{\partial \psi }{\partial x}+\frac{\partial \left(
2\chi \right) }{\partial y}\right] \left( \rho _{1}+\rho _{2}\right) -\frac{1%
}{2}i\frac{\partial }{\partial y}\left( \rho _{1}-\rho _{2}\right)
\end{eqnarray*}%
We have 
\[
J_{+}=\frac{1}{2}i\left( \rho _{1}+\rho _{2}\right) \partial _{+}\left[ \psi
-\left( 2i\chi \right) \right] -\frac{1}{2}i\partial _{+}\left( \rho
_{1}-\rho _{2}\right) 
\]%
\[
J_{-}=-\frac{1}{2}i\left( \rho _{1}+\rho _{2}\right) \partial _{-}\left[
\psi +\left( 2i\chi \right) \right] -\frac{1}{2}i\partial _{-}\left( \rho
_{1}-\rho _{2}\right) 
\]

\bigskip

We \textbf{NOTE} that at SELF-DUALITY we have 
\[
\omega =-4\sinh \psi 
\]%
and it results 
\begin{eqnarray*}
J_{+} &=&\frac{1}{2}i\left( \rho _{1}+\rho _{2}\right) \partial _{+}\left[
\psi -\left( 2i\chi \right) \right] -\frac{1}{2}i\partial _{+}\omega \\
J_{-} &=&-\frac{1}{2}i\left( \rho _{1}+\rho _{2}\right) \partial _{-}\left[
\psi +\left( 2i\chi \right) \right] -\frac{1}{2}i\partial _{-}\omega
\end{eqnarray*}%
We try to connect this with the pure self-dual state%
\begin{eqnarray*}
\rho _{1}+\rho _{2} &=&2\cosh \psi \\
\omega &=&-4\sinh \psi \\
\partial _{+}\omega &=&-4\cosh \psi \left( \partial _{+}\psi \right) \\
\partial _{-}\omega &=&-4\cosh \psi \left( \partial _{-}\psi \right)
\end{eqnarray*}%
then%
\begin{eqnarray*}
J_{+} &=&\frac{1}{2}i\left( \rho _{1}+\rho _{2}\right) \partial _{+}\left[
\psi -\left( 2i\chi \right) \right] -\frac{1}{2}i\partial _{+}\omega \\
&=&\frac{1}{2}i2\cosh \psi \left( \partial _{+}\psi \right) -\frac{1}{2}%
i2\cosh \psi \partial _{+}\left( 2i\chi \right) -\frac{1}{2}i\left[ -4\cosh
\psi \left( \partial _{+}\psi \right) \right] \\
&=&2\cosh \psi \left( \partial _{+}\chi \right) \\
&&+i\cosh \psi \left( \partial _{+}\psi \right) +2i\cosh \psi \left(
\partial _{+}\psi \right)
\end{eqnarray*}

\textbf{NOTE}

Assume that%
\begin{eqnarray*}
\partial _{+}\psi &=&0 \\
\partial _{+}\omega &=&0
\end{eqnarray*}

For monopolar vortices with circular symmetry, $\psi $ and $\omega $ are
constant along streamlines 
\begin{eqnarray*}
J_{+} &=&\left( \rho _{1}+\rho _{2}\right) \partial _{+}\chi \\
J_{-} &=&\left( \rho _{1}+\rho _{2}\right) \partial _{-}\chi \\
J_{0} &=&\omega
\end{eqnarray*}%
where 
\[
\rho _{1}+\rho _{2}=2\cosh \psi =2\sqrt{1+\omega ^{2}} 
\]

\textbf{END}

\bigskip

We \emph{note} that when $\omega =0$ the expression $\rho _{1}+\rho _{2}$ is 
\emph{not} zero.

In the \ "second quantisation - like" terminology,  $\rho _{1}$ comes from $%
\phi _{1}$ and represents the creation of vorticity and $\rho _{2}$
represents rarefaction of vortices then their sum is a sort of total effort
in these two actions. They have no reason to be individually $0$ and shows
that in this theory even the vacuum, $\omega \equiv 0$ is obtained from an
effort of densification followed by an equal effort of rarefaction of
vortices, such that the final state is unchanged, is void of vorticity.

\bigskip

\textbf{NOTE}

Let us make a comparison with the structure of the current in the \emph{%
Abelian, non-relativistic, CS, }$4^{th}$ order (nonlinear gauged Schrodinger
eq.) or Abelian version for Euler, or \emph{Liouville}.

The only source of rotational in the expressions of $J^{x,y}$ is 
\[
J^{i}\sim \left( \rho _{1}+\rho _{2}\right) \varepsilon ^{ij}\partial
_{j}\psi 
\]%
but we also have other terms which are gradients 
\[
\frac{\partial \left( 2\chi \right) }{\partial y}\left( \rho _{1}+\rho
_{2}\right) -i\frac{\partial }{\partial y}\left( \rho _{1}-\rho _{2}\right) 
\]%
In the case of the SD states for Euler, we have 
\begin{eqnarray*}
&&\left[ \frac{\partial \left( 2\chi \right) }{\partial y}-i\frac{\partial
\psi }{\partial y}\right] \cosh \psi  \\
&=&-i\cosh \psi \frac{\partial }{\partial y}\left( i2\chi +\psi \right) 
\end{eqnarray*}

The part $\left( \rho _{1}+\rho _{2}\right) \varepsilon ^{ij}\partial
_{j}\psi $ is basically the physical velocity $\widehat{\mathbf{e}}%
_{z}\times \mathbf{\nabla }\psi =\mathbf{v}$.

The part $\frac{\partial \left( 2\chi \right) }{\partial y}\left( \rho
_{1}+\rho _{2}\right) -i\frac{\partial }{\partial y}\left( \rho _{1}-\rho
_{2}\right) $ looks like a pinch of point-like vortices.

\textbf{END}

\bigskip

\subsubsection{Connection between the Euler FT current and the equations of
motion of the initial point-like vortex model}

We look for a different form for the current in \textbf{EULER} case. At
self-duality 
\begin{eqnarray*}
\rho _{1}+\rho _{2} &=&2\cosh \psi \\
\rho _{1}-\rho _{2} &=&2\sinh \psi
\end{eqnarray*}
and

\textbf{Version 1} 
\begin{eqnarray*}
\frac{J_{x}}{\rho _{1}+\rho _{2}} &\simeq &\left[ -\frac{\partial }{\partial
y}\left( \ln \rho _{1}\right) +2\frac{\partial \chi }{\partial x}-i\frac{1}{%
2\cosh \psi }2\cosh \psi \frac{\partial \psi }{\partial x}\right] H \\
&\simeq &\left( -\frac{\partial \psi }{\partial y}+2\frac{\partial \chi }{%
\partial x}-i\frac{\partial \psi }{\partial x}\right) H \\
&=&\left[ \frac{\partial }{\partial x}\left( 2\chi \right) -i\left( \frac{%
\partial }{\partial x}-i\frac{\partial }{\partial y}\right) \psi \right] H
\end{eqnarray*}
We can use the definitions 
\begin{eqnarray}
\partial _{+} &=&\partial _{x}+i\partial _{y}=2\frac{\partial }{\partial
z^{\ast }}  \label{4387} \\
\partial _{-} &=&\partial _{x}-i\partial _{y}=2\frac{\partial }{\partial z} 
\nonumber
\end{eqnarray}

\bigskip

\textbf{Version 2}

We introduce the vorticity 
\begin{eqnarray*}
\rho _{1}-\rho _{2} &=& \\
&=&2\sinh \psi \\
&=&-2\omega
\end{eqnarray*}
and we get 
\begin{eqnarray*}
\frac{J_{x}}{\rho _{1}+\rho _{2}} &\simeq &\left[ -\frac{1}{\rho _{1}}\frac{%
\partial \rho _{1}}{\partial y}+2\frac{\partial \chi }{\partial x}-i\frac{1}{%
\rho _{1}+\rho _{2}}\frac{\partial }{\partial x}\left( \rho _{1}-\rho
_{2}\right) \right] H \\
&=&\left[ -\frac{\partial }{\partial y}\left( \ln \rho _{1}\right) +\frac{%
\partial \left( 2\chi \right) }{\partial x}-i\frac{1}{\rho _{1}+\rho _{2}}%
\frac{\partial \left( -2\omega \right) }{\partial x}\right] H \\
&\simeq &\left[ ii\frac{\partial \psi }{\partial y}+\frac{\partial \left(
2\chi \right) }{\partial x}-i\frac{1}{\rho _{1}+\rho _{2}}\frac{\partial
\left( -2\omega \right) }{\partial x}\right] H \\
&=&\left[ \left( \frac{\partial \left( 2\chi \right) }{\partial x}+i\frac{%
\partial i\psi }{\partial y}\right) -i\frac{1}{\rho _{1}+\rho _{2}}\frac{%
\partial \left( -2\omega \right) }{\partial x}\right] H
\end{eqnarray*}

\textbf{Comment for this NOTE}

This current is composed of two parts 
\[
\frac{J_{x}}{\rho _{1}+\rho _{2}}\simeq \left[ J_{x}^{\left( 1\right)
}+J_{x}^{\left( 2\right) }\right] H 
\]

The first term contains the derivative 
\begin{eqnarray*}
J_{x}^{\left( 1\right) } &=&2\frac{\partial \left( i\psi +2\chi \right) }{%
\partial z^{\ast }} \\
&=&\partial _{+}\left( i\psi +2\chi \right)
\end{eqnarray*}
and the first part is practically zero since the phase $\psi $ is the
streamfunction and it is constant on circles. But the phase $\chi $ should
be growing along the streamlines.

The second term is 
\[
J_{x}^{\left( 2\right) }=-i\frac{1}{\rho _{1}+\rho _{2}}\frac{\partial
\left( -2\omega \right) }{\partial x} 
\]
This looks like the gradient of the vorticity.

For the \textbf{Euler} fluid we have 
\begin{eqnarray*}
J_{x}^{\left( 2\right) } &=&-i\frac{1}{\rho _{1}+\rho _{2}}\frac{\partial
\left( -2\omega \right) }{\partial x} \\
&=&-i\frac{1}{2\cosh \psi }\frac{\partial }{\partial x}\left( 2\sinh \psi
\right) \\
&=&-i\frac{\partial \psi }{\partial x}
\end{eqnarray*}

This should be added to the first current 
\[
J_{x}^{\left( 1\right) }+J_{x}^{\left( 2\right) }=\left( \frac{\partial
\left( 2\chi \right) }{\partial x}+i\frac{\partial \left( i\psi \right) }{%
\partial y}\right) -i\frac{\partial \psi }{\partial x} 
\]

\bigskip

\section{The $2D$ plasma in a strong magnetic field}

This is described by the equation%
\begin{equation}
\Delta \psi +\frac{1}{2}\sinh \left( \psi \right) \left[ \cosh \left( \psi
\right) -1\right] =0  \label{usplus}
\end{equation}%
(see however the Second Part for a certain ambiguity in the application of
the Bogomolnyi procedure, originating from the absence of a topological
constraint on the residual energy term. This is due to the triviality of the
first homotopy group of the manifold of the $su\left( 2\right) $ algebra).

\subsection{The large scale vortices in a $2D$ plasma with tokamak parameters%
}

We have obtained large scale vortical flows in the meridional section of a
tokamak, described by the equation (\ref{usplus}). They correspond to
previous similar flows identified by analytical methods which are based on
the technics developed for the Larichev-Reznik modon.

Numerical solution for $L=307$ : monopolar vortex 
\begin{figure}[!htb]
\hspace*{-1.cm} \centering
\begin{minipage}[t]{0.4\linewidth}
    \centering
    \includegraphics[height=3cm]{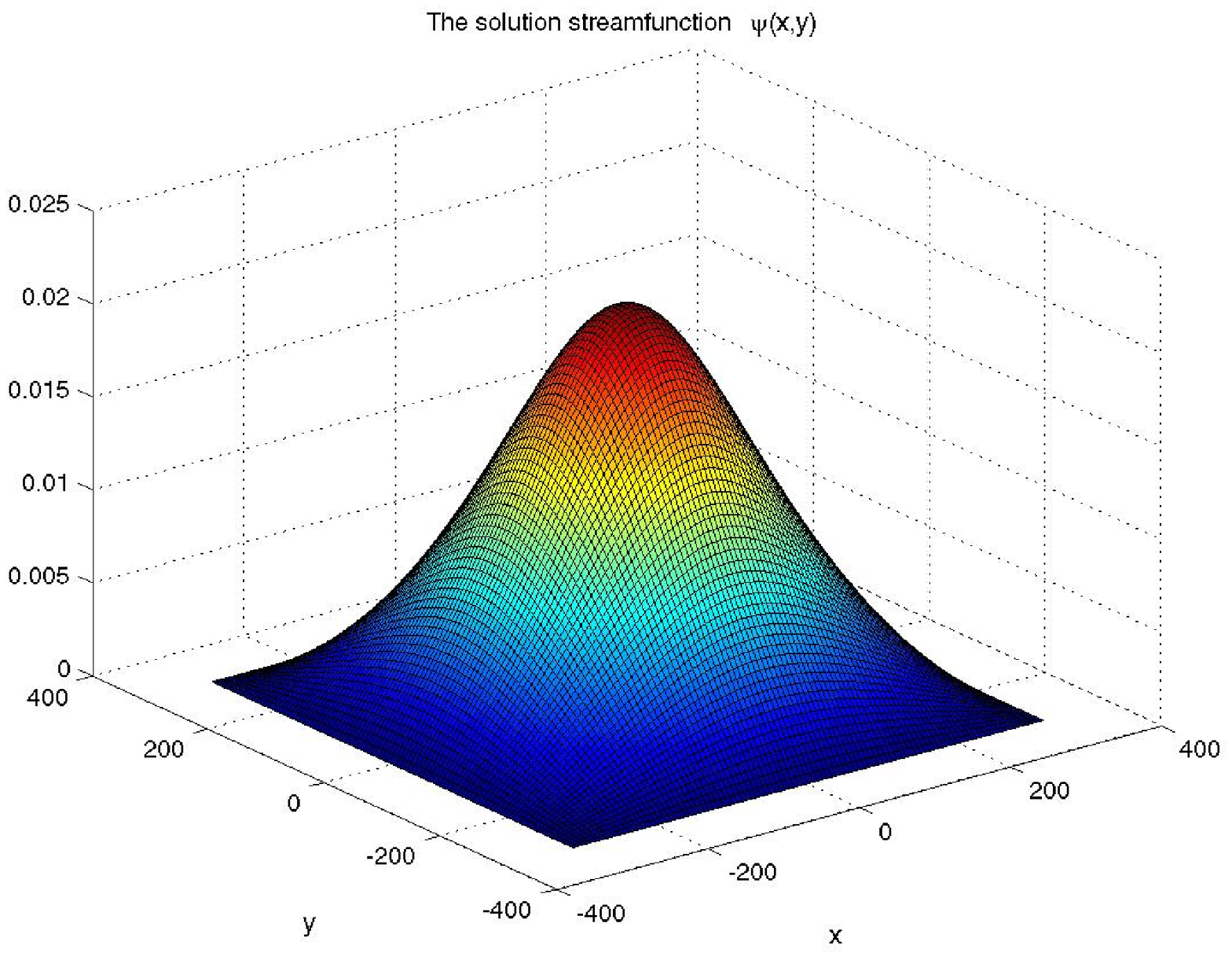}
   \end{minipage}
\hspace{0.1\textwidth} 
\begin{minipage}[t]{0.4\linewidth}
    \centering
    \includegraphics[height=3cm]{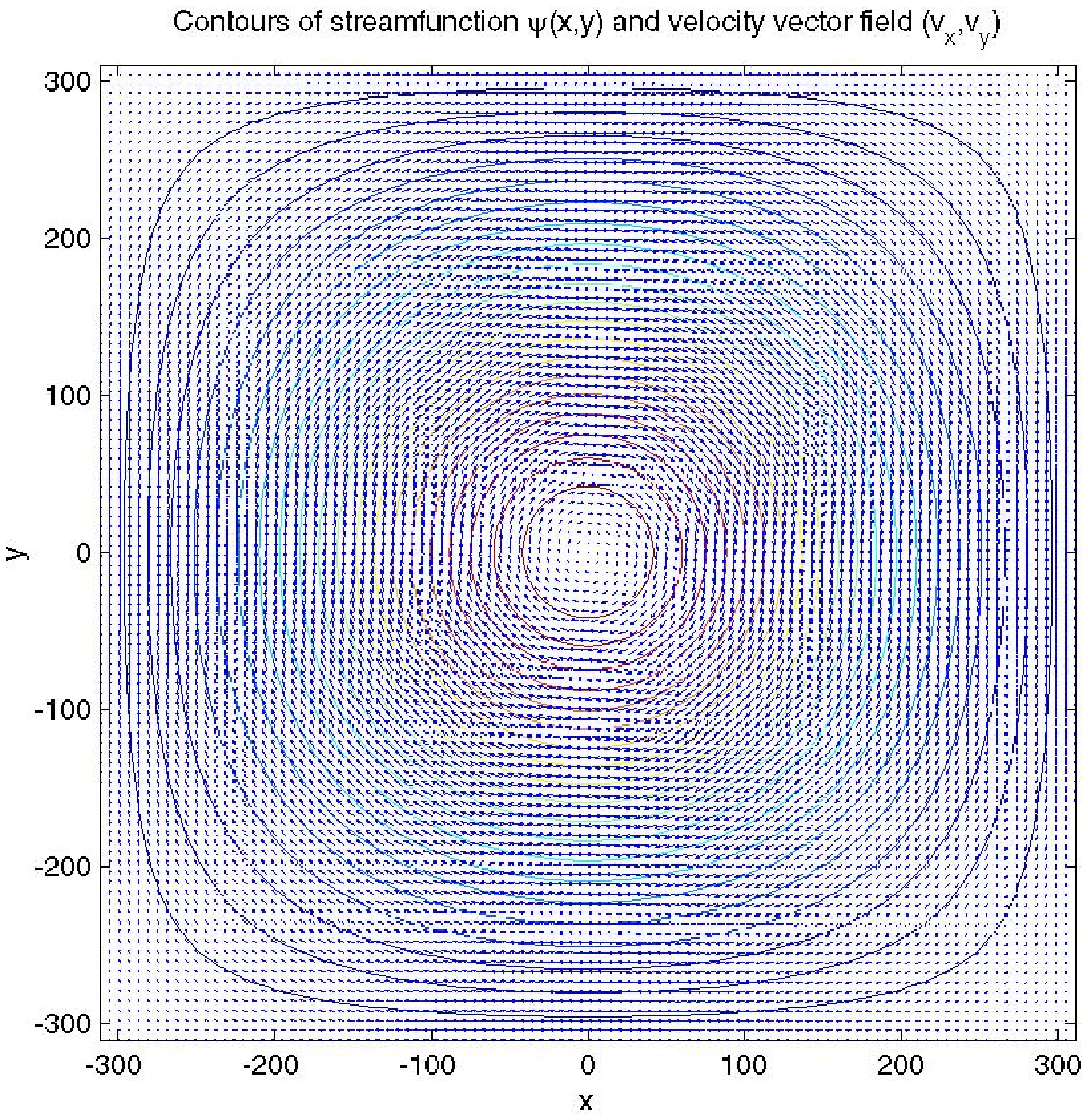}
   \end{minipage}
\caption{The streamfunction ($\protect\varphi/B$) and the velocity, $v_{%
\protect\theta}(x,y)$}
\end{figure}
Physical parameters: $\rho _{s}=0.003\;\left( m\right)$, $%
L^{phys}=a=1\;\left( m\right)$. After normalization $L=\frac{a}{\rho _{s}}=%
\frac{1}{0.003}\simeq 330$. The unit of velocity is $c_{s}=9.79\times 10^{3}%
\sqrt{T_{e}\left( eV\right) }\;\left( m/s\right)$

Numerical solution for $L=307$: dipolar vortex. 
\begin{figure}[!htb]
\hspace*{-1cm} \centering
\begin{minipage}[t]{0.4\linewidth}
    \centering
    \includegraphics[height=4cm]{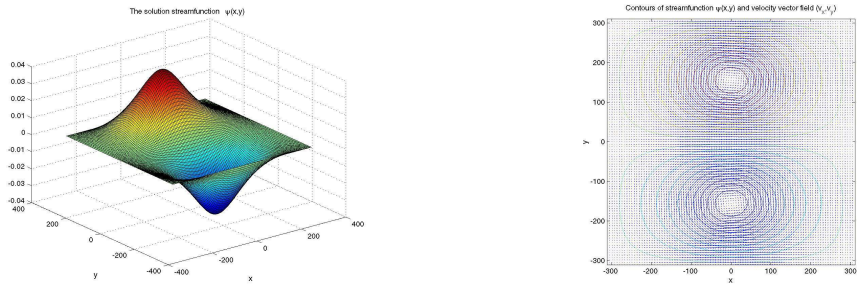}
   \end{minipage}
\hspace{0.1\textwidth} 
\begin{minipage}[t]{0.4\linewidth}
    \centering
    \includegraphics[height=4cm]{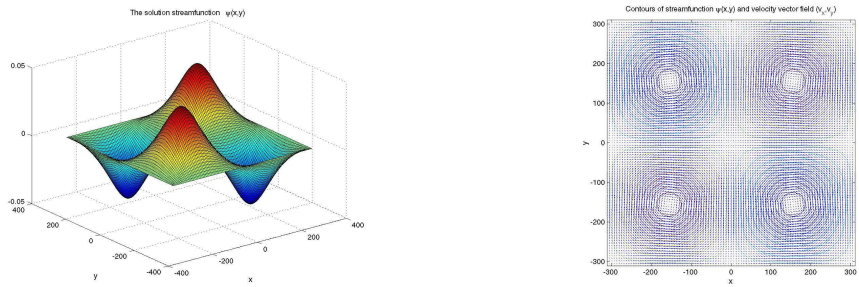}
   \end{minipage}
\caption{The streamfunction ($\protect\varphi/B$) and the velocity, $v_{%
\protect\theta}(x,y)$}
\end{figure}

Numerical solution for $L=307$: quadrupolar vortex. 
\begin{figure}[!htb]
\hspace*{-1cm} \centering
\begin{minipage}[t]{0.4\linewidth}
    \centering
    \includegraphics[height=4cm]{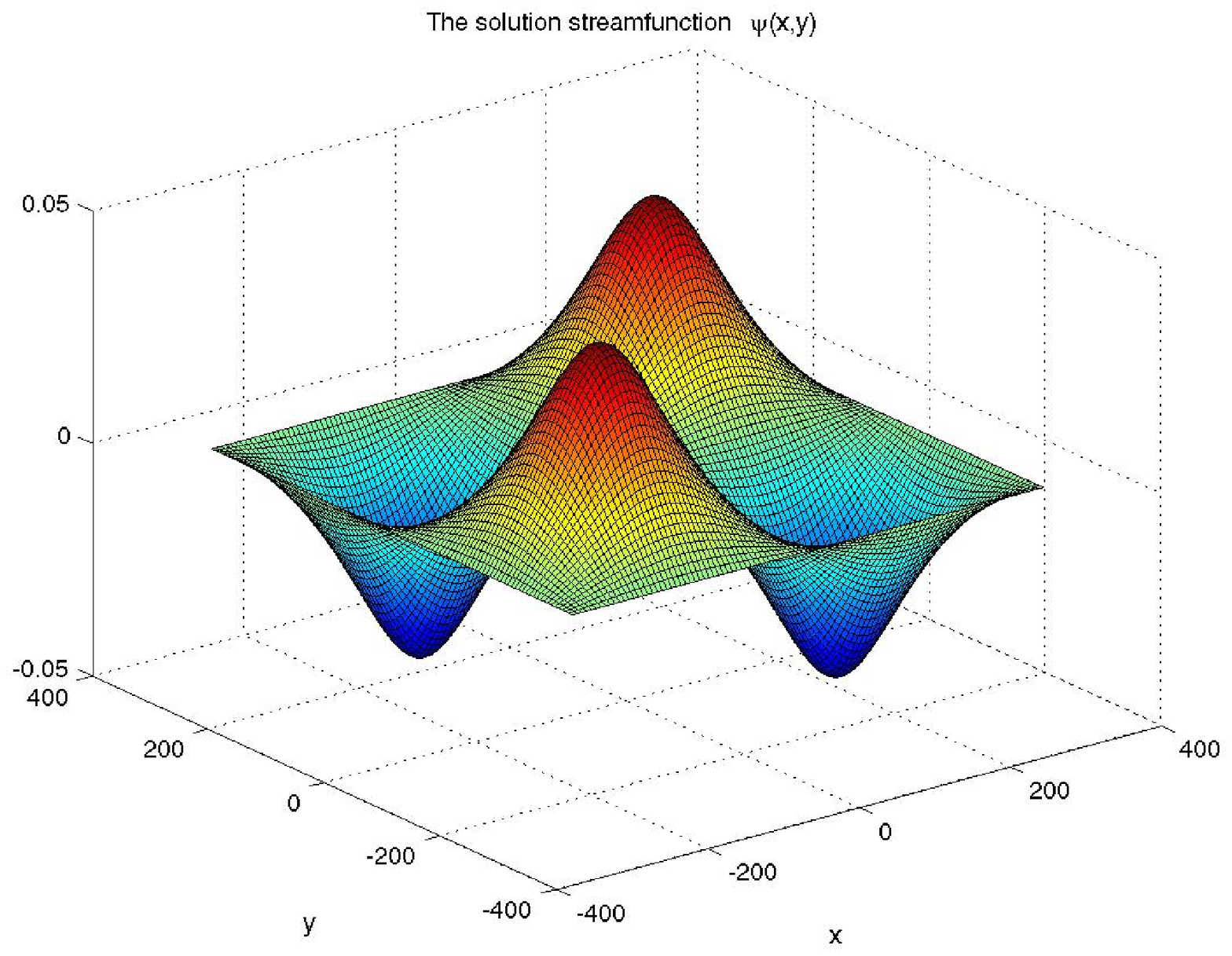}
   \end{minipage}
\hspace{0.1\textwidth} 
\begin{minipage}[t]{0.4\linewidth}
    \centering
    \includegraphics[height=4cm]{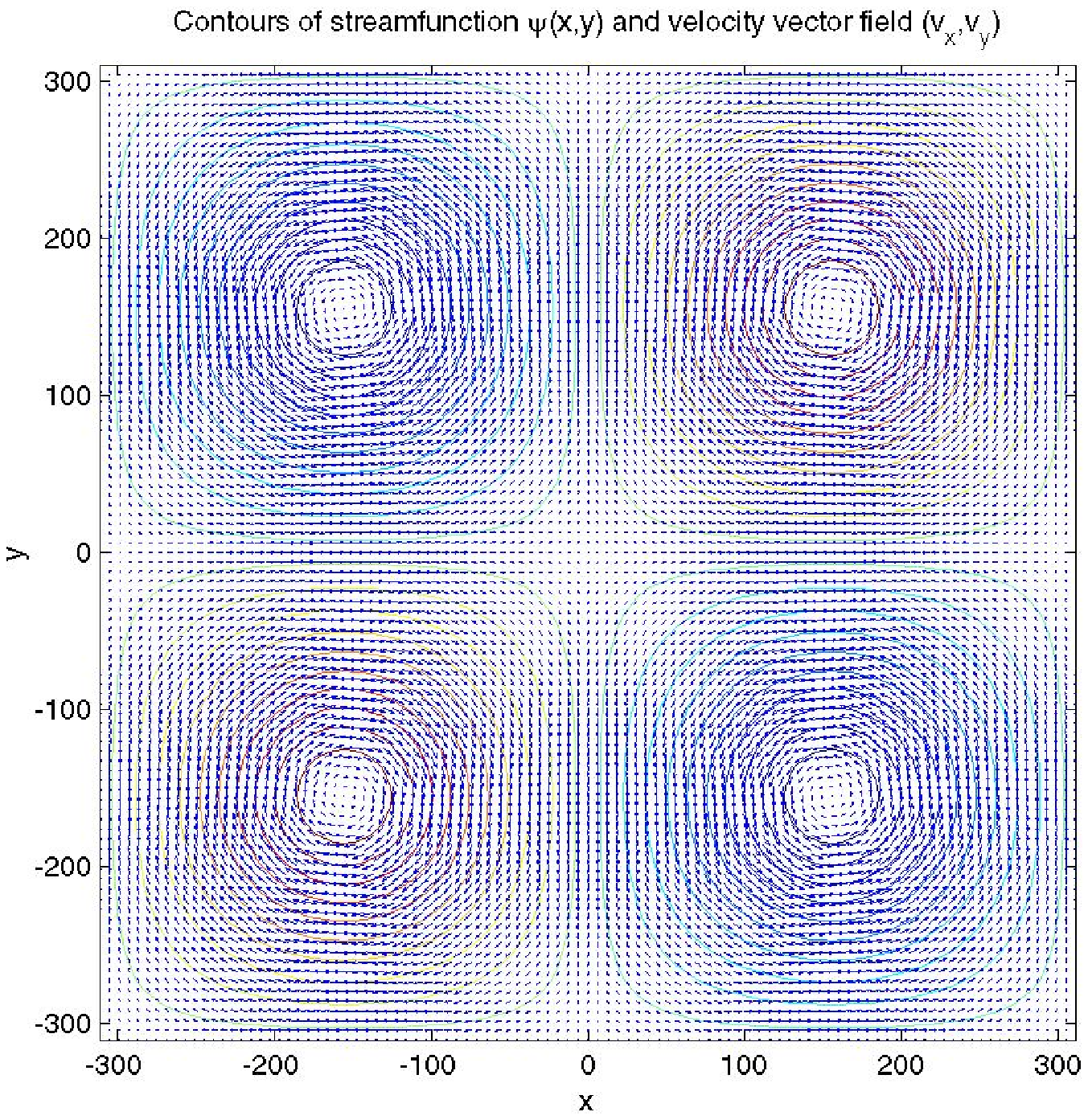}
   \end{minipage}
\caption{The streamfunction ($\protect\varphi/B$) and the velocity, $v_{%
\protect\theta}(x,y)$}
\end{figure}

The amplitudes of the flows is in general small (but not incompatible with
the previous analytic results) and should be corrected for the redefinition
of the Larmor radius.

\subsection{Self-organisation of the drift turbulence}

Compare the result from our equation with the result of a
statistical/variational theory of drift wave turbulence for the
Hasegawa-Wakatani equation.

\begin{figure}[!htb]
\hspace{-1cm} \centering
\begin{minipage}[t]{0.3\linewidth}
    \centering
    \includegraphics[height=3cm]{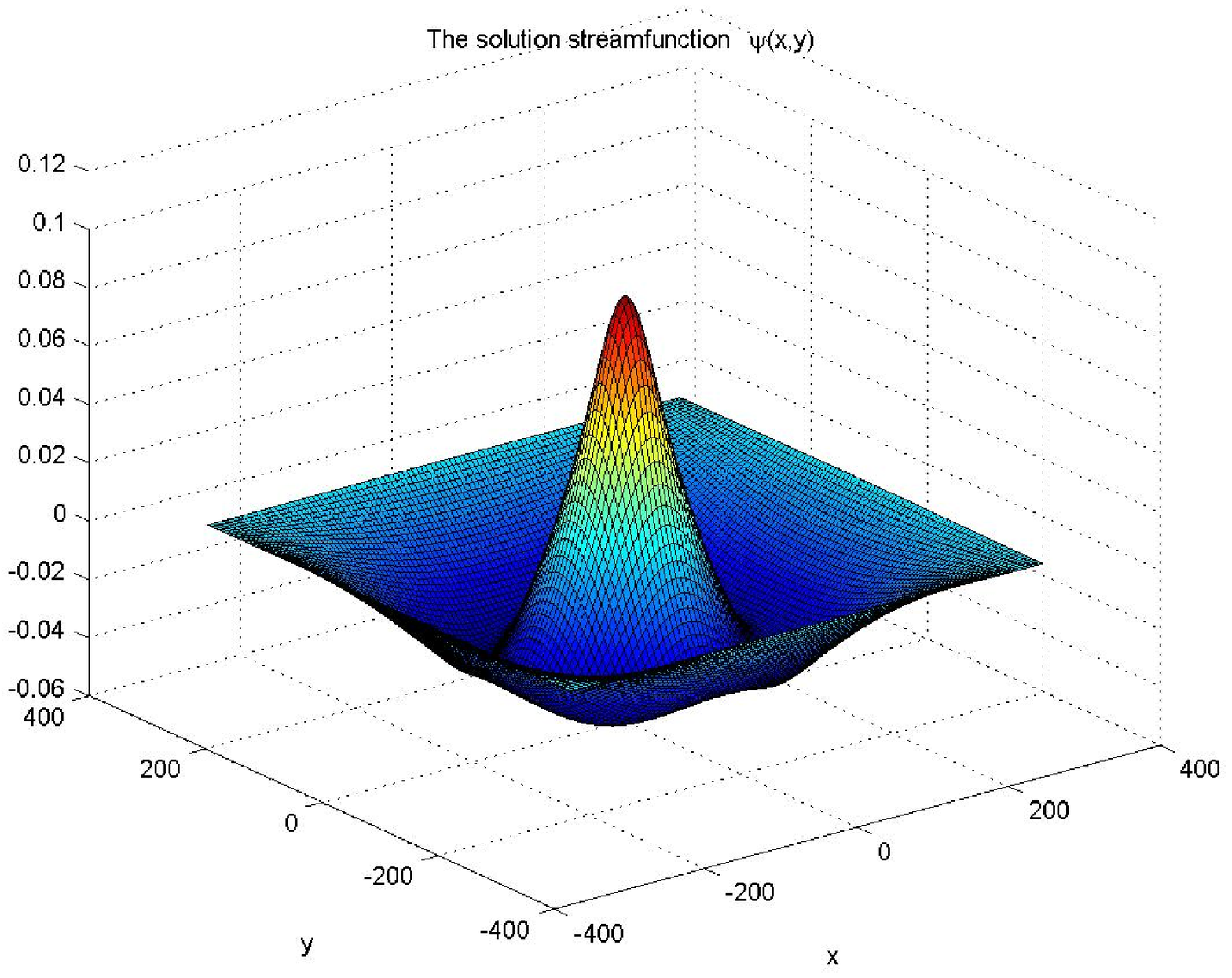}
   \end{minipage}
\hspace{0.03\textwidth} 
\begin{minipage}[t]{0.3\linewidth}
    \centering
    \includegraphics[height=3cm]{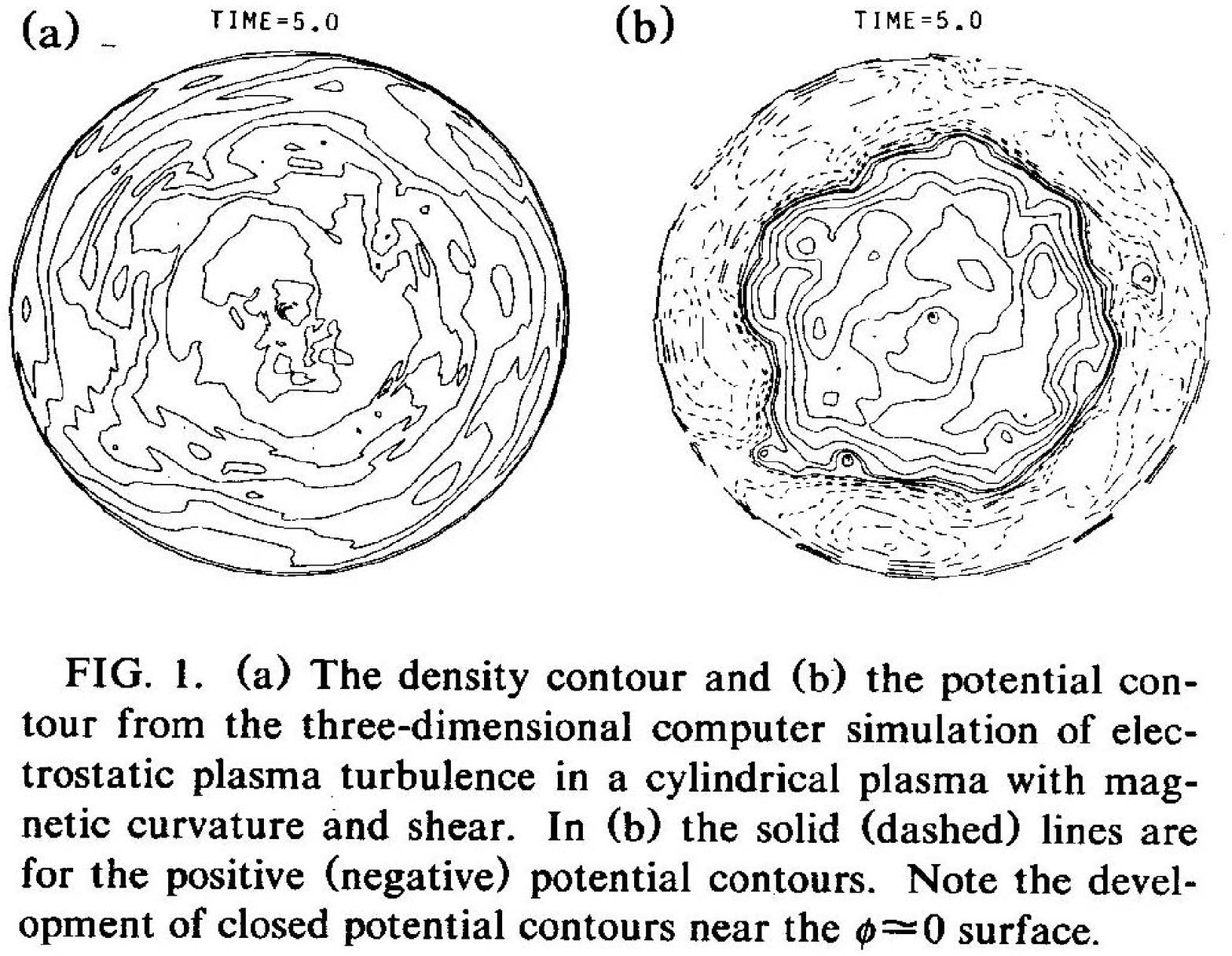}
   \end{minipage}
\hspace{0.03\textwidth} 
\begin{minipage}[t]{0.3\linewidth}
    \centering
    \includegraphics[height=3cm]{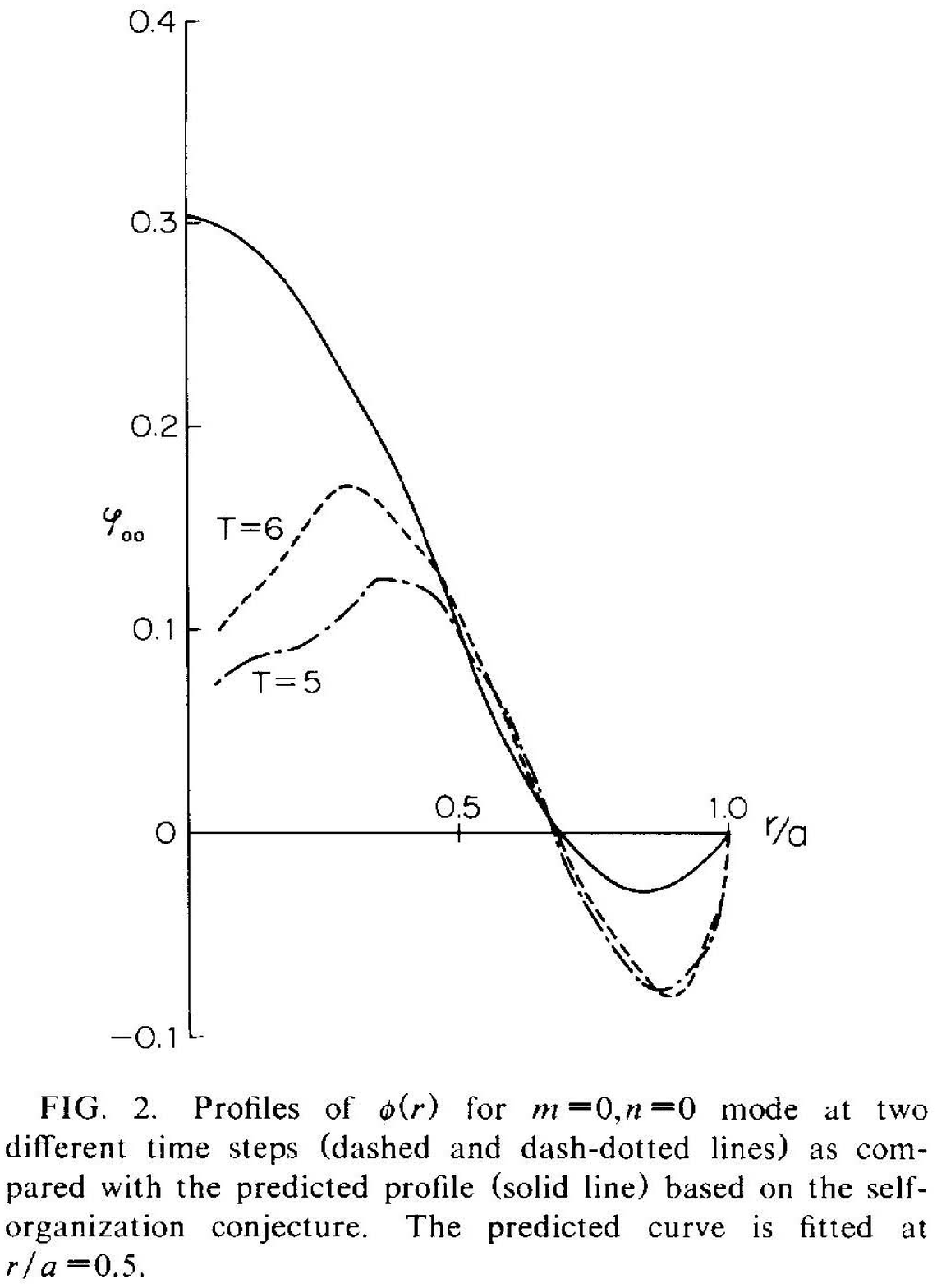}
   \end{minipage}
\caption{A theory of self-organization of turbulence (Hasegawa-Wakatani)
leads to this radial profile for the potential}
\end{figure}
\FloatBarrier

\subsection{The LH transition}

We obtain as solution of the Eq.(\ref{usplus}) profiles of velocity with a
deep drop at the edge. They are similar with the measured profiles observed
after the transition L to H.

\emph{We must note that the following results are not exact solutions of the
Eq.(\ref{usplus}). They are approximative solutions,or, in other terms, they
verify Eq.(\ref{usplus}) only when a larger tolerance is accepted in the
GIANT code. Nevertheless, they appear persistently and we can suppose that
they correspond not to exact minimum action states, but are local minima of
the action or they may have a slow drift in the space of system's
configurations.}

\bigskip

\begin{figure}[tbh]
\centering
\begin{minipage}[t]{0.95\textwidth}
\centerline{\includegraphics[height=6cm]{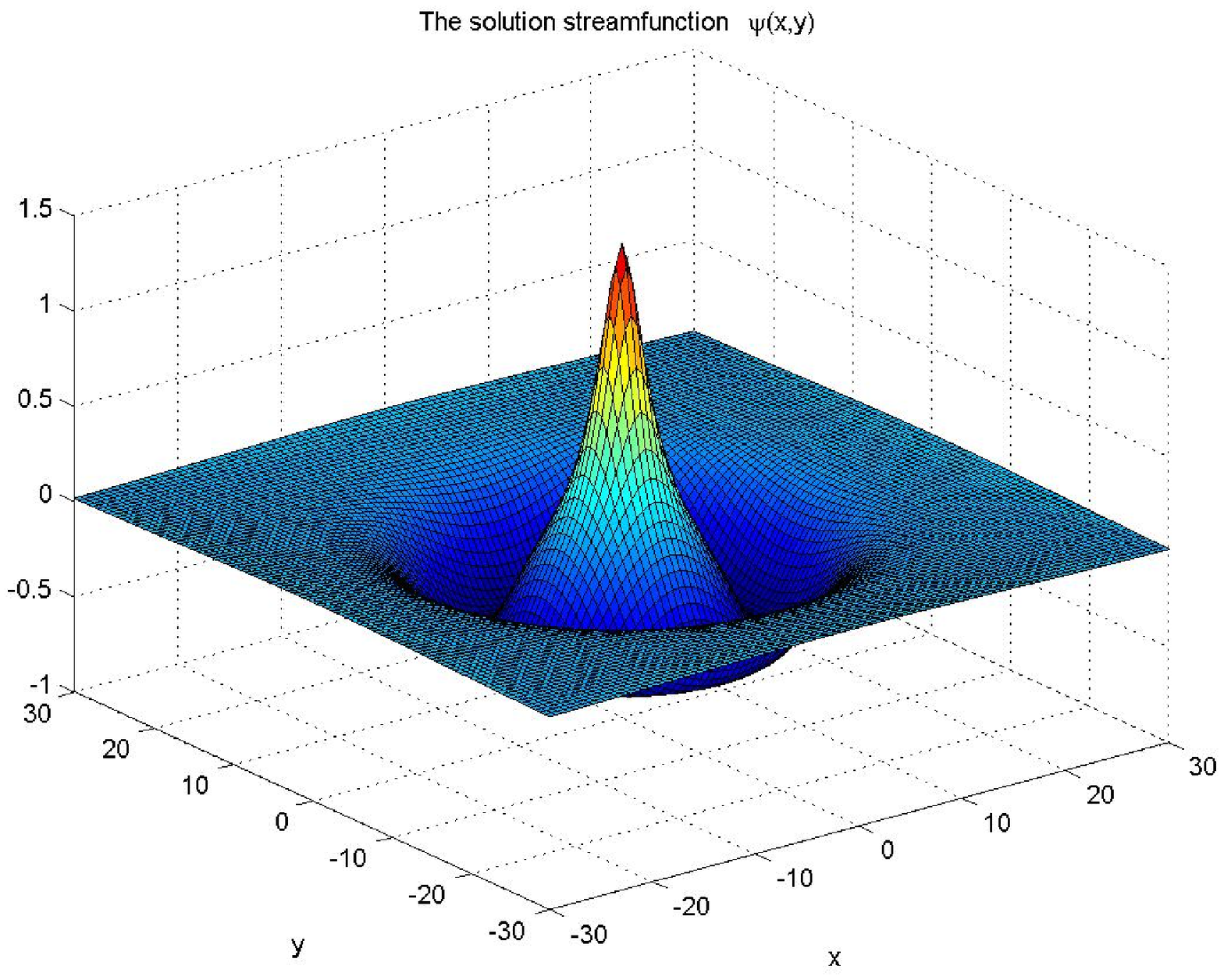}}
\vfill
\centerline{\includegraphics[height=6cm]{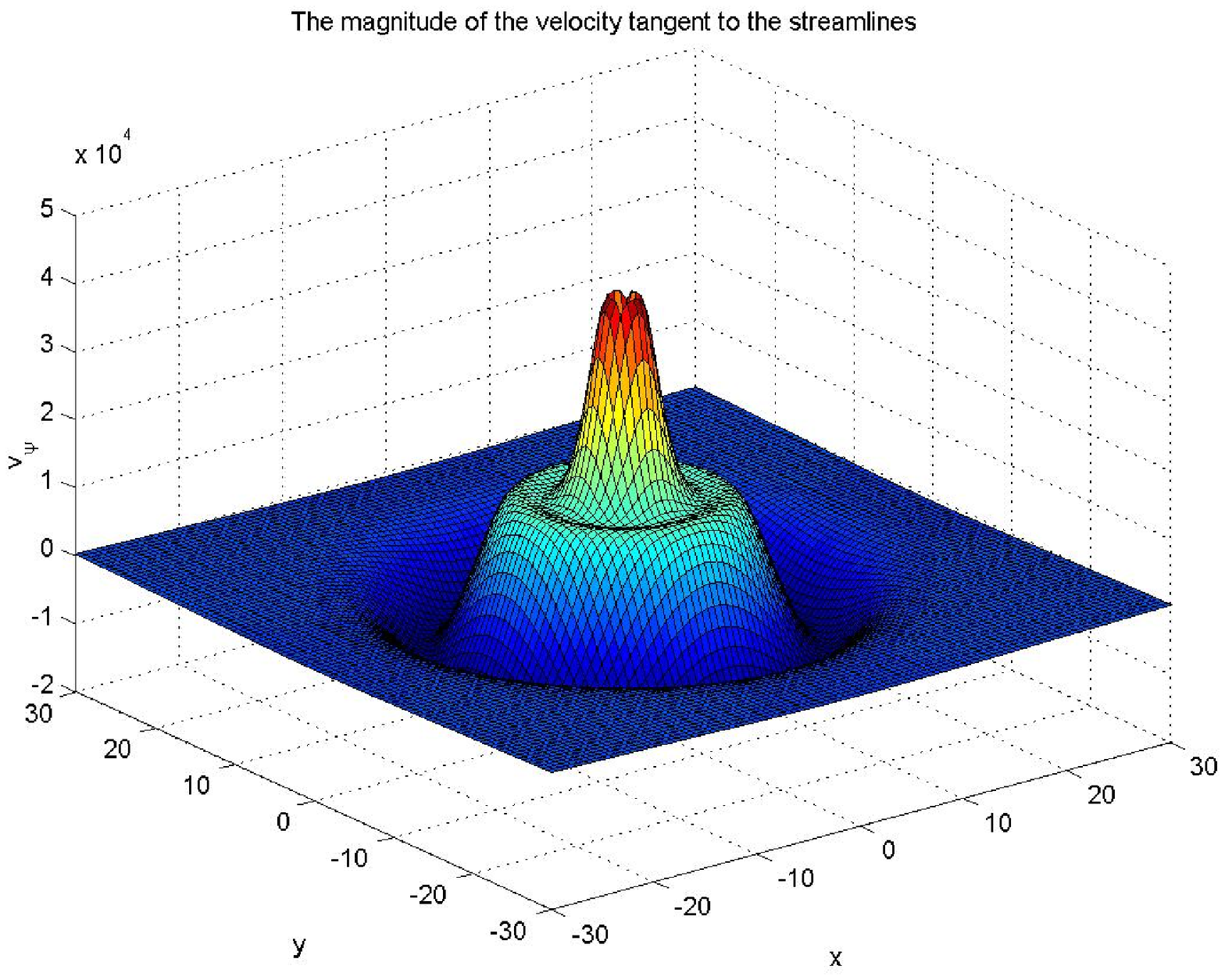}}
\vfill
\centerline{\includegraphics[height=6cm]{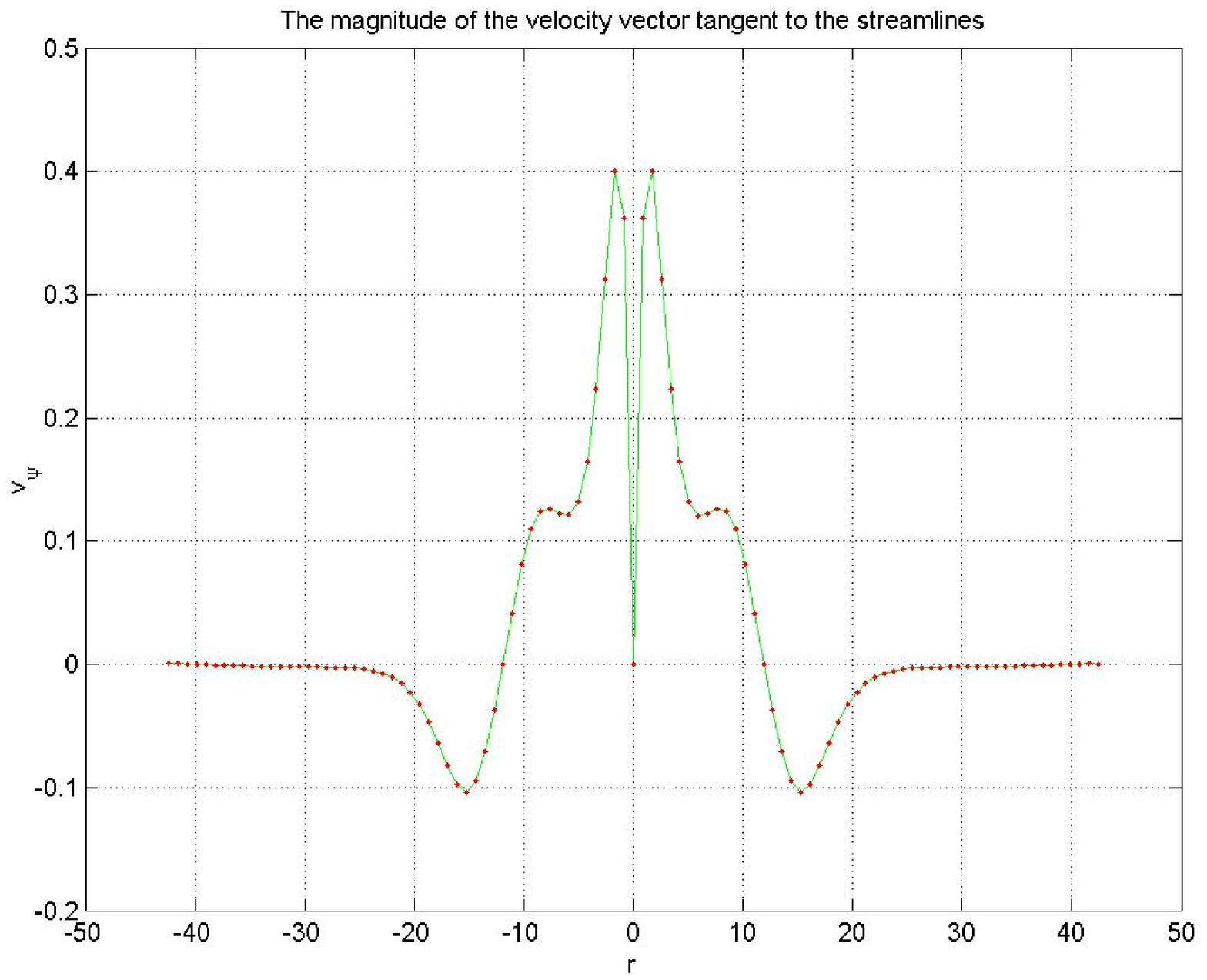}}
\caption{Density series30, set\_1.02. $p=1$ and $L=30.$ Here $ampuh=1.02$. 
The streamfunction $\psi(x,y)$ (f53),
the azimuthal velocity $v_{\theta}(x,y)$ (f54) and
$v_{\theta} diag$ (f55).}
\label{f55}
\end{minipage}
\end{figure}
\clearpage

\begin{figure}[tbph]
\centerline{\includegraphics[height=7cm]{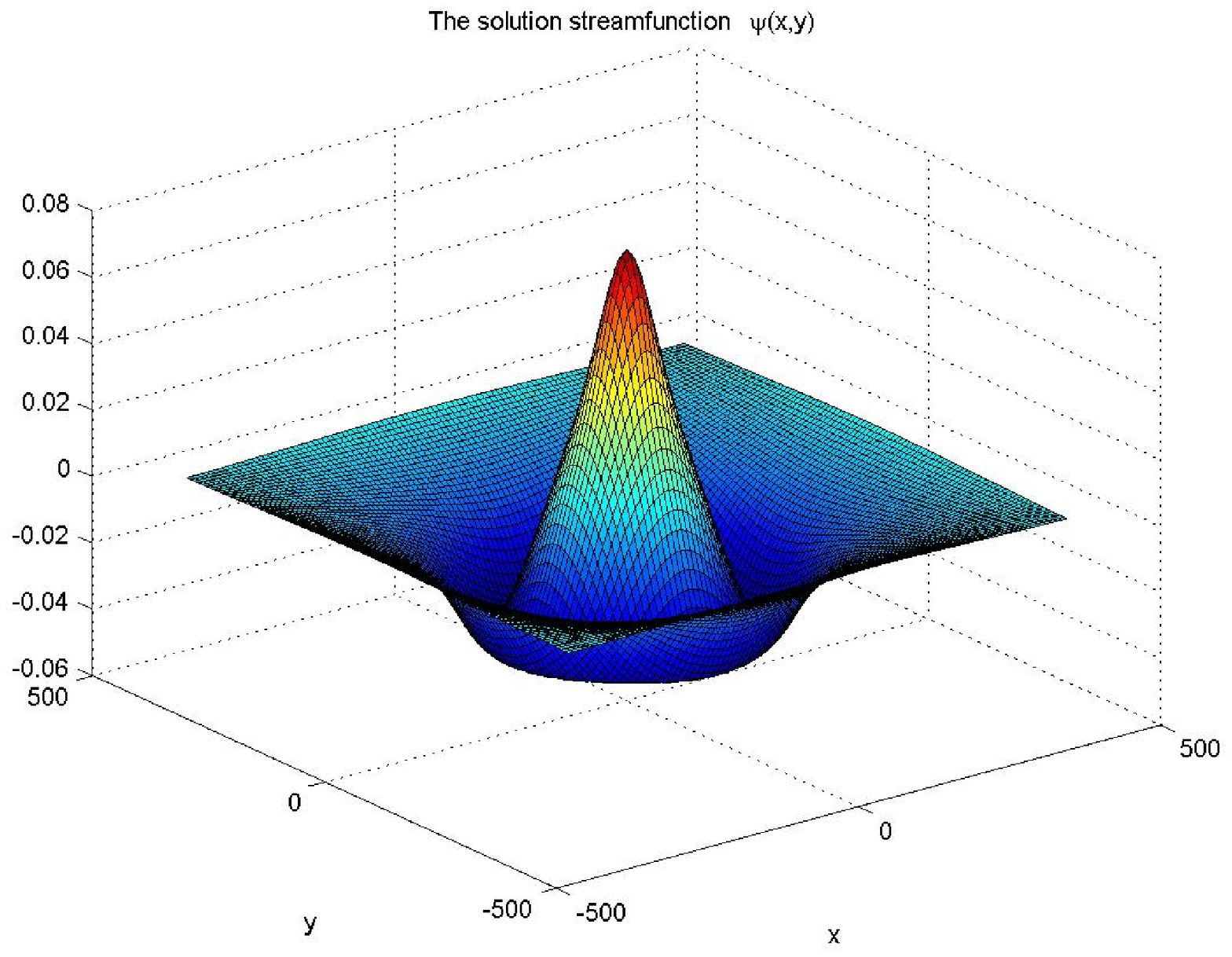}}
\caption{Density, set15\_1. $p=1$ and $L=411.$ Streamfunction $\protect\psi $
(f26).}
\label{f26}
\end{figure}
\begin{figure}[tbph]
\centerline{\includegraphics[height=7cm]{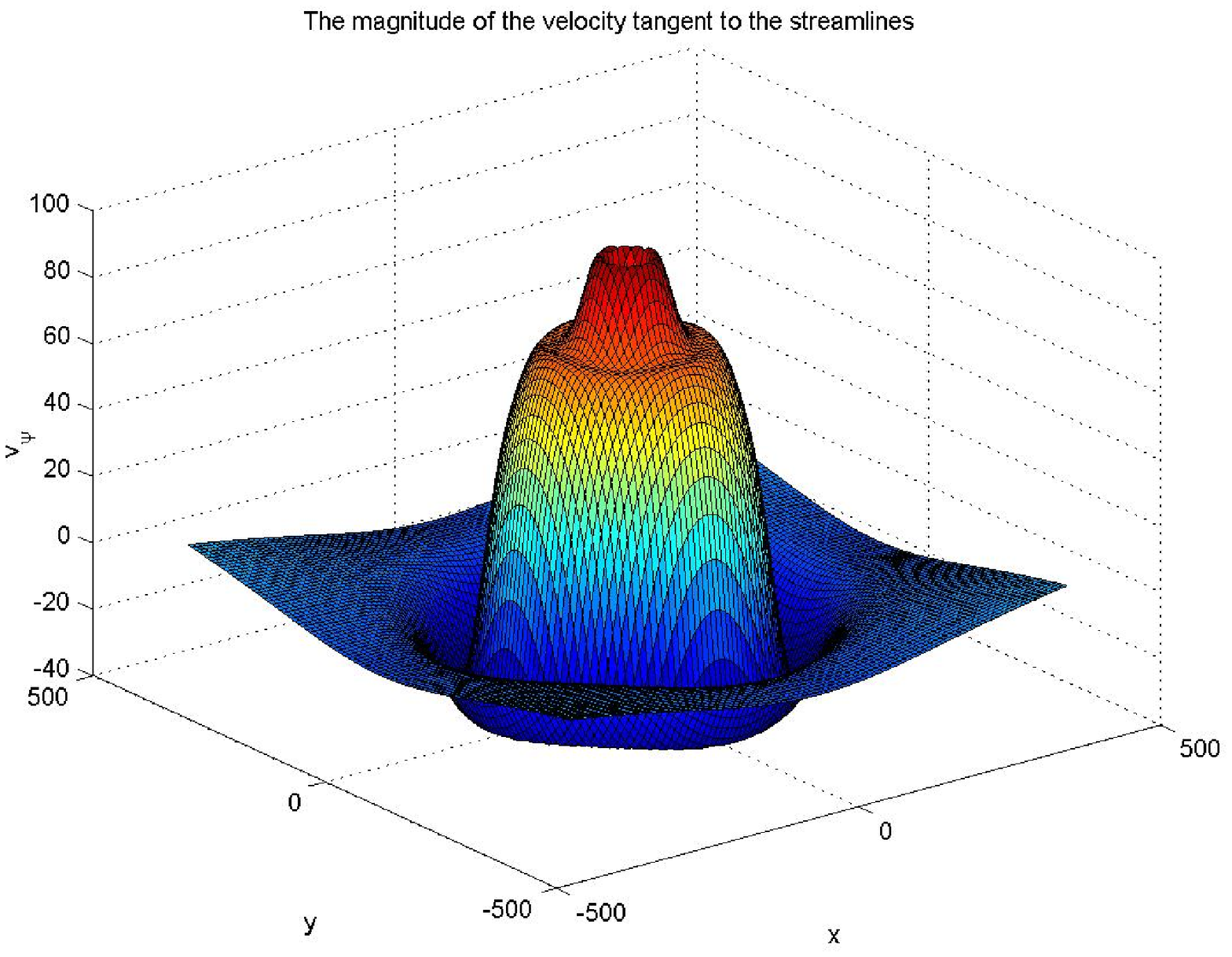}}
\caption{density, set15\_1. $p=1$ and $L=411.$ $v_{\protect\theta }$(f27). }
\label{f27}
\end{figure}
\begin{figure}[tbph]
\centerline{\includegraphics[height=7cm]{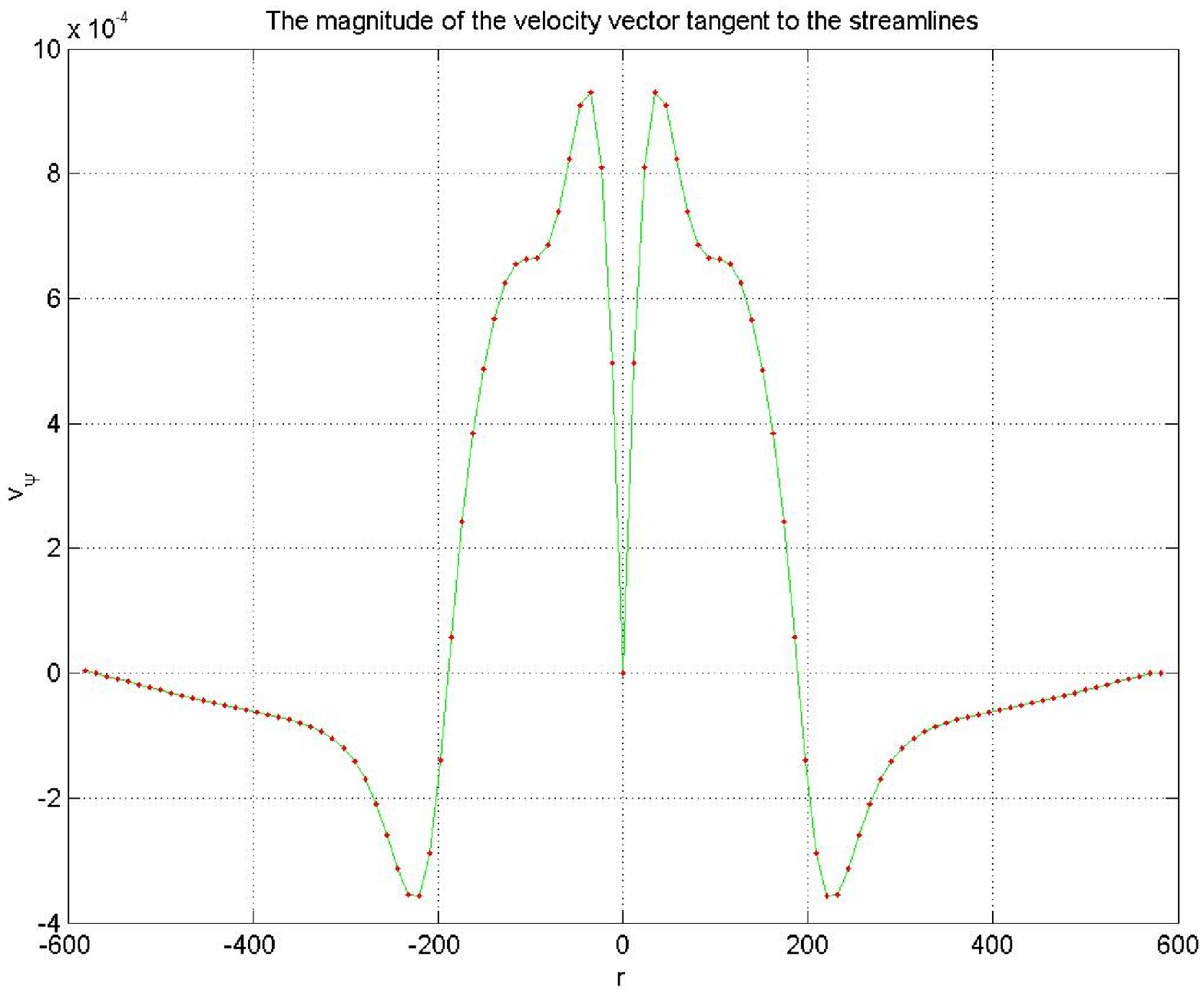}}
\caption{density, set15\_1. $p=1$ and $L=411.$ $v_{\protect\theta }$ (f28).}
\label{f28}
\end{figure}

\bigskip

\FloatBarrier

\begin{figure}[tbph]
\centerline{\includegraphics[height=7cm]{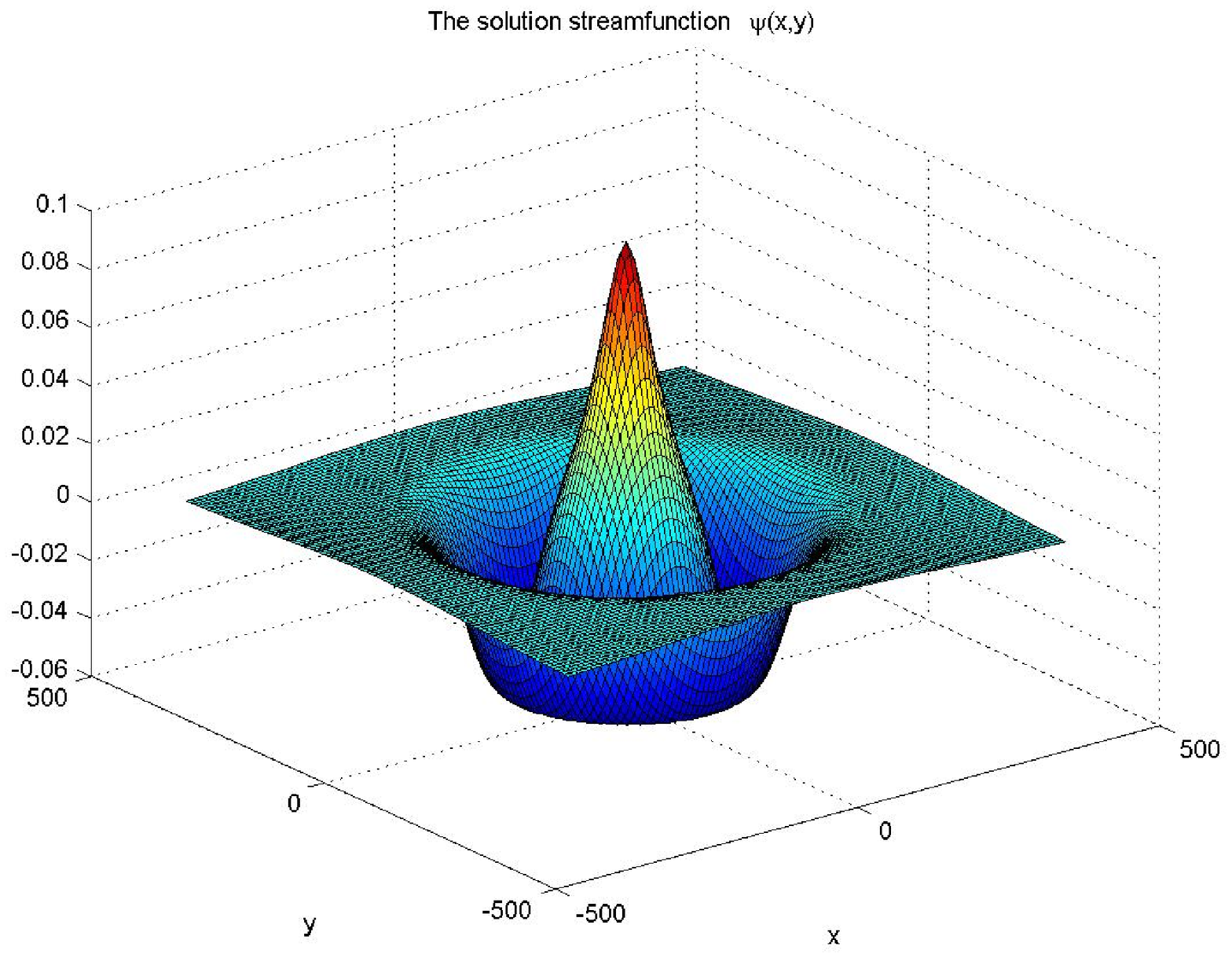}}
\caption{Density, set15\_5. $p=1$ and $L=411.$ Streamfunction $\protect\psi $
(f29).}
\label{f29}
\end{figure}
\begin{figure}[tbph]
\centerline{\includegraphics[height=7cm]{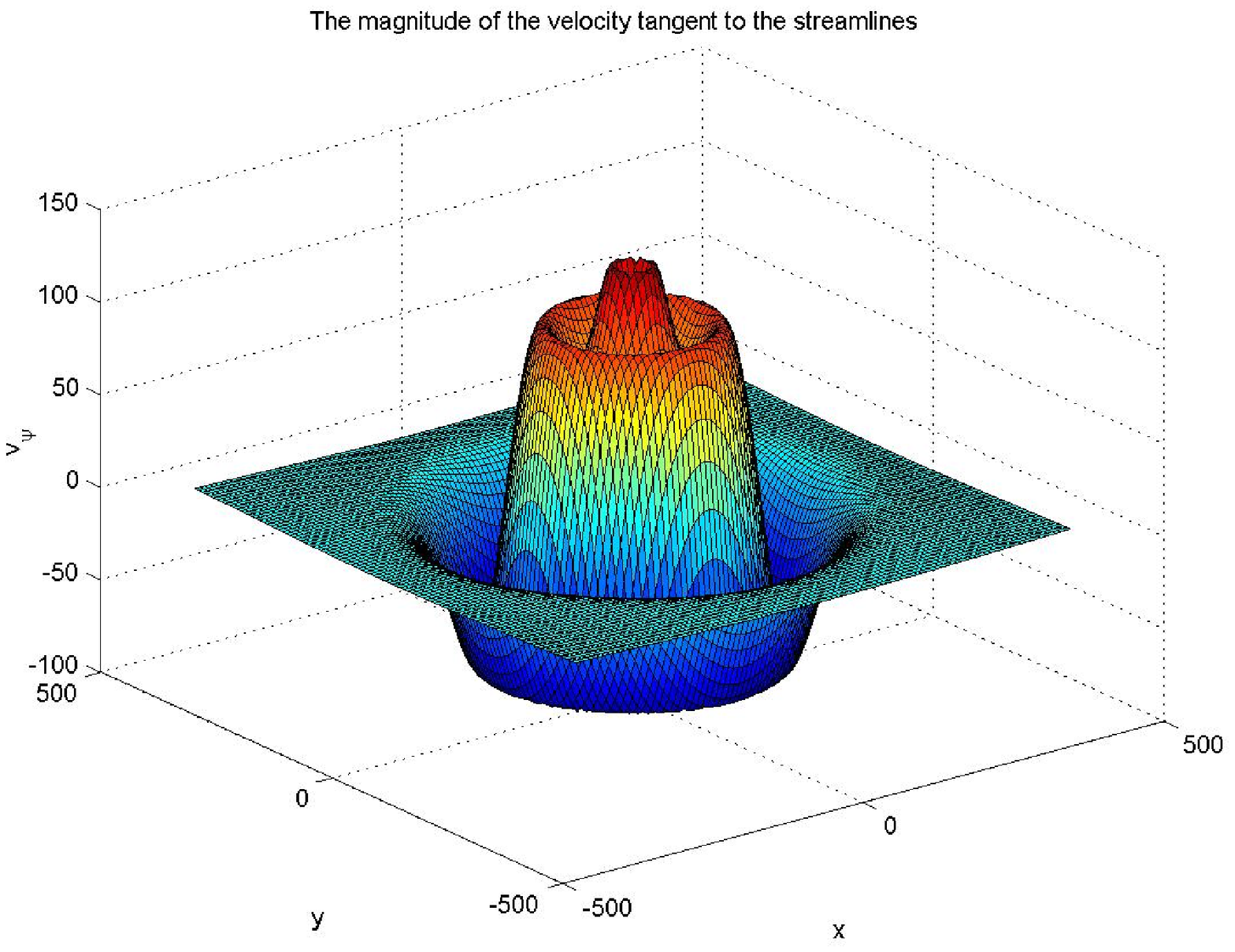}}
\caption{density, set15\_5. $p=1$ and $L=411.$ $v_{\protect\theta }$(f30). }
\label{f30}
\end{figure}
\begin{figure}[tbph]
\centerline{\includegraphics[height=7cm]{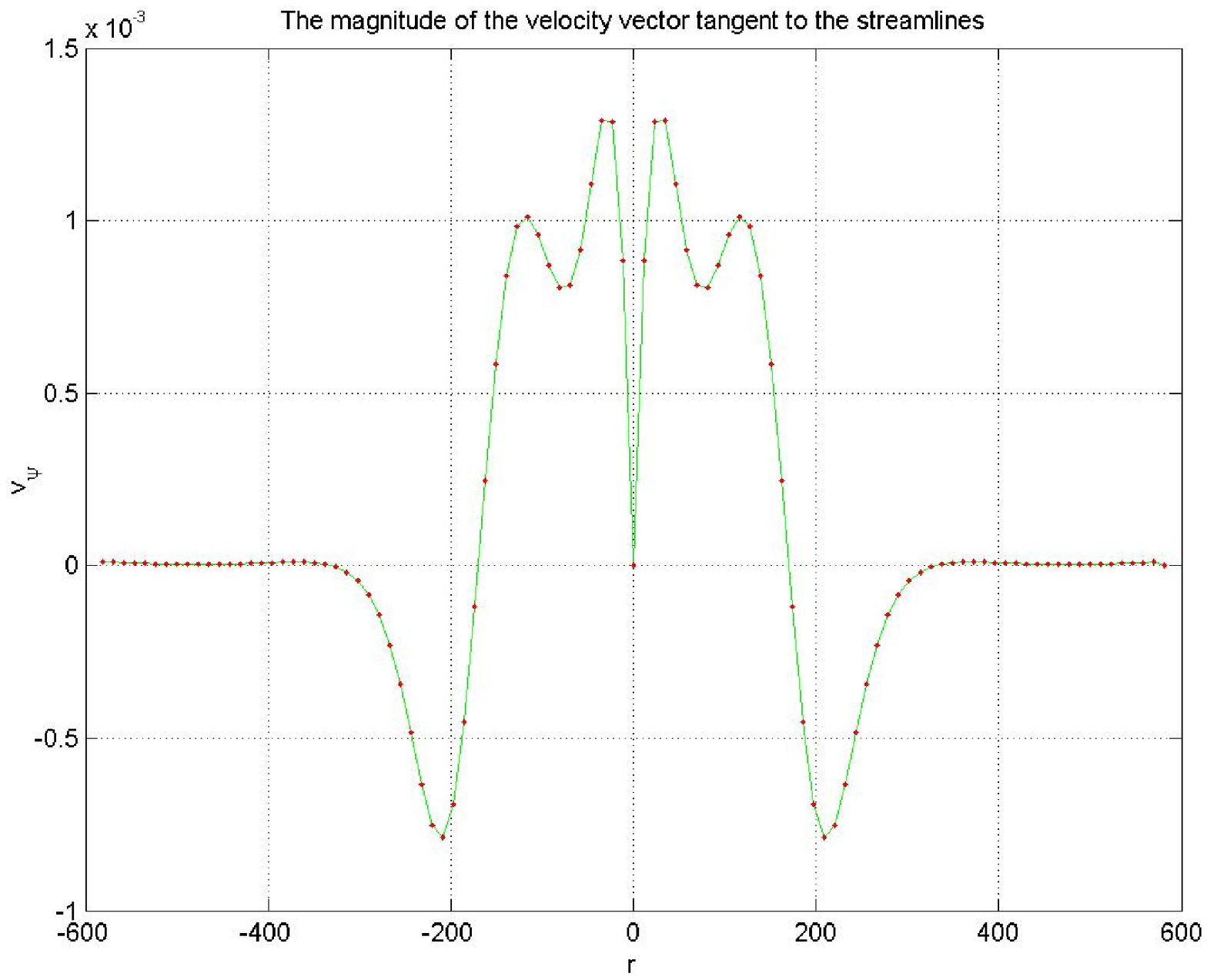}}
\caption{density, set15\_5. $p=1$ and $L=411.$ $v_{\protect\theta }$ (f31).}
\label{f31}
\end{figure}

\bigskip

\subsubsection{Physical parameters for the H states}

The input parameters are 
\begin{eqnarray*}
a &=&1.25\;\left( m\right) \\
B_{T} &=&2.5\,\left( T\right)
\end{eqnarray*}
From this we can calculate the ion cyclotron frequency 
\begin{eqnarray*}
\Omega _{ci} &=&\frac{eB_{T}}{m_{i}} \\
&=&9.58\times 10^{3}B_{T}\left( Gs\right) \\
&=&9.58\times 10^{3}\times 25000 \\
&=&239.5\times 10^{6}\left( s^{-1}\right)
\end{eqnarray*}

These are fixed parameters. We need to choose a value for the ion
temperature, that will fix the Larmor radius. Take 
\[
T=2\,(KeV) 
\]
then 
\begin{eqnarray*}
c_{s} &=&9.79\times 10^{3}\sqrt{T_{e}\left( eV\right) } \\
&=&4.38\times 10^{5}\,\left( m/s\right)
\end{eqnarray*}
\begin{eqnarray*}
\rho _{s} &=&\frac{c_{s}}{\Omega _{i}} \\
&=&0.0018\,\left( m\right)
\end{eqnarray*}
This gives 
\begin{eqnarray*}
L &=&\frac{a}{\rho _{s}} \\
&=&682.35
\end{eqnarray*}

The following table shows to what extent we can expect change of the
effective (adimensional) length of the integration domain

\begin{center}
$%
\begin{tabular}{|c|c|c|}
\hline
$T\,\left( KeV\right) $ & $\rho _{s}\,\left( m\right) $ & $L$ \\ \hline
$0.5$ & $9.14\times 10^{-4}$ & $1367$ \\ 
$1$ & $0.0013$ & $967$ \\ 
$2$ & $0.0018$ & $683$ \\ 
$3$ & $0.0022$ & $558$ \\ 
$4$ & $0.0026$ & $483$ \\ 
$5$ & $0.0029$ & $432$ \\ 
$6$ & $0.0032$ & $395$ \\ 
$7$ & $0.0034$ & $365$ \\ 
$8$ & $0.0037$ & $341$ \\ 
$9$ & $0.0039$ & $322$ \\ 
$10$ & $0.0041$ & $306$ \\ \hline
\end{tabular}
$
\end{center}

\bigskip

A case with $T=5000\,\left( eV\right) $, where $\rho _{s}\simeq 0.003$ and 
\[
L\simeq 411 
\]
From the numerical simulations, we have the velocity at the bottom of the
region of negative values, close to the border (from the figure f28) 
\[
v_{\theta ,low}\simeq 4\times 10^{-4} 
\]
The unit of velocity is 
\begin{eqnarray*}
c_{s} &=&9.79\times 10^{3}\sqrt{T_{e}} \\
&\simeq &7\times 10^{5}\,\left( m/s\right)
\end{eqnarray*}
Then the physical velocity is 
\[
v_{\theta ,low}^{phys}=280\,\left( m/s\right) 
\]
The radial electric field that can generate this rotation is 
\[
\frac{E_{r}\left( V/m\right) }{B_{T}\left( T\right) }=v_{\theta
}^{phys}\left( m/s\right) 
\]
or 
\[
E_{r}=280\times 2.5=700\,\left( V/m\right) 
\]

\subsubsection{Calculation of the radial electric field at the edge}

The input parameters are from JET 
\begin{eqnarray*}
a &=&1.\;\left( m\right) \\
B_{T} &=&2.5\,\left( T\right)
\end{eqnarray*}%
From this we can calculate the ion cyclotron frequency 
\begin{eqnarray*}
\Omega _{ci} &=&\frac{eB_{T}}{m_{i}} \\
&=&239.5\times 10^{6}\left( s^{-1}\right)
\end{eqnarray*}%
and 
\[
c_{s}=9.79\times 10^{3}\sqrt{T_{e}} 
\]%
\[
\rho _{s}=\frac{c_{s}}{\Omega _{i}} 
\]%
We must use a single value for $T_{e}$ instead of a radial profile. Then the
value of $\rho _{s}$ will not be precise. We adopt the point of view that
the sonic Larmor radius may be modified by various effects leading to 
\[
\frac{1}{\rho _{s}^{2}}\rightarrow \frac{1}{\left( \rho _{s}^{eff}\right)
^{2}}\equiv \frac{1}{\rho _{s}^{2}}\left( 1-\frac{v_{d}}{u}\right) 
\]%
\begin{equation}
v_{d}=\frac{\rho _{s}c_{s}}{L_{n}}  \label{vddef}
\end{equation}%
We assume that the system evolves to a state \ where two parameters become
comparable in magnitude: the velocity of plasma rotation $u$ and the
diamagnetic velocity $v_{d}$%
\[
u\sim v_{d} 
\]%
This is achieved in a structured way, \emph{i.e.} the various elements
behind these parameters evolve individually but in a corralated manner to
achieve a new state of equilibrium. The gradient of the density profile $%
L_{n}$ evolves such that the diamagnetic velocity $v_{d}$ increases on a
certain region of plasma, which may be connected with the strong gradient at
the pedestal. The plasma poloidal rotation velocity $u$ increases due to the
redistribution of vorticity leading to accumulation of vorticity \emph{%
elements} toward the center of plasma. It is possible that these parameters
attain a relative magnitude such that the effective Larmor radius $\rho
_{s}^{eff}$ becomes sufficiently large. This makes possible a new
equilibrium of rotation velocity via a new equilibrium vorticity
distribution, as a quasi-stationary state. We can take several possible
values for them 
\[
\begin{array}{cccccc}
v_{d}/u & \rho _{s}^{eff}/\rho _{s} & L=a/\rho _{s}^{eff} & \left\vert
v_{\theta bottom}\right\vert & \left\vert v_{\theta
bottom}^{phys}\right\vert \left( m/s\right) & \left\vert E_{r}\right\vert
(kV/m) \\ 
0.2 & 1.118 & 273 & 1\times 10^{-3} & 979 & 2.45\, \\ 
0.4 & 1.29 & 236 & 1\times 10^{-3} & 979 & 2.45 \\ 
0.85 & 2.58 & 95 & 2.2\times 10^{-3} & 2152 & 5.38 \\ 
0.95 & 4.47 & 63 & 0.02 & 19580 & 48.95%
\end{array}%
\]

It stops increasing $\rho _{s}^{eff}$ for a certain ratio $v_{d}/u$.

See below for a more detailed discussion on this topic.

\subsection{The density pinch}

\subsubsection{Introduction}

The vorticity $\omega =\mathbf{\nabla }_{\perp }^{2}\phi $ and the particle
density $n\left( r,t\right) $ are strongly connected in $2D$ tokamak plasma.
Then we must look to the intrinsic evolution of the vorticity profile since
from this we can draw conclusions about the density pinch. When we apply the
Bogomolnyi procedure to the action functional for $2D$ plasma and atmosphere
we find equations describing the stationary states which formally would
correspond to \emph{self-duality}, plus an additional energy which does not
have a topological character. We can assume that the equations describe
states which are actually quasi-stationary and that the additional (or
residual) energy term provides a estimation of the possibility that the
system still continues to evolve. This assumption is valid if we can prove
that the evolution is slow, which means that the system behaves
adiabatically. Although this perspective is not clear yet, we will take it
as a basis to investigate the evolution of the vorticity profiles in
tokamak, under the field theoretical formulation.

We first recall that in the model of J.B. Taylor of natural current profiles
in tokamak it is identified a quantity playing the role of a \emph{magnetic
temperature} for the statistical ensemble of current filaments. This
temperature is related with the peaking factor of the profile and it is
shown that when the magnetic temperature reaches a critical value, $T_{m}^{c}
$, the profile of the current shows a high degree of concentration, in a
form of a single current filament on the magnetic axis. This is valid for
his statistical (maximum entropy) theory whose analytical result is the
Liouville equation for the flux function. However the theory does not
provide a physical reason that a possible evolution of the system would
consist of an evolution of the magnetic temperature toward the critical
value.

In the field theoretical formulation of the same physical model (which also
derive the Liouville equation) the physical parameters are $\kappa $ (the
coefficient of the Chern-Simons term in the Lagrangian) and the electric
charge $e$ (which is actually the elementary current $j_{0}$ of the discrete
model). The later can be absorbed into the definition of the potential since
the connection between the matter field $\phi $ and the interaction (gauge)
potential $A^{\mu }$ needs not be parametrized in any physical way, being
established at the level of the discrete model. Therefore the quantity that
is still present is $\kappa $. On the other hand, comparing the Taylor model
with the field theory model, we find the relationship between the parameters,%
\[
\left\vert \kappa \right\vert =\frac{16\pi e^{2}}{J_{0}^{2}a^{2}c}\frac{1}{%
T_{m}/T_{m}^{c}-1} 
\]%
from which we see that a very peaked current profile in the form of a
concentrated filament, \emph{i.e.} 
\[
T_{m}\rightarrow T_{m}^{c} 
\]%
corresponds in field theory to very high values of $\kappa $.

While in the statistical model we cannot find an intrinsic reason for $T_{m}$
to approach (from above) the critical value $T_{m}^{c}$, we can look for the
presence of an equivalent tendency in the field theoretical model, asking
when there can be reasons for $\kappa $ to increase adiabatically its value.

In the field theoretical model for the $2D$ plasma, where $\kappa $ is
identified as the sound speed%
\[
\kappa \equiv c_{s}=\rho _{s}\Omega _{ci} 
\]%
we can assume that $\kappa $ behaves as $\rho _{s}$, since we take the
external confining magnetic field as constant, $B^{ext}=$const. Then an
increase of $\kappa $ can result from an increase in $\rho _{s}$.

Physically this is not acceptable when the variation of the temperature is
not considered, as was one of our basic assumption. Then what can provide us
with an increase of the Larmor radius?

A careful consideration of the physical model underlying the field
theoretical formulation may clarify this problem.

The discrete system is based on the idea that elements of vorticity interact
on distances of the order of the Larmor radius $\rho _{s}$ (the inverse of
the mass of the photon in our theory) and this parameter is taken constant
throughout. However in the plasma there is another `field' that has a strong
connection with the vorticity: the density $n\left( \mathbf{r},t\right) $.
Due to the compressibility of the ion polarization drift velocity, the
density is not constant in the CHM plasma (as opposed to the Euler fluid
case). The spatial variation of the density (\emph{i.e.} $L_{n}\neq \infty $%
) is the cause for the diamagnetic flow with the velocity $v_{dia}=\rho
_{s}c_{s}/L_{n}$. The interaction between point-like vortices of the
discrete model adapts itself to the presence of two velocities: one is the
velocity $u$, simply associated with the vorticity, with which the cluster
of point-like vortices (\emph{i.e.} the physical vorticity) is in the
relationship $\mathbf{\omega }=\mathbf{\nabla \times u}$. The other is the
diamagnetic velocity of the plasma, $v_{dia}$ essentially induced by density
gradients. The distance of interaction between two elementary vortices is
modified with the factor%
\[
\left( 1-\frac{v_{dia}}{u}\right) ^{-1} 
\]%
since this has been proven that affects directly the Larmor radius,
replacing the physical parameter $\rho _{s}$ with an effective one, $\rho
_{s}^{eff}$,%
\[
\frac{1}{\left( \rho _{s}^{eff}\right) ^{2}}=\frac{1}{\rho _{s}^{2}}\left( 1-%
\frac{v_{dia}}{u}\right) 
\]%
It is this quantity that appears systematically when vortical motion is
examined in tokamak plasma, as illustrated below by the treatment for a
drift wave vortex.

Then we should accept that our field theoretical formulation, based on a
constant $\rho _{s}$ can only be valid on small patches, and there, with
actually the \emph{effective} Larmor radius. Then we have that the parameter 
$\kappa $, the coefficient of the Chern-Simons term in Lagrangian, has a
\textquotedblleft slow\textquotedblright\ spatial variation coming from the
spatial variation of $\rho _{s}^{eff}$, which, in turn is given by the
difference between the plasma \emph{rotation} velocity $u$ and the ion \emph{%
diamagnetic} velocity $v_{dia}$.

The first conclusion that we can draw at this moment is that an evolution of
the density profile that leads to an increase of the gradient (\emph{i.e.\ }$%
L_{n}$ smaller) will increase the diamagnetic velocity and then $v_{dia}$
tends to approach from below the plasma rotation velocity $u$. In
consequence the factor $\left( 1-\frac{v_{dia}}{u}\right) $ becomes smaller
and the effective Larmor radius increases. This is translated in the field
theoretical model in an increase of $\kappa $. If we follow the analogy
presented above, according to which the increase of $\kappa $ is equivalent
to $T_{m}\searrow T_{m}^{c}$, we find that there is an enhanced
clusterization of the elements of vorticity, with a possible evolution
toward a single filament in the center.

This process is not stationary since it has a positive feedback: the
clusterization of the vorticity toward the center drags the density
(basically from Ertel's theorem, but the process is more complex) and the
evolution of the density increases the gradient and consequently the
diamagnetic velocity. The effective Larmor radius increases and the
parameter $\kappa $ increases which still enhances the clusterization
process for the vortical elements.

\bigskip

However there is a limit to this process, coming exactly from the same
reason for which the vorticity and the density evolve together and exhibit
an inward pinch in tokamak.

The limit consists of the fact that the factor $1-v_{dia}/u$ ends up by
saturating the process of pinching, due to a too large increase of the
effective Larmor radius.

When the effective Larmor radius is too big, the interaction between the
elements of vorticity is no more of short range but can be considered of
long range. Or, this is the case of the Euler fluid, where the range of
interaction is Coulombian (\emph{i.e.} $ln$). For the Euler fluid there is
no compressibility of the background density, the density and the vorticity
are decoupled and the density cannot follow the vorticity. The
compressibility of the ion polarization drift is proportional with the \emph{%
inverse} of the square of the effective Larmor radius and this diminishes
accordingly. The fluid becomes less Hasegawa-Mima and more Euler. Actually
the state where $v_{dia}=u$ corresponds exactly to the two fluids:
Hasegawa-Mima (because the density is adiabatic and then $v_{dia}=u$) and
Euler (since there the range of interaction is infinite, Coulombian). But
the directions from which the two fluids arrive at this state are completely
different.

This intuitive representation of the density pinch (via the vorticity pinch)
may be useful. However the field theory introduces an additional factor in
this model: the fact that intrinsically the \emph{quasi-self-dual} solutions
have an adiabatic evolution related with the existence of a residual energy
which is \emph{not} of a topological nature.

For the Eq.(\ref{usplus}) the residual energy puts severe limitations to the
increase of the normalized value of the streamfunction. Any $\psi $ that is
greater than $\sim 3$ is strongly inhibited by a severe increase of this
residual energy. Or, the concentration in the center of the vorticity is in
general accompanied by an increase of $\psi $. We should note however that
for a normalized domain of integration $L=L^{phys}/\rho _{s}^{eff}$ of the
order greater than $\sim 10$ the value of $\psi $ is already smaller than $1$%
, and the limitations imposed by the residual energy are far. Moreover, the
graph of this residual energy shows that there is a favorable slow evolution
for the small values of $\psi $ since close to $\psi =0$ the energy is
decreasing with increasing $\psi $.

We still have to work on that.

\subsubsection{The role of the effective Larmor radius $\protect\rho %
_{s}^{eff}$}

In \textbf{Stationary vortices drift waves (Nycander)} it is derived the
equation 
\[
\Delta \phi =\left( \frac{1}{u}\frac{n_{0}^{\prime }}{n_{0}}+\frac{1}{\tau
\left( x\right) }\right) \phi 
\]%
where 
\[
\tau \left( x\right) \equiv \frac{T_{e}\left( x\right) }{T_{0}}
\]%
and the normalizations are usual: $\rho _{s}$ , $\Omega _{i}$ and $e/T_{e}$
for space, time, potential. From here we derive, taking constant temperature 
\begin{eqnarray*}
\frac{1}{u}\frac{n_{0}^{\prime }}{n_{0}}+\frac{1}{\tau \left( x\right) } &=&%
\frac{1}{u^{phys}\left( \frac{\Omega _{i}}{\rho _{s}}\right) }\frac{1}{n_{0}}%
\frac{dn_{0}}{dr^{phys}/\rho _{s}}+1 \\
&=&\frac{\rho _{s}^{2}}{\Omega _{i}}\left( \frac{dn_{0}}{n_{0}dr}\right)
^{phys}\frac{1}{u}+1 \\
&=&1+\frac{c_{s}\rho _{s}}{\Omega _{i}}\frac{1}{L_{n}}\frac{1}{u} \\
&=&1-\frac{v_{d}}{u}
\end{eqnarray*}%
where we defined 
\[
v_{d}\equiv -\frac{c_{s}\rho _{s}}{\Omega _{i}}\frac{1}{L_{n}}
\]%
such that when 
\[
L_{n}<0
\]%
we have 
\[
v_{d}>0
\]%
\textbf{The two velocities, }$v_{d}$\textbf{\ and }$u$\textbf{\ are in the
same direction (positive) when the density decreases radially, which is
normal in tokamak.} Therefore we have that 
\[
1-\frac{v_{d}}{u}>0
\]%
if the velocity of plasma rotation is higher than the diamagnetic rotation.
When the diamagnetic rotation becomes comparable with the plasma rotation
velocity, just slightly lower, this factor is almost zero and the \emph{%
effective Lamor radius} becomes infinite.

\bigskip 

\section{Conclusions}

This work is intended to provide arguments in favor of validity and
usefulness of the Field Theoretical approach to the description of fluids
and plasmas in the evolution towards asymptotic, coherent structures.

The basis material consisting of definition of the Lagrangians, equations of
motion and Self-Duality states are presented elsewhere and is not mentioned
here. The applications are only shortly presented.

We focused on various instruments, like the field-theoretical currents,
energy and action functional in close proximity of the self-duality. We have
noted that there are important aspects that can be examined using these
instruments, specific to the field theory: 

\begin{itemize}
\item the existence of the current of point-like vortices, which can explain
the concentration of vorticity

\item the slowing down of the evolution in the cuasi-asymptotic state, where
the number of vortices is still not the minimum that is obtained as the pure
extremum of the action at self-duality

\item the importance of parameters like the coefficient of the Chern-Simons
factor, in the adiabatic change of the regimes;

\item the role of the effective spatial dimension, which is adjusted
spontaneously via the gradients of the vorticity field.
\end{itemize}

Much remains to be investigated but there are already reasons to develop the
field theoretical formulations in a large class of fluid and plasma problems.

\begin{acknowledgement}
This work has been supported by the Romanian Ministry of Education and
Research under Ideas Exploratory Research Project No.557
\end{acknowledgement}

\end{document}